\documentclass[a4paper,11pt]{article}
\pdfoutput=1
\usepackage{jheppub}
\usepackage{amsmath,amssymb,euscript}
\usepackage{slashed}
\usepackage{xspace}
\usepackage{color}
\usepackage{accents}
\usepackage{hyperref}
\usepackage{epsfig}
\usepackage{xcolor}
\usepackage{verbatim}
\usepackage{multirow}
\usepackage{booktabs,graphicx}
\usepackage{mathtools}
\usepackage{tabulary}
\usepackage{dsfont}
\usepackage{parskip}
\usepackage{slashbox}
\usepackage{tcolorbox}
\usepackage{pifont}
\usepackage{graphicx}
\usepackage{booktabs}
\usepackage[normalem]{ulem}
\usepackage{wasysym}
\usepackage{subcaption}
\usepackage{textcomp}

\newcommand{\commentout}[1]{}

\newcolumntype{K}[1]{>{\centering\arraybackslash}p{#1}}

\emergencystretch=\maxdimen
\definecolor{darkturquoise}{rgb}{0.0, 0.81, 0.82}
\definecolor{amaranth}{rgb}{0.9, 0.17, 0.31}
\definecolor{awesome}{rgb}{1.0, 0.13, 0.32}
\definecolor{ballblue}{rgb}{0.13, 0.67, 0.8}
\definecolor{blue(munsell)}{rgb}{0.0, 0.5, 0.69}
\definecolor{cerulean}{rgb}{0.0, 0.48, 0.65}
\definecolor{darkcyan}{rgb}{0.0, 0.55, 0.55}
\definecolor{darklavender}{rgb}{0.45, 0.31, 0.59}
\definecolor{darkmagenta}{rgb}{0.55, 0.0, 0.55}
\definecolor{deepfuchsia}{rgb}{0.76, 0.33, 0.76}
\definecolor{emerald}{rgb}{0.31, 0.78, 0.47}
\definecolor{goldenpoppy}{rgb}{0.99, 0.76, 0.0}
\definecolor{green(pigment)}{rgb}{0.0, 0.65, 0.31}
\definecolor{jazzberryjam}{rgb}{0.65, 0.04, 0.37}
\definecolor{jade}{rgb}{0.0, 0.66, 0.42}
\definecolor{navy}{rgb}{0.0, 0.0, 0.5}
\definecolor{princetonorange}{rgb}{1.0, 0.56, 0.0}
\definecolor{richelectricblue}{rgb}{0.03, 0.57, 0.82}

\newcommand{\tev}{\, \mathrm{TeV}}

\newcommand{\GeV}{\, \mathrm{GeV}}

\renewcommand{\aa}{(aa)}
\newcommand{\ab}{(ab)}
\newcommand{\ba}{(ba)}

\newcommand{\ac}{(ac)}
\newcommand{\ca}{(ca)}
\newcommand{\bc}{(bc)}
\newcommand{\cb}{(cb)}
\newcommand{\aat}{(\widetilde{\phantom{aa}}\!\!\!\!\!\!aa)}
\newcommand{\abt}{(\widetilde{\phantom{aa}}\!\!\!\!\!\!ab)}
\newcommand{\bat}{(\widetilde{\phantom{aa}}\!\!\!\!\!\!ba)}

\newcommand{\act}{(\widetilde{\phantom{aa}}\!\!\!\!\!\!ac)}
\newcommand{\cat}{(\widetilde{\phantom{aa}}\!\!\!\!\!\!ca)}
\newcommand{\bct}{(\widetilde{\phantom{aa}}\!\!\!\!\!\!bc)}
\newcommand{\cbt}{(\widetilde{\phantom{aa}}\!\!\!\!\!\!cb)}

\newcommand{\aqqm}{A_{qq}^{(-)}}
\newcommand{\atqqm}{\widetilde{A}_{qq}^{(-)}}
\newcommand{\aqqth}{A_{qq}^{(3)}}
\newcommand{\atqqth}{\widetilde{A}_{qq}^{(3)}}
\newcommand{\Fqqm}{F_{qq}^{(-)}}

\newcommand{\Fqqth}{F_{qq}^{(3)}}

\newcommand{\bsg}{\ensuremath{{B\to X_s\gamma}}\xspace}
\newcommand{\bsmm}{\ensuremath{{B_s\to \mu^+\mu^-}}\xspace}

\newcommand{\wilson}{\texttt{wilson}\xspace}
\newcommand{\EOS}{\texttt{EOS}\xspace}

\begin{document}
\title{The Flavor of UV Physics}

\author[1]{Sebastian Bruggisser,}
\author[1]{Ruth Sch\"afer,}
\author[2]{Danny van Dyk,}
\author[1]{Susanne Westhoff\,}

\affiliation[1]{Institute for Theoretical Physics, Heidelberg University, 69120 Heidelberg, Germany}
\affiliation[2]{Physik Department, Technische Universit\"at M\"unchen, 85748 Garching, Germany}

\emailAdd{bruggisser@thphys.uni-heidelberg.de}
\emailAdd{schaefer\_r@thphys.uni-heidelberg.de}
\emailAdd{danny.van.dyk@gmail.com}
\emailAdd{westhoff@thphys.uni-heidelberg.de}

		\begin{flushright}
		P3H-21-002\\
        EOS-2021-01\\
        TUM-HEP 1304/20\\
		\end{flushright}
		
\abstract{New physics not far above the TeV scale should leave a pattern of virtual effects in observables at lower energies. What do these effects tell us about the flavor structure of a UV theory? Within the framework of Standard Model Effective Field Theory (SMEFT), we resolve the flavor structure of the Wilson coefficients in a combined analysis of top-quark and $B$-physics observables. Our fit to LHC and $b$-factory measurements shows that combining top and bottom observables is crucial to pin down possible sources of flavor symmetry breaking from UV physics. Our analysis includes the full analytic expansion of SMEFT coefficients in Minimal Flavor Violation and a detailed study of SMEFT effects in $b\to s$ flavor transitions.
}

\maketitle
\flushbottom

\clearpage


\section{Introduction}
\noindent In times of abundant data from collider experiments, precision measurements guide our search for physics beyond the Standard Model (BSM). If new particles exist at energies above the TeV scale, we expect to detect their virtual effects in observables at the LHC and at low-energy experiments, provided that these new particles couple sufficiently strongly to the Standard Model (SM). In general, any such UV completion of the Standard Model should manifest itself as a pattern of effects in several observables. To analyze these effects in a model-independent and gauge-invariant way, we consider the Standard Model as an Effective Field Theory (SMEFT) below the TeV scale and parametrize BSM effects as a series of local operators~\cite{Buchmuller:1985jz,Grzadkowski:2010es} (see Ref.~\cite{Brivio:2017vri} for a review). Global fits of the corresponding Wilson coefficients to data allow us to exploit the correlations of BSM effects that are characteristic of a specific UV theory.

Recent analyses aim at interpreting global SMEFT fits in terms of concrete UV extensions of the Standard Model~\cite{Dawson:2020oco,David:2020pzt,Ellis:2020unq} or classify UV completions by exploiting the symmetries of the effective theory~\cite{Zhang:2020jyn,Chakrabortty:2020mbc}. Whatever the approach, the key question is: What can we learn from data about the possible structure of UV physics? By resolving the parameter space of SMEFT Wilson coefficients, we strive to pin down features of a UV theory that are in overall agreement with data. Such features are, for instance, the charges of SM particles under a new force, the Lorentz and gauge structure of new interactions, or global symmetries. In this work we focus on the flavor structure. We want to establish whether new interactions must be flavor-universal or might introduce new sources of flavor symmetry breaking.

What do SMEFT fits tell us about the flavor structure of a UV theory?

In the Standard Model the main source of flavor symmetry breaking is the top Yukawa coupling. CKM mixing in weak interactions is physical because the top Yukawa coupling sets a reference direction in flavor space~\cite{Feldmann:2008ja}. Phenomenologically, the top quark has played a crucial role in pinning down the flavor structure of the Standard Model: Flavor-changing neutral currents among down quarks are driven by virtual effects of the top quark in loops; they have been studied in great detail in rare meson decays~\cite{Buchalla:1995vs}. Flavor-changing neutral currents among up quarks are suppressed by the small flavor breaking in the down-quark sector~\cite{Eilam:1990zc,Mele:1998ag}; the non-observation of exotic top decays at the LHC is in agreement with this prediction. The apparent absence of additional flavor violation has led to the concept of Minimal Flavor Violation (MFV)~\cite{Buras:2000dm,DAmbrosio:2002vsn}, which extrapolates the flavor structure of the Standard Model to new interactions. Alternative scenarios postulate flavor universality among all three quark generations or among light quarks~\cite{Barbieri:2012uh,Efrati:2015eaa,AguilarSaavedra:2018nen,Ellis:2020unq,Faroughy:2020ina}.

To pin down the flavor structure of Wilson coefficients in the SMEFT, we rely on observations of flavor violation among SM particles. Flavor-changing rare meson decays or meson mixing are extremely sensitive to BSM effects~\cite{Hewett:1993em,Grzadkowski:2008mf,Kamenik:2011dk,Drobnak:2011aa,Brod:2014hsa,Aebischer:2018iyb,Silvestrini:2018dos,Aebischer:2020dsw} and measured with high precision. Although such observables tell us much about flavor in BSM physics, they often probe combinations of several Wilson coefficients. This leaves blind directions in the SMEFT parameter space, which can be resolved by connecting flavor observables with high-energy observables at the LHC, Tevatron or LEP~\cite{Fox:2007in,Efrati:2015eaa,Cirigliano:2016nyn,Alioli:2017ces,Biekotter:2018rhp,Bissmann:2019gfc,Falkowski:2019hvp,Aoude:2020dwv,Bissmann:2020mfi}.

In our analysis we exploit the connection between rare $B$ decays and top-quark production at the LHC in SMEFT. This choice is motivated by the high experimental precision in both sectors, as well as the key role of the top in probing quark flavor transitions. Moreover, top production at the LHC provides us with a plethora of observables~\cite{Zhang:2010dr,Kamenik:2011dk,Faller:2013gca,Rontsch:2014cca,Rontsch:2015una,Bylund:2016phk}, which are known to effectively explore the SMEFT parameter space~\cite{Buckley:2015lku,Brivio:2019ius,Hartland:2019bjb,Durieux:2019rbz}. The connection between top and bottom observables is established by matching the SMEFT coefficients onto the Weak Effective Theory (WET)~\cite{Aebischer:2015fzz,Jenkins:2017jig,Dekens:2019ept,Hurth:2019ula} and evolving the WET coefficients to the scale of $B$ physics using the Renormalization Group (RG)~\cite{Aebischer:2017gaw,Jenkins:2017dyc}. In this framework the relative BSM effects in bottom and top observables strongly depend on the flavor structure of the SMEFT coefficients. This renders the top-bottom connection highly sensitive to the flavor structure of the underlying UV theory. In this work we will assume Minimal Flavor Violation for the SMEFT coefficients, but our strategy could equally well be applied to other scenarios of flavor breaking.

The goal of our work is to pin down the flavor structure of the SMEFT coefficients in MFV in a combined fit of top and bottom observables. In Sec.~\ref{sec:framework}, we introduce the relevant set of SMEFT operators and expand the Wilson coefficients in MFV in terms of spurions. In Sec.~\ref{sec:flavor}, we propagate the flavor structure to the WET coefficients and perform a detailed analysis of SMEFT effects in $b\to s$ transitions. In particular, we study the interplay of tree-level and one-loop effects in rare meson decays. In Sec.~\ref{sec:top}, we analyze the flavor structure of SMEFT contributions in top observables and find an unexpectedly good sensitivity to flavor breaking in electroweak top processes. Sec.~\ref{sec:results} is devoted to our statistical analysis of SMEFT coefficients in a combined top-flavor fit. We show that the combination of top and bottom observables strongly constrains flavor breaking in BSM interactions and how their interplay resolves otherwise blind directions in flavor space. We conclude in Sec.~\ref{sec:conclusions} with an outlook to future research directions opened up by our analysis strategy.


\section{Effective operators and their flavor structure}\label{sec:framework}
\noindent In theories where the mass scale $\Lambda$ of BSM physics lies above the reach of current particle physics experiments, we can describe observable effects of such BSM effects by an effective quantum field theory with Standard Model particles as dynamical degrees of freedom. In our work, we use two kinds of effective theories: at energy scales above the weak scale $\mu = m_Z$, relevant for top and Higgs processes, we work with the Standard Model Effective Field Theory (SMEFT)~\cite{Buchmuller:1985jz}. At the weak scale, we match the SMEFT onto the Weak Effective Theory (WET)~\cite{Aebischer:2017gaw,Jenkins:2017jig}, which is the appropriate framework to describe new physics effects in flavor observables at low scales $\mu = m_b$.

In the SMEFT the effective Lagrangian reads
\begin{align}\label{eq:smeft-lagrangian}
\mathcal{L}_{\rm SMEFT} = \sum_a \frac{C_a}{\Lambda^2}\,O_a + \sum_b \left( \frac{C_b}{\Lambda^2}\,{^\ddagger O_b} + h.c. \right) + \dots,
\end{align}
where the sum is over all dimension-six operators with external quark fields. The dots denote higher dimensional operators, which we do not consider in our analysis. We also do not include operators with leptons, assuming that new physics couples dominantly to quarks. All operators are defined in the Warsaw basis~\cite{Grzadkowski:2010es}. Here and below in Eqs.~\eqref{eq:2q-operators} and \eqref{eq:4q-operators} non-hermitian operators are marked as $^\ddagger O$.

\subsection{SMEFT operators}\label{sec:operators}
In our analysis we focus on 23 operators that capture all effects of effective quark interactions in the top and bottom observables we consider. Our set of operators includes 11 operators with two quark fields
\begin{align}\label{eq:2q-operators}
O_{\phi q}^{(1),kl} & = (\phi^\dagger \stackrel{\longleftrightarrow}{iD_\mu} \phi)(\overline{Q}^k\gamma^\mu Q^l) & ^\ddagger O_{uB}^{kl} & = (\overline{Q}^k\sigma^{\mu\nu} U^l)\,\widetilde{\phi}\,B_{\mu\nu} \\\nonumber
O_{\phi q}^{(3),kl} & = (\phi^\dagger \stackrel{\longleftrightarrow}{iD_\mu^I} \phi)(\overline{Q}^k\gamma^\mu\tau^I Q^l) & ^\ddagger O_{uW}^{kl} & = (\overline{Q}^k\sigma^{\mu\nu} U^l)\,\tau^I\widetilde{\phi}\,W^I_{\mu\nu} \\\nonumber
O_{\phi u}^{kl} & = (\phi^\dagger \stackrel{\longleftrightarrow}{iD_\mu} \phi)(\overline{U}^k\gamma^\mu U^l) & ^\ddagger O_{uG}^{kl} & = (\overline{Q}^k\sigma^{\mu\nu} T^A U^l)\,\widetilde{\phi}\,G^A_{\mu\nu} \\\nonumber
O_{\phi d}^{kl} & = (\phi^\dagger \stackrel{\longleftrightarrow}{iD_\mu} \phi)(\overline{D}^k\gamma^\mu D^l) & ^\ddagger O_{dB}^{kl} & = (\overline{Q}^k\sigma^{\mu\nu} D^l)\,\widetilde{\phi}\,B_{\mu\nu} \\\nonumber
^\ddagger O_{\phi ud}^{kl} & = (\widetilde{\phi}^\dagger \stackrel{\longleftrightarrow}{iD_\mu} \phi)(\overline{U}^k\gamma^\mu D^l) & ^\ddagger O_{dW}^{kl} & = (\overline{Q}^k\sigma^{\mu\nu} D^l)\,\tau^I\phi\,W^I_{\mu\nu} \\\nonumber
& & ^\ddagger O_{dG}^{kl} & = (\overline{Q}^k\sigma^{\mu\nu} T^A D^l)\,\widetilde{\phi}\,G^A_{\mu\nu}
\end{align}
and 12 operators with four quark fields
\begin{align}\label{eq:4q-operators}
O_{qq}^{(1),klmn} & = (\overline{Q}^k\gamma^\mu Q^l)(\overline{Q}^m\gamma_\mu Q^n) & O_{uu}^{klmn} & = (\overline{U}^k\gamma^\mu U^l)(\overline{U}^m\gamma_\mu U^n) \\\nonumber
O_{qq}^{(3),klmn} & = (\overline{Q}^k\gamma^\mu \tau^I Q^l)(\overline{Q}^m\gamma_\mu \tau^I Q^n) & O_{dd}^{klmn} & = (\overline{D}^k\gamma^\mu D^l)(\overline{D}^m\gamma_\mu D^n) \\\nonumber
O_{qu}^{(1),klmn} & = (\overline{Q}^k\gamma^\mu Q^l)(\overline{U}^m\gamma_\mu U^n) & O_{ud}^{(1),klmn} & = (\overline{U}^k\gamma^\mu U^l)(\overline{D}^m\gamma_\mu D^n) \\\nonumber
O_{qu}^{(8),klmn} & = (\overline{Q}^k\gamma^\mu T^A Q^l)(\overline{U}^m\gamma_\mu T^A U^n) & O_{ud}^{(8),klmn} & = (\overline{U}^k\gamma^\mu T^A U^l)(\overline{D}^m\gamma_\mu T^A D^n) \\\nonumber
O_{qd}^{(1),klmn} & = (\overline{Q}^k\gamma^\mu Q^l)(\overline{U}^m\gamma_\mu U^n) & ^\ddagger O_{quqd}^{(1),klmn} & = (\overline{Q}^{k}_I\,U^l)\,\epsilon_{IJ}\,(\overline{Q}^{m}_J\,D^n) \\\nonumber
O_{qd}^{(8),klmn} & = (\overline{Q}^k\gamma^\mu T^A Q^l)(\overline{U}^m\gamma_\mu T^A U^n) &
^\ddagger O_{quqd}^{(8),klmn} & = (\overline{Q}^{k}_I\,T^A U^l)\,\epsilon_{IJ}\,(\overline{Q}^{m}_J\,T^A D^n)\,.
\end{align}
The generators of weak and strong interactions are denoted as $\tau^I$ and $T^A$. The quark fields are interaction eigenstates of left-handed quark doublets $Q^k$, right-handed up-type quarks $U^k$, and right-handed down-type quarks $D^k$. The indices $k,l,m,n \in \{1,2,3\}$ denote the three quark generations.

In general, the Wilson coefficients $C^{kl},\,C^{klmn}$ of the operators $O^{kl},\,O^{klmn}$ with different flavor indices are independent parameters. However, for the four-quark operators $O_{qq}^{(1)}$, $O_{qq}^{(3)}$, $O_{uu}^{(1)}$ and $O_{dd}^{(1)}$ they fulfill
\begin{align}
    C^{klmn} = C^{mnkl},
\end{align}
because the two quark bilinears are identical in their gauge and Lorentz structure. In the Lagrangian in Eq.~\eqref{eq:smeft-lagrangian} the sum over generations is implicit. We include independent degrees of freedom, \emph{i.e.}, we keep $C^{klmn}$ but discard $C^{mnkl}$ in the sum over flavor indices. We also assume that the effective Lagrangian preserves CP symmetry, which implies that all Wilson coefficients are real.

In a concrete extension of the Standard Model the flavor structure of the Wilson coefficients is determined by the fundamental interactions of the new particles. By \emph{flavor structure} we mean a specific pattern of the Wilson coefficients in flavor space, \emph{i.e.}, the form of the $3\times 3$ matrix $C^{kl}$ or the $3 \times 3\times 3 \times 3$ tensor $C^{klmn}$. Any concrete flavor structure leads to a certain pattern of effects in low-energy observables. One of our main goals is to analyze these effects in top and bottom observables and to deduce the possible flavor structures of a UV theory by comparing the predictions to data. For concreteness we consider the scenario of Minimal Flavor Violation, which is phenomenologically motivated by the strong constraints from existing measurements. Minimal Flavor Violation is based on the assumption that the Yukawa couplings are the only sources of flavor symmetry breaking in and beyond the Standard Model. Scenarios with extra sources of flavor breaking can lead to a different phenomenology; we comment on such scenarios in the course of our analysis. In the remainder of this section, we briefly review the main concept of MFV and derive the resulting flavor degrees of freedom in the effective Lagrangian.

\subsection{Minimal flavor violation}\label{sec:mfv}
\noindent In the Standard Model, the gauge interactions of quarks respect the flavor symmetry
\begin{align}
G_F = U(3)_Q \times U(3)_U\times U(3)_D\,,
\end{align}
which, however, is broken by the Yukawa couplings $Y_U,Y_D$. Under the assumption of MFV, the Yukawa couplings are also the only sources of flavor symmetry breaking in extensions of the Standard Model~\cite{Buras:2000dm,DAmbrosio:2002vsn}. By treating them as \emph{spurions}, \emph{i.e.}, as fictitious fields transforming under $G_F$ as
\begin{align}
Y_U:\ (3,\overline{3},1)\,,\qquad Y_D:\ (3,1,\overline{3})\,,
\end{align}
we can build flavor structures for quark bilinears that are $G_F$ singlets. We define three spurions as complex $3\times 3$ matrices in flavor space, transforming under $G_F$ as
\begin{align}
\mathcal{A}_Q:\ (3\times \overline{3},1,1),\qquad \mathcal{A}_U:\ (1,3\times \overline{3},1),\qquad \mathcal{A}_D:\ (1,1,3\times \overline{3}).
\end{align}
Expanded in terms of the Yukawa matrices and using $3\times \overline{3} = 1+8$, they read
\begin{align}\label{eq:mfv-expansion}
\mathcal{A}_Q & = a\, {\bf 1} + b\,Y_U Y_U^\dagger + c\,Y_D Y_D^\dagger + \dots\\\nonumber
\mathcal{A}_U & = a\, {\bf 1} + b\,Y_U^\dagger Y_U + c\,Y_U^\dagger Y_D Y_D^\dagger Y_U + \dots\\\nonumber
\mathcal{A}_D & = a\, {\bf 1} + b\,Y_D^\dagger Y_U Y_U^\dagger Y_D + c\,Y_D^\dagger Y_D + \dots
\end{align}
Here $a,b,c$ are free parameters for each of the spurions, and we have kept only the leading terms in $Y_U,Y_D$. The flavor structures of all relevant quark bilinears can now be expressed as
 \begin{align}\label{eq:bilinears}
 (\mathcal{A}_Q)_{kl}\ & (\overline{Q}^k\gamma_\mu\,Q^l) & (\mathcal{A}_U)_{kl}\ & (\overline{U}^k\gamma_\mu\, U^l) \\\nonumber
 (\mathcal{A}_Q Y_U)_{kl}\ & (\overline{Q}^k\sigma_{\mu\nu} U^l) & (\mathcal{A}_D)_{kl}\ & (\overline{D}^k\gamma_\mu\, D^l) \\\nonumber
(\mathcal{A}_Q Y_D)_{kl}\ & (\overline{Q}^k\sigma_{\mu\nu} D^l) & (Y_U^\dagger \mathcal{A}_Q Y_D)_{kl}\ & (\overline{U}^k\gamma_\mu\, D^l)\,.
 \end{align}
The Yukawa couplings are diagonalized by unitary matrices $\mathcal{U}_{R,L}$ and $\mathcal{D}_{R,L}$, so that
\begin{align}\label{eq:yukawa-rot}
Y_U = \mathcal{U}_L Y_u\, \mathcal{U}_R^\dagger\,,\quad Y_D = \mathcal{D}_L Y_d\, \mathcal{D}_R^\dagger\,,\quad V = \mathcal{U}_L^\dagger \mathcal{D}_L\,.
\end{align}
Here $Y_u = \text{diag}(y_u,y_c,y_t)$, $Y_d = \text{diag}(y_d,y_s,y_b)$ are the physical Yukawa couplings, and $V$ is the CKM matrix. Without losing generality, we work in the \emph{up mass basis} where the gauge eigenstates of left-handed quarks are aligned with the mass eigenstates of up-type quarks,
\begin{align}\label{eq:up-alignment}
Q^k = \begin{pmatrix} u_L^k \\V_{kl} d^l_L \end{pmatrix},\quad U^k = u_R^k\,,\quad D^k = d_R^k\,.
\end{align}
Here $u^k$ and $d^k$ are the mass eigenstates. For the Yukawa matrices this implies
\begin{align}
 Y_U = Y_u\,, \quad Y_D = V Y_d\,.
\end{align}
We will refer to this setup as \emph{up-alignment}. Notice that the choice of alignment does not affect observables as long as $Y_U$ and $Y_D$ are the only sources of flavor breaking. Working in the down mass basis and aligning the Wilson coefficients with the down-type Yukawa couplings would lead to the same physical effects.

For operators with two quark fields, we can directly infer the flavor structure of the Wilson coefficients from Eq.~\eqref{eq:mfv-expansion}. For instance, for the coefficient of $O_{\phi u}^{kl}$ the most general flavor structure in MFV is
\begin{align}\label{eq:cphiu}
C_{\phi u}^{kl} & = a_{\phi u}\,\delta_{kl} + b_{\phi u}\, (Y_u^\dagger Y_u)_{kl} + c_{\phi u}\, (Y_u^\dagger VY_d (VY_d)^\dagger Y_u)_{kl} + \dots\\\nonumber
& = a_{\phi u}\,\delta_{kl} + b_{\phi u}\,y_t^2 \,\delta_{k3}\,\delta_{l3} + c_{\phi u} y_t^2y_b^2 V_{tb}V_{tb}^\ast\delta_{k3}\delta_{l3} + \mathcal{O}(y_c^2)\\\nonumber
& = a_{\phi u}\,\delta_{kl} + b_{\phi u}\,y_t^2 \,\delta_{k3}\,\delta_{l3} + \mathcal{O}(y_b^2)\,.
\end{align}
In the second line, we have assumed up-alignment and have only kept the leading contributions in the top and bottom Yukawa couplings and neglected the Yukawa couplings of light quarks. In the third line, we have also neglected the contribution $y_t^2 y_b^2$ to the Wilson coefficient $C_{\phi u}^{33}$, which does not introduce a new flavor pattern, but merely adds a correction to the leading $y_t^2$ term. We will use this approximation throughout this work. In Tab.~\ref{tab:2q-mfv}, we show the Wilson coefficients of all two-quark operators from Eq.~\eqref{eq:2q-operators} in MFV with up-alignment. Notice that the flavor parameters $a,b,c$ are defined specifically for each operator, \emph{i.e.}, $a$ denotes $a_{\phi q}^{(1)}$, $a_{\phi q}^{(3)}$ \emph{etc.}, respectively.
\begin{table}[tp]\begin{center}
\renewcommand{\arraystretch}{1.2}
\setlength{\tabcolsep}{1.7mm}
\begin{tabular}{c|c|c|c|c|c|c|c}
    \toprule
    & $C_{\phi q}^{(1)}$ & $C_{\phi q}^{(3)}$ & $C_{\phi u}$ & $C_{\phi d}$ & $C_{\phi ud}$ & $C_{uX}$ & $C_{dX}$\\\midrule
    $ii$ & $a$ & $a$ & $a$ & $a$ & 0 & 0 & 0 \\
    $33$ & $a+b y_t^2$ & $a+b y_t^2$ & $a+b y_t^2$ & $a$ & $(a +b y_t^2)y_by_tV_{tb}$ & $(a+b y_t^2)y_t$ & $(a +b y_t^2)y_bV_{tb}$\\
    $ki$ & $cy_b^2V_{kb}V_{ib}^\ast$ & $cy_b^2V_{kb}V_{ib}^\ast$ & 0 & 0 & 0 & 0 & 0\\
    $i3$ & $cy_b^2V_{ib}V_{tb}^\ast$ & $cy_b^2V_{ib}V_{tb}^\ast$ & 0 & 0 & 0 & $c y_b^2y_tV_{ib}V_{tb}^\ast$ & $ay_{b}V_{ib}$\\
\midrule
    \# & 3 & 3 & 2 & 1 & 1 & 2 & 2\\\bottomrule
\end{tabular}
\end{center}
\caption{Wilson coefficients of two-quark SMEFT operators in MFV with up-alignment. The flavor indices are $i \in \{1,2\}$ and $k \in \{1,2,3\}$ with $k\neq i$, and $X = B,W,G$ denotes the respective gauge field. Subleading contributions in $y_b$ and contributions of $\mathcal{O}(y_c)$ or smaller have been neglected. The last line shows the number of independent degrees of freedom in this expansion of the flavor structures.}
\label{tab:2q-mfv}
\end{table}

 Due to the large top Yukawa coupling, contributions of $\mathcal{O}(y_t^n)$ can be sizeable. However, in two-quark operators they do not lead to new degrees of freedom in the observables we consider: Either the flavor structure is identical to the leading contribution in $y_t$, or it is suppressed by light-quark Yukawa couplings and its effect in the observables we consider is negligible. We do not include these higher powers of $y_t$ explicitly, but they could be restored in reinterpretations of our numerical results.\footnote{For a systematic treatment of $\mathcal{O}(y_t^n)$ contributions, see Refs.~\cite{Feldmann:2008ja,Kagan:2009bn}.}
 
The requirement of MFV imprints itself as a pattern on the Wilson coefficients in flavor space. For operators with a single right-handed down-type quark, the effective coupling is suppressed by the small bottom Yukawa coupling.
Flavor-changing neutral currents (FCNCs) are suppressed by CKM mixing. At tree level, FCNCs among left-handed down-type quarks are proportional to $y_t^2$, while FCNCs among up-type quarks are further suppressed by $y_b^2$. As a result of this expansion, each Wilson coefficient $C^{kl}$ in MFV can be parametrized by at most $3$ parameters, as we show in the last line of Tab.~\ref{tab:2q-mfv}.

\begin{table}[tp]
\renewcommand{\arraystretch}{1.3}
\resizebox{\textwidth}{!}{
\begin{tabular}{c|l|l|l}
    \toprule
     & $C_{qq}^{(1)},\,C_{qq}^{(3)}$ & $C_{uu}$ & $C_{dd}$\\\midrule
     $iiii$ & $\aa +\aat$ & $\aa +\aat$ & $\aa +\aat$\\
     $iijj$ & $\aa$ & $\aa$ & $\aa$ \\
     $ijji$ & $\aat$ & $\aat$& $\aat$ \\
     $33ii$ & $\aa+\ba y_t^2$
     & $\aa+\ba y_t^2$
     & $\aa$ \\
     $3ii3$ & $\aat+\bat y_t^2$ & $\aat+\bat y_t^2$ & $\aat$ \\
     $3333$ & $\aa+\aat + 2 (\ba +\bat) y_t^2 +\mathcal{O}(y_t^4)$ & $\aa+\aat + 2 (\ba + \bat) y_t^2 +\mathcal{O}(y_t^4)$ & $\aa+\aat$\\
     $iikl$ & $\ac y_b^2V_{kb}V_{l b}^\ast $ & 0 & 0 \\
     $il k i$ & $\act y_b^2 V_{kb} V_{l b}^\ast$ & 0 & 0 \\
     $33kl$ & $(\ac+\bc y_t^2)y_b^2V_{kb}V_{l b}^\ast$ & 0 & 0 \\
     $3lk3$ & $(\act+\bct y_t^2)y_b^2 V_{k b} V_{l b}^\ast$ & 0 & 0 \\\midrule
     \# & 9 & 5 & 2\\\bottomrule
     \end{tabular}}
     
     \vspace*{0.3cm}
     
     \renewcommand{\arraystretch}{1.3}
     \setlength{\tabcolsep}{2mm}
     \resizebox{\textwidth}{!}{
     \begin{tabular}{c|l|l|l|l}
    \toprule
     & $C_{qu}^{(1)},\,C_{qu}^{(8)}$ & $C_{qd}^{(1)},\,C_{qd}^{(8)}$ & $C_{ud}^{(1)},\,C_{ud}^{(8)}$ & $C_{quqd}^{(1)},\,C_{quqd}^{(8)}$\\\midrule
     $iiii$ & $\aa$ & $\aa$ & $\aa$ & 0\\
         $iijj$ & $\aa$ & $\aa$ & $\aa$ & 0 \\
     $ii33$ & $\aa+\ab y_t^2$ & $\aa$ & $\aa$ & 0 \\
     $33ii$ & $\aa+\ba y_t^2$ & $\aa+\ba y_t^2$ & $\aa+\ba y_t^2$ & 0\\
     $\!\begin{array}{l} 3333 \\
     \phantom{ }\\
     \phantom{ }\end{array}$ & $\!\begin{array}{l}\aa \\+ (\ab+\ba+\aat)y_t^2\\  + \mathcal{O}(y_t^4)\end{array}$ & $\!\begin{array}{l}\aa + \ba y_t^2\\
     \phantom{ }\\
     \phantom{ }\end{array}$ & $\!\begin{array}{l}\aa + \ba y_t^2\\
     \phantom{ }\\
     \phantom{ }\end{array}$ & $\!\begin{array}{l}\big(\aa+\aat + \ab y_t^2\\
     + (\ba+\abt+\bat) y_t^2\big)\\ \times y_by_tV_{tb} + \mathcal{O}(y_t^4)\end{array}$ \\
     $klii$ & $\ca y_b^2V_{kb}V_{l b}^\ast$ & $\ca y_b^2V_{kb}V_{l b}^\ast$ & 0 & 0 \\
     $ij33$ & $(\ca+\cb y_t^2)y_b^2V_{ib}V_{jb}^\ast$ & $(\ca+\aat)y_b^2 V_{ib} V_{jb}^\ast$ 
     & 0 & 0 \\
     $\!\begin{array}{l} i333 \\
     \phantom{ }\end{array}$ & $\!\begin{array}{l}\big(\ca + (\cb + \cat)y_t^2\big) \\ \times y_b^2V_{ib}V_{tb}^\ast + \mathcal{O}(y_t^4)\end{array}$ &
      $\!\begin{array}{l}\big(\ca + \aat +\abt y_t^2\big)\\ \times y_b^2 V_{ib}V_{tb}^\ast \end{array}$
      & $\!\begin{array}{l} 0 \\
     \phantom{ }\end{array}$ & 
      $\!\begin{array}{l}(\aat + \abt y_t^2) y_b y_t V_{ib}\\
      \phantom{ }\end{array}$\\
     $\!\begin{array}{l} 3i33 \\
     \phantom{ }\end{array}$ & $\!\begin{array}{l}\big(\ca + ( \cb + \act)y_t^2\big)\\\times y_b^2 V_{tb}V_{ib}^\ast + \mathcal{O}(y_t^4)\end{array}$
      & $\!\begin{array}{l} \big(\ca + \aat + \bat y_t^2\big)\\ \times y_b^2 V_{tb} V_{ib}^\ast\end{array}$ & $\!\begin{array}{l} 0 \\
     \phantom{ }\end{array}$ & $\!\begin{array}{l} 0 \\
     \phantom{ }\end{array}$ \\
     $33i3$ & 0 & 0 & 0 & $(\aa + \ba y_t^2) y_b y_t V_{ib}$
     \\\midrule
     \# & 8 & 6 & 2 & 3
     \\\bottomrule
\end{tabular}}
\caption{Wilson coefficients of four-quark SMEFT operators in MFV with up-alignment. The flavor indices are $i,j \in \{1,2\}$ with $i\neq j$ and $k,l \in \{ 1,2,3\}$ with $k\neq l \neq i$.
 Subleading contributions in $y_b$ and contributions of $\mathcal{O}(y_c)$ or smaller have been neglected. Terms of $\mathcal{O}(y_t^4)$ are indicated where they introduce additional flavor parameters. The last line of each table shows the number of independent flavor degrees of freedom in the Wilson coefficients up to $\mathcal{O}(y_b^2)$.  The Wilson coefficients for all other combinations of flavor indices vanish in our expansion.}
\label{tab:4q-mfv}
\end{table}

For operators $O^{klmn}$ with four quark fields, the flavor structure factorizes into two quark-antiquark bilinears. In general there are two ways to form these bilinears, namely $(kl)(mn)$ and $(kn)(ml)$. Using Eq.~\eqref{eq:bilinears}, we obtain the flavor structures
\begin{align}\label{eq:4f-2}
\big[ (\mathcal{A}_Q)_{kl} (\mathcal{A}_Q)_{mn} + (\widetilde{\mathcal{A}}_Q)_{kn}(\widetilde{\mathcal{A}}_Q)_{ml} \big]\ & (\overline{Q}^k\gamma^\mu\,Q^l)(\overline{Q}^m\gamma_\mu\,Q^n) \\\nonumber
\big[ (\mathcal{A}_U)_{kl}(\mathcal{A}_U)_{mn} + (\widetilde{\mathcal{A}}_U)_{kn}(\widetilde{\mathcal{A}}_U)_{ml} \big]\ & (\overline{U}^k\gamma^\mu\,U^l)(\overline{U}^m\gamma_\mu\,U^n) \\\nonumber
\big[ (\mathcal{A}_D)_{kl}(\mathcal{A}_D)_{mn} + (\widetilde{\mathcal{A}}_D)_{kn}(\widetilde{\mathcal{A}}_D)_{ml} \big]\ & (\overline{D}^k\gamma^\mu\,D^l)(\overline{D}^m\gamma_\mu\,D^n) \\\nonumber
\big[(\mathcal{A}_Q)_{kl}(\mathcal{A}_U)_{mn} + (\widetilde{\mathcal{A}}_Q Y_U)_{kn}(Y_U^\dagger \widetilde{\mathcal{A}}_Q^\dagger)_{ml}\big]\ & (\overline{Q}^k\gamma^\mu\,Q^l)(\overline{U}^m\gamma_\mu\,U^n) \\\nonumber
\big[(\mathcal{A}_Q)_{kl}(\mathcal{A}_D)_{mn}  + (\widetilde{\mathcal{A}}_Q Y_D)_{kn}(Y_D^\dagger \widetilde{\mathcal{A}}_Q^\dagger)_{ml} \big]\ & (\overline{Q}^k\gamma^\mu\,Q^l)(\overline{D}^m\gamma_\mu\,D^n) \\\nonumber
\big[(\mathcal{A}_U)_{kl}(\mathcal{A}_D)_{mn} + (Y_U^\dagger \widetilde{\mathcal{A}}_Q Y_D)_{kn}(Y_D^\dagger \widetilde{\mathcal{A}}_Q^\dagger Y_U)_{ml}\big] \ & (\overline{U}^k\gamma^\mu\,U^l)(\overline{D}^m\gamma_\mu\,D^n) \\\nonumber
\big[ (\mathcal{A}_Q Y_U)_{kl}(\mathcal{A}_Q Y_D)_{mn} + (\widetilde{\mathcal{A}}_Q Y_D)_{kn}(\widetilde{\mathcal{A}}_Q Y_U)_{ml} \big] \ & (\overline{Q}^k\,U^l)(\overline{Q}^m\,D^n)\,,
\end{align}
where we have denoted the two flavor contractions $(kl)(mn)$ and $(kn)(ml)$ by $\mathcal{A}$ and $\widetilde{\mathcal{A}}$ and suppressed the gauge structure of the operators. Using Eq.~\eqref{eq:4f-2} and the expansions of quark bilinears from Tab.~\ref{tab:2q-mfv}, we determine the flavor structure of the four-quark coefficients $C^{klmn}$. We define the corresponding flavor parameters $\aa$, $\aat$ \emph{etc.} by expanding the spurions as
\begin{align}
    (\mathcal{A}_Q)_{kl} (\mathcal{A}_Q)_{mn} & = \aa \delta_{kl}\delta_{mn} + \ba y_t^2 \delta_{k3}\delta_{l3}\delta_{mn} + \ab y_t^2 \delta_{kl}\delta_{m3}\delta_{n3} + \dots\\\nonumber
    (\widetilde{\mathcal{A}}_Q)_{kn} (\widetilde{\mathcal{A}}_Q)_{ml} & = \aat \delta_{kn}\delta_{ml} + \bat y_t^2 \delta_{k3}\delta_{n3}\delta_{ml} + \abt y_t^2 \delta_{kn}\delta_{m3}\delta_{l3} + \dots,
\end{align}
and similarly for the remaining flavor structures. In Tab.~\ref{tab:4q-mfv}, we show the Wilson coefficients of all four-quark operators from Eq.~\eqref{eq:4q-operators} in MFV.

We show only the leading combinations of flavor parameters, as they appear in the Lagrangian of Eq.~\eqref{eq:smeft-lagrangian}.
Higher powers of $y_t$ are included in those Wilson coefficients where they change the flavor structure compared to the leading contribution in $y_t$.
Compared to the two-quark operators in Tab.~\ref{tab:2q-mfv}, the flavor structure of four-quark coefficients in MFV is much richer, with up to 9 independent flavor degrees of freedom. As for two-quark operators, FCNC interactions of left-handed up-type quarks $\ca,\cat,\cb,\cbt$ are suppressed by $y_b^2$ and CKM elements. All interactions with right-handed down quarks are proportional to $y_b$.


\section{Flavor effects in bottom observables}\label{sec:flavor}
The main goal of our analysis is to combine top-quark observables at the LHC with flavor observables in SMEFT, to decipher possible flavor structures of the underlying UV theory and to resolve existing blind directions in top fits. In the top sector, we build on the results from a recent global analysis of LHC observables~\cite{Brivio:2019ius}. On the flavor side, we do not aim at a comprehensive analysis, but rather select a small set of observables that probe new directions in the SMEFT parameter space. We focus on $b\to s$ decays, which are loop-suppressed in the Standard Model and thus particularly sensitive to new physics. The dominant top-quark loop allows us to probe top interactions through virtual effects in $b\to s$ transitions. Later in Sec.~\ref{sec:results}, we combine the results with tree-level effects in top production at the LHC.

We begin this section by discussing the Weak Effective Theory for $b\to s \gamma$ and $b\to s \ell^+\ell^-$ decays at the parton level. In Sec.~\ref{sec:b-observables} we discuss WET effects in $b\to s$ observables as they arise from a general SM-invariant UV completion. In Sec.~\ref{sec:matching}, we describe the matching of SMEFT to WET, focusing on the flavor structure of the Wilson coefficients in WET.

\subsection{Weak effective theory of $b\to s$ transitions}\label{sec:wet}
At low energies $E \ll m_Z$, parton-level interactions of the five light quarks $u,d,s,c,b$, photons and gluons are described by the Lagrangian
\begin{align}
    \mathcal{L} = \mathcal{L}_\text{QED} + \mathcal{L}_\text{QCD} + \mathcal{L}_\text{WET}\,.
\end{align}
Here $\mathcal{L}_\text{QED}$ and $\mathcal{L}_\text{QCD}$ describe electromagnetic and strong interactions in the Standard Model, while $\mathcal{L}_\text{WET}$ describes the effective theory of weak interactions at low energies, both within and beyond the Standard Model. The WET Lagrangian reads
\begin{align}\label{eq:wet-lag}
    \mathcal{L}_\text{WET} = \sum_a \mathcal{C}_a\,\mathcal{O}_a + h.c. + \dots
\end{align}
with $\{\mathcal{O}_a\}$ representing a complete basis of operators at mass dimension six, as for instance in Ref.~\cite{Jenkins:2017dyc}. In our analysis we use the \EOS software~\cite{EOS} to compute the flavor observables, so we define the relevant WET operators in the 
\EOS notation~\cite{Aebischer:2017ugx}. The translation into other WET bases is possible by means of the \texttt{Wilson
Coefficients eXchange Format}~\cite{Aebischer:2017ugx}. Higher-dimensional operators, denoted by dots, will be neglected because the effects in $b\to s$ observables are typically suppressed as $m_b/m_W$ or stronger.

The Wilson coefficients $\mathcal{C} = \mathcal{C}(m_b)$ and the hadronic matrix elements of the operators are
evaluated at the mass scale of the bottom quark, unless stated otherwise. They encode the sum of SM contributions and potential new physics effects. 

The FCNC processes $b\to s \gamma$, $b\to s g$ and $b\to s\ell^+\ell^-$ are described by the WET operators
\begin{align}\label{eq:wet-SM+flipped}
\mathcal{O}_{7^{(')}}  & = \frac{4G_F }{\sqrt{2}} V_{tb}V_{ts}^\ast \frac{e}{16\pi^2}m_b\left(\overline{s}\, \sigma^{\mu\nu} P_{R(L)} b\right)F_{\mu\nu}  \\\nonumber
\mathcal{O}_{8^{(')}}  & = \frac{4G_F}{\sqrt{2}}V_{tb}V_{ts}^\ast \frac{g_s}{16\pi^2}m_b\left(\overline{s}\, \sigma^{\mu\nu}T^A P_{R(L)} b\right)G_{\mu\nu}^A                  \\\nonumber
\mathcal{O}_{9^{(')}}  & = \frac{4G_F}{\sqrt{2}} V_{tb}V_{ts}^\ast \frac{e^2}{16\pi^2}\left(\overline{s}\,\gamma_\mu P_{L(R)} b\right)\left(\overline{\mu}\gamma^\mu \mu\right) \\\nonumber
\mathcal{O}_{10^{(')}} & = \frac{4G_F}{\sqrt{2}} V_{tb}V_{ts}^\ast \frac{e^2}{16\pi^2}\left(\overline{s}\,\gamma_\mu P_{L(R)} b\right)\left(\overline{\mu}\gamma^\mu\gamma_5 \mu\right).
\end{align}
In the Standard Model, the operators $\mathcal{O}_{7}$ and $\mathcal{O}_{8}$ dominate $b\to s\gamma$ and $b\to sg$, while $\mathcal{O}_{9}$ and $\mathcal{O}_{10}$ dominate in the branching ratios of $b\to s\ell\ell$ processes. Numerically the SM Wilson coefficients of these operators are~\cite{EOS}
\begin{align}\label{eq:sm-wc}
    \mathcal{C}^{\text{SM}}_7 \sim -0.3\,, \qquad \mathcal{C}^{\text{SM}}_8 \sim -0.2\,,\qquad \mathcal{C}^{\text{SM}}_9 \sim +4.2\,, \qquad \mathcal{C}_{10}^{\text{SM}} \sim -4.3\,.
\end{align}
The operators $\mathcal{O}_{a'}$ involve quarks with different chiralities compared to the SM operators. In the Standard Model, the Wilson coefficients $\mathcal{C}_{7'}$ and $\mathcal{C}_{8'}$ are suppressed by $m_s/m_b$ compared to $\mathcal{C}_7$ and $\mathcal{C}_8$. The operators $\mathcal{O}_{9'}$ and $\mathcal{O}_{10'}$ are absent because weak interactions are left-chiral. In the SMEFT context, the WET operators in Eq.~\eqref{eq:wet-SM+flipped} are generated from tree-level or one-loop matching of the SMEFT operators listed in Eqs.~\eqref{eq:2q-operators} and \eqref{eq:4q-operators}. The prefactors in Eq.~\eqref{eq:wet-SM+flipped} are part of the operator definition; they ensure that the Wilson coefficients are dimensionless and normalized to the loop-induced SM contribution.

Beyond the dipole operators and semileptonic operators discussed above, two four-quark operators in the WET can contribute to $b\to s$ transitions both in the Standard Model and beyond:
\begin{align}
\mathcal{O}_{1^{(')}} & = \frac{4G_F}{\sqrt{2}} V_{tb}V_{ts}^\ast \left(\overline{c} \gamma^\mu P_{L(R)} T^A b\right)\left(\overline{s} \gamma_\mu P_L T^A c\right)\,\\\nonumber
\mathcal{O}_{2^{(')}} & = \frac{4G_F}{\sqrt{2}} V_{tb}V_{ts}^\ast \left(\overline{c} \gamma^\mu P_{L(R)}  b\right)\left(\overline{s} \gamma_\mu P_L c\right).
\end{align}
In the Standard Model both operators are generated by a $W$ boson exchange at tree-level and their Wilson coefficients are sizeable,
\begin{equation}
    \mathcal{C}^{\text{SM}}_2 \sim 1\,, \qquad \mathcal{C}^{\text{SM}}_1 \sim -0.3\,.
\end{equation}
The two four-quark operators contribute to $b\to s\gamma$ and $b\to s\ell\ell$ processes through two mechanisms: mixing with other WET operators under the renormalization group, and non-local contributions to the observables.

\paragraph{Operator mixing} Interpreting $b \to s$ observables in terms of SMEFT coefficients requires the evolution of the WET coefficients from the matching scale $\mu \sim m_Z$ down to the bottom mass scale $\mu \sim m_b$ by means of the renormalization group in QCD. Under the RG evolution, the four-quark operators $\mathcal{O}_1$ and $\mathcal{O}_2$ mix into the dipole operators $\mathcal{O}_{7}$ and $\mathcal{O}_{8}$ and the semileptonic operator $\mathcal{O}_9$ and similarly $\mathcal{O}_{1'}$ and $\mathcal{O}_{2'}$ mix into $\mathcal{O}_{7'}$, $\mathcal{O}_{8'}$ and $\mathcal{O}_{9'}$. The operators $\mathcal{O}_{10}$ and $\mathcal{O}_{10'}$ do not receive mixing contributions from four-quark operators, due to the axial-vector nature of their lepton currents.
    
The operator mixing in WET has a significant effect on the phenomenology of $b\to s\ell\ell$ processes. In the Standard Model, mixing generates approximately $50\%$ of $\mathcal{C}_9(m_b)$,  about $10\%$ of $\mathcal{C}_7(m_b)$, and about $3\%$ of $\mathcal{C}_8(m_b)$. Mixing effects in BSM contributions will be discussed in Sec.~\ref{sec:matching}.
    
\begin{figure}[t]
    \centering
        \includegraphics[page=1,scale=1.1]{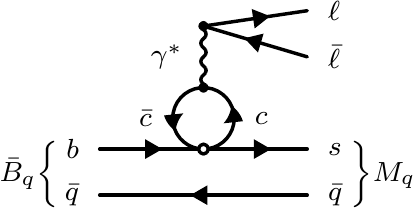}\quad
        \includegraphics[page=2,scale=1.1]{draft/figures/feynmangraphs.pdf}\quad
        \includegraphics[page=3,scale=1.1]{draft/figures/feynmangraphs.pdf}
    \caption{Sketch of the pollution of $b\to s\ell\ell$ processes by hadronic intermediate states.
    Here $M_q = \bar{K}^{(\ast)},\phi,\dots$ denotes a meson with quark content $\bar{q}s$.
    On the left we show the partonic representation;
    in the middle and on the right we show possible hadronisations of the partonic picture.  The unfilled circle represents an insertion of an $(\bar{s}c)(\bar{c} b)$ WET operator.}
    \label{fig:non-local-contributions}
\end{figure}

\paragraph{Non-local operator contributions} 
In observables, we cannot distinguish between ${b\to s \{\gamma,\ell^+\ell^-\}}$ and $b\to s \left[c\bar{c} \to \{\gamma,\ell^+\ell^-\}\right]$ decay amplitudes. In WET the former are local interactions captured by the operators from Eq.~\eqref{eq:wet-SM+flipped}. The latter correspond to non-local operator contributions, arising from the time-ordered product of the electromagnetic Lagrangian and the WET Lagrangian, see \emph{e.g.} Ref.~\cite{Beneke:2001at}. The numerical effect of non-local contributions on observables strongly depends on the invariant mass (squared) of the di-lepton system, $q^2$. In particular, if $q^2$ is close to the masses of hadronic $c\bar{c}$ bound states like $J/\psi$ and $\psi(2S)$, non-local contributions are resonantly enhanced. This enhancement readily dominates over the rare and genuinely short-distance process. For instance, in $B\to K \mu^+\mu^-$ decays the branching ratio $\mathcal{B}(B\to J/\psi K)$ exceeds the off-resonance bin $1\,\GeV^2 \leq q^2 \leq 6\,\GeV^2$ by a factor of $\sim 3\cdot 10^3$~\cite{Zyla:2020zbs}.
Moreover, the fall-off of the resonance tails can significantly distort the $q^2$ dependence of the short-distance contribution to $B\to K \mu^+\mu^-$. The shape description of these resonances introduces presently the largest systematic theory uncertainty in exclusive $b\to s\ell^+\ell^-$ decays and is a subject of active research~\cite{Lyon:2014hpa,Ciuchini:2015qxb,Capdevila:2017ert,Jager:2017gal,Bobeth:2017vxj,Arbey:2018ics,Gubernari:2020eft}.
A pictorial representation of the effect is shown in Fig.~\ref{fig:non-local-contributions}.

Non-local contributions to $b\to s$ transitions can be described to leading order in the electromagnetic coupling $\alpha_e$ by two process-dependent effects:
\begin{itemize}
    \item[1.] constant shifts $\Delta \mathcal{C}$: $\qquad \qquad \qquad \qquad \quad \mathcal{C}_{7^{(')}} \to \mathcal{C}_{7^{(')}}^\text{eff} = \mathcal{C}_{7^{(')}} + \Delta\mathcal{C}_{7^{(')}}$
    \item[2.] momentum-dependent shifts $\ \Delta \mathcal{C}(q^2)$: $\quad \mathcal{C}_{9^{(')}}  \to \mathcal{C}_{9^{(')}}^\text{eff} = \mathcal{C}_{9^{(')}} + \Delta\mathcal{C}_{9^{(')}}(q^2)$.
\end{itemize}
The non-local contributions feature an explicit dependence on the renormalization scale $\mu$, which largely compensates the scale dependence due to local operator mixing. It is common to consider the ``effective'' Wilson coefficients $\mathcal{C}^{\text{eff}}$, which are more stable under the RG evolution than the individual non-local contributions $\Delta \mathcal{C}$.

Several of the four-quark SMEFT operators in Eq.~\eqref{eq:4q-operators} match onto the WET operators $\mathcal{O}_1$ and $\mathcal{O}_2$ and thus affect the processes $b\to s\gamma$ and $b\to s\ell^+\ell^-$ through operator mixing and non-local contributions. Moreover, there exist 18 further $(\bar{s} c)(\bar{c} b)$ four-quark operators and many more operators with other flavor structures in the WET.
In general, all of them can contribute to $b\to s\gamma$ and $b\to s\ell^+\ell^-$ through a similar mechanism as $\mathcal{O}_1$ and $\mathcal{O}_2$. The non-local contributions of these operators, notably the momentum-dependent shifts $\Delta\mathcal{C}_{9^{(')}}(q^2)$, complicate the interpretation of flavor data within the SMEFT.
For our analysis, we therefore choose observables that are largely unaffected by non-local contributions of four-quark WET operators.

Besides these four-quark operators, there are six additional $b\to s\ell^+\ell^-$ operators with (pseudo-)scalar and tensor structure. Scalar and pseudo-scalar $b\to s\ell^+\ell^-$ interactions (similar to Higgs penguin contributions in the Standard Model) arise only from semileptonic SMEFT operators, even at the one-loop level. Since SMEFT operators with leptons are not part of our set of interest (see Eqs.~\eqref{eq:2q-operators} and \eqref{eq:4q-operators}), (pseudo-)scalar WET operators do not occur in our analysis. WET operators with a tensor structure are only induced from the matching of SMEFT operators of mass dimension larger than six~\cite{Alonso:2014csa,Cata:2015lta}. Since we confine our analysis to dimension-six operators, tensor operators in WET are also absent in our analysis.

\subsection{Observables in $B$ meson decays}\label{sec:b-observables}
For our statistical analysis, we select flavor observables that are sensitive to the WET coefficients in Eq.~\eqref{eq:wet-SM+flipped}. Natural candidates are the FCNC decays $\bsmm$, $\bsg$, and $B\to K^{(\ast)}\mu^+\mu^-$. In this section, we provide a comprehensive discussion of SMEFT contributions to these rare $B$ decays. In our numerical analysis we work with $\bsmm$ and $\bsg$. 

Within the WET, the branching ratio for $\bsmm$ is given by
\begin{align}\label{eq:bsmm-br}
    	\mathcal{B}(B_s\to\mu^+\mu^-)\times 10^{9} & = \left[3.57 - 1.71\, \mathcal{C}_{10} + 0.21\, \mathcal{C}_{10}^2\right] \times \big(1 \pm 1.2\%\big|_{f_{B_s}} \pm 1.5\%\big|_{\text{CKM}}\big)\,,
\end{align}
where we have absorbed the SM contributions $\mathcal{C}_a^{\text{SM}}$ into the constant term.
We have produced this expression using the \EOS software~\cite{EOS}. Details and a discussion of the theory uncertainties are relegated to App.~\ref{app:polynomials}. 
Non-local contributions to \bsmm from WET four-quark operators are absent at the leading order in $\alpha_e$.
$\mathcal{B}(\bsmm)$ is precisely predicted in and beyond the Standard Model, and it is therefore a clean probe of SMEFT contributions to $\mathcal{C}_{10^{(')}}$.

Within the WET, the branching ratio of \bsg is given by~\cite{EOS}
\begin{align}\label{eq:bsg-br}
	\mathcal{B}(B\to X_s\gamma) \times 10^{4} & = \big[3.26 
	-15.17
	\,\mathcal{C}_7
	-0.77\, \mathcal{C}_8 + 1.66\, \mathcal{C}_7 \mathcal{C}_8 
	+ 1.36\, \mathcal{C}_{7'} \mathcal{C}_{8'}\\\nonumber
	& \qquad\qquad\qquad\, + 18.03\,(\mathcal{C}_7^2 + \mathcal{C}_{7'}^2) + 0.20\, \mathcal{C}_8^2 +0.09\, \mathcal{C}_{8'}^2\big] \times \big(1 \pm 5\%\big)\,,
\end{align}
where again we have absorbed the SM contributions $\mathcal{C}_a^{\text{SM}}$ into the constant term.
The BSM contributions of $\mathcal{O}_7$ and $\mathcal{O}_{7'}$ are
particularly large, due to a chiral enhancement of $m_W/m_b$ compared to the SM contributions. The chiral enhancement of
$\mathcal{O}_{7^{(')}}$ is also at work in the interference terms $\mathcal{C}_7\mathcal{C}_8$ and $\mathcal{C}_{7'}\mathcal{C}_{8'}$ and can affect the sensitivity to $\mathcal{C}_{8^{(')}}$. Unlike in the Standard Model, quadratic operator contributions to the observables are
numerically relevant and can even dominate over linear new physics effects. Taking them into account is indispensable for a reliable
interpretation of \bsg in SMEFT.

Contributions to $\mathcal{B}(\bsg)$ arising from the full basis of WET four-quark operators are presently unknown. However,
they can be accounted for by reinterpreting the constraints on $\mathcal{C}_{7^{(')}}$.
For details we refer the reader to App.~\ref{app:polynomials}.

Within the WET, observables arising in $B\to K^{(\ast)}\mu^+ \mu^-$ decays are sensitive to $\mathcal{O}_{7}$, $\mathcal{O}_{9}$ and $\mathcal{O}_{10}$ in the Standard Model and in addition to $\mathcal{O}_{7'}$, $\mathcal{O}_{9'}$ and $\mathcal{O}_{10'}$ in the SMEFT. Together with \bsmm and \bsg, they would allow us to probe the complete set of $b\to s$ operators.
However, as discussed in Sec.~\ref{sec:wet}, non-local operator contributions to ${B\to K^{(\ast)}\mu^+ \mu^-}$ decays introduce
momentum-dependent shifts $\Delta \mathcal{C}_{9^{(')}}(q^2)$, which make it difficult to extract new physics contributions to
$\mathcal{C}_{9^{(')}}$ in a model-independent manner. We therefore do not include $B\to K^{(\ast)}\mu^+\mu^-$ observables in our analysis.

For our combined statistical analysis of top and bottom observables it is important that the theory predictions of the flavor observables are
uncorrelated and introduce only few hadronic quantities that can be treated as nuisance parameters. 
The branching ratios of \bsmm and \bsg fulfill both requirements. Due to their dependence on $\mathcal{C}_{10}$ and $\mathcal{C}_{7^{(')}}$,
$\mathcal{C}_{8^{(')}}$, respectively, they probe different directions in the SMEFT parameter space. This makes them good candidates to break
blind directions in a global fit.

Including further flavor observables quickly enlarges the complexity of a global analysis. In this pilot study we confine ourselves to \bsmm and \bsg, keeping in mind that a more comprehensive analysis of flavor observables will yield valuable additional information on the SMEFT parameter space.

In combined fits of flavor and high-energy data, nuisance parameters from hadronic uncertainties in flavor observables are a serious limiting factor. Ultimately, it would be very helpful if the likelihood function of WET coefficients from existing global analyses of flavor data became available. Working directly with likelihood functions would make the combined analysis of different low-energy data sets more robust and efficient. It would also allow us to include more flavor observables, thus greatly contributing to fully resolving the SMEFT parameter space.

\subsection{SMEFT-to-WET matching in minimal flavor violation}\label{sec:matching}
\noindent To understand how the flavor structure of a UV completion imprints itself onto $B$ physics observables, we investigate how the flavor pattern in SMEFT coefficients translates to WET coefficients. In what follows we discuss the flavor structure of SMEFT-to-WET matching at NLO~\cite{Dekens:2019ept} in detail for MFV.
Our approach is equally applicable to any other flavor pattern.

\begin{table}[tp]\begin{center}
\renewcommand{\arraystretch}{1.3}
\setlength{\tabcolsep}{1.5mm}
\begin{tabular}{c|c|c|c|c}
\toprule
  & $\mathcal{C}_7$ & $\mathcal{C}_8$ & $\mathcal{C}_{9}$ & $\mathcal{C}_{10}$ \\
\midrule
$\ C_{\phi q}^{(1)}$ & $-$ & $-$ & $b_{\phi q}^{(1)}\, y_t^2$  & $b_{\phi q}^{(1)}\, y_t^2$ \\
$\ C_{\phi q}^{(3)}$ & $-$ & $-$ & $b_{\phi q}^{(3)}\, y_t^2$  & $b_{\phi q}^{(3)}\, y_t^2$ \\
$C_{dB}$ & $b_{dB}\, y_t^2$ & $-$ & $-$ & $-$ \\
$C_{dW}$ & $b_{dW}\, y_t^2$ & $-$ & $-$ & $-$ \\
$C_{dG}$ & $-$ & $b_{dG}\, y_t^2$ & $-$ & $-$ \\
 \bottomrule[1pt]
\end{tabular}
\end{center}
\caption{Tree-level SMEFT-to-WET matching relations in MFV with up-alignment. Shown are the flavor structures of the SMEFT contributions $C(m_Z)$ to the WET coefficients $\mathcal{C}(m_Z)$. Effects of $\mathcal{O}(m_s/m_b)$ or smaller have been neglected, including contributions to $\mathcal{C}_{a'}$, which first appear at $\mathcal{O}(y_s/y_b)$.}
\label{tab:smeft-to-wet-tree}
\end{table}

\begin{table}[tp]\begin{center}
\renewcommand{\arraystretch}{1.3}
\setlength{\tabcolsep}{1.5mm}
\begin{tabular}{c|c|c|c|c}
\toprule
  & $\quad\mathcal{C}_7$ & $\quad\mathcal{C}_8$ & $\quad\mathcal{C}_9$ & $\quad\mathcal{C}_{10}$ \\
\midrule
$C_{\phi q}^{(1)}$ & ($A_{\phi q}^{(1)}$, $b_{\phi q}^{(1)} y_t^2$) & ($-$, $b_{\phi q}^{(1)} y_t^2$) & ($A_{\phi q}^{(1)}$, $b_{\phi q}^{(1)} y_t^2$) & ($A_{\phi q}^{(1)}$, $b_{\phi q}^{(1)} y_t^2$) \\
$C_{\phi q}^{(3)}$ & ($A_{\phi q}^{(3)}$, $b_{\phi q}^{(3)} y_t^2$) & ($A_{\phi q}^{(3)}$, $b_{\phi q}^{(3)} y_t^2$) & ($A_{\phi q}^{(3)}$, $b_{\phi q}^{(3)} y_t^2$) & ($A_{\phi q}^{(3)}$, $b_{\phi q}^{(3)} y_t^2$)  \\
$C_{\phi u}$ & $-$ & $-$ & ($A_{\phi u}$, $-$) & ($A_{\phi u}$, $-$) \\
$C_{\phi d}$ & $-$&$-$&$-$&$-$ \\
$C_{\phi ud}$ & ($A_{\phi ud}\,y_t$, $-$) & ($A_{\phi ud}\,y_t$, $-$) & $-$ & $-$ \\
$C_{uB}$ & ($A_{uB}\,y_t$, $-$) & $-$ & ($A_{uB}\,y_t$, $-$) & ($A_{uB}\,y_t$, $-$) \\
$C_{uW}$ & ($A_{uW}\,y_t$, $-$) & ($A_{uW}\,y_t$, $-$) & ($A_{uW}\,y_t$, $-$) & ($A_{uW}\,y_t$, $-$) \\
$C_{uG}$ & $-$ & ($A_{uG}\,y_t$, $-$) & $-$ & $-$ \\
$C_{dB}$ & ($-$, $b_{dB}\,y_t^2$) & ($-$, $b_{dB}\,y_t^2$) & $-$ & $-$ \\
$C_{dW}$ & ($A_{dW}$, $b_{dW}\,y_t^2$) & ($A_{dW}$, $b_{dW}\,y_t^2$) & $-$ & $-$ \\
$C_{dG}$ & $-$ & ($b_{dG}\,y_t^2$, $-$) & $-$ & $-$ \\
 \bottomrule[1pt]
\end{tabular}
\end{center}
\caption{One-loop SMEFT-to-WET matching relations in MFV with up-alignment for operators with two quarks. Shown are the flavor structures of the SMEFT contributions $C(m_Z)$ to the WET coefficients $\mathcal{C}(m_Z)$. In each entry $(x,y)$, $x$ is the $tW$ loop contribution, and $y$ is from a $sZ$ or $bZ$ loop (see text). Effects of $\mathcal{O}(m_s/m_b)$ or smaller have been neglected, in particular, contributions to $\mathcal{C}_{a'}$.}
\label{tab:smeft-to-wet-loop}
\end{table}

For two-quark operators, the relevant flavor structures in MFV are presented in Tab.~\ref{tab:smeft-to-wet-tree} for tree-level matching and in Tab.~\ref{tab:smeft-to-wet-loop} for one-loop matching. Contributions $C^{33}$ from the third generation always scale as
\begin{align}\label{eq:3rd-generation}
    A_{xy} = a_{xy} + b_{xy} y_t^2.
\end{align}
For four-quark operators, we present the flavor structures in one-loop matching in Eqs.~\eqref{eqn:4q-matching} and~\eqref{eq:smeft-to-wet-four-quark-matching}. We focus on the dominant terms in the MFV expansion.\footnote{The one-loop matching conditions of Ref.~\cite{Dekens:2019ept} exclude terms with a relative suppression of $\mathcal{O}(m_s/m_b)$ and smaller, so we do not include these contributions in our numerical analysis.}

In MFV, all SMEFT contributions to $\mathcal{L}_{\rm WET}$
are proportional to the CKM elements\footnote{
  The CKM elements and their extraction from low-energy observables are also susceptible to
  BSM contributions, which are not considered here. A global analysis of bottom observables
  will inevitably need to take them into account, and a promising solution has been put forward
  in Ref.~\cite{Descotes-Genon:2018foz}.
} $V_{ts}^\ast V_{tb}$, which has been absorbed into the WET operator definitions in Eq.~\eqref{eq:wet-SM+flipped}. As a consequence, the matching conditions do not involve CKM elements. The $V_{ts}^\ast V_{tb}$ scaling can have three origins: for operators with left-handed down quarks in a basis with up-alignment, CKM elements occur in the rotation of down-quark interaction eigenstates in SMEFT to mass eigenstates in WET. For operators with right-handed down-quark FCNCs, the CKM elements are part of the flavor structure in MFV, see Tab.~\ref{tab:2q-mfv}. Finally, in one-loop matching the CKM dependence can also be due to the SM weak charged currents.

\begin{figure}[t]
    \centering
    \resizebox{\textwidth}{!}{
    \subcaptionbox{\label{fig:loop-matching:a}}[.25\textwidth]
    {\includegraphics[page=4,scale=1.1]{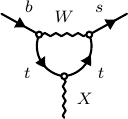}}
	\subcaptionbox{\label{fig:loop-matching:b}}[.25\textwidth]
	{\includegraphics[page=5,scale=1.1]{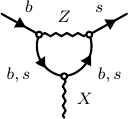}}
	\subcaptionbox{\label{fig:loop-matching:c}}[.25\textwidth]
	{\includegraphics[page=6,scale=1.1]{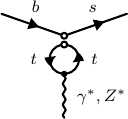}}
	\subcaptionbox{\label{fig:loop-matching:d}}[.25\textwidth]
	{\includegraphics[page=7,scale=1.1]{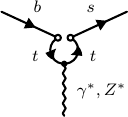}}
	}
    \caption{Feynman diagrams for SMEFT-to-WET matching in $b\to s$ transitions at one-loop level.
    On the left we show possible two-quark operator insertions (circles) in a $tW$ loop (a) and a $sZ$ or $bZ$ loop (b). Here $X$ is either an on-shell photon or gluon $(\mathcal{C}_7,\mathcal{C}_8)$, or an off-shell photon or $Z$ boson coupling to a muon pair $(\mathcal{C}_9,\mathcal{C}_{10})$.
    On the right we show possible four-quark operator insertions. The two circles specify the quark currents.
    Contributions to $\mathcal{C}_9$ and $\mathcal{C}_{10}$ are due to $F_{qq}^{(-)}$ and $F_{qu}^{(1)}$ (c), as well as $F_{qq}^{(3)}$ (d).
    }
    \label{fig:loop-matching}
\end{figure}

\textbf{Two-quark operator contributions}~Two-quark SMEFT operators contribute to the matching already at tree level. To determine the matching contribution of each SMEFT coefficient $C^{kl}$, we have summed over all flavor indices $k,l$ and used the flavor structures shown in Tab.~\ref{tab:2q-mfv}.
At tree level, only operators with unsuppressed $bs$ couplings contribute. At one-loop level, matching contributions
of two-quark operators can arise from two different sources: operator insertions in a $tW$ loop as shown in Fig.~\ref{fig:loop-matching:a},
and operator insertions in an $sZ$ or $bZ$ loop as shown in Fig.~\ref{fig:loop-matching:b}.

 In the $tW$ loop, the $b\to s$ flavor change is due to the CKM structure of the weak charged current and only SMEFT coefficients $C^{33}$
contribute. In the $sZ$ loop, the $b\to s$ flavor change is caused by the SMEFT operator, and all three generations of left-handed quarks
contribute. We show this in Tab.~\ref{tab:smeft-to-wet-loop}, where each entry $(x,y)$ contains the flavor structure from the $tW$ loop ($x$)
and from the $sZ$ and $bZ$ loops ($y$). The complete matching condition involves a linear combination of these two flavor structures.

Depending on the flavor, gauge, and chiral structure, the various SMEFT contributions can be enhanced or suppressed compared to the SM contribution. In some cases they can even be absent due to CKM unitarity. Here we discuss the underlying mechanisms through a series of examples.

The SMEFT dipole operators with right-handed down quarks, $O_{dX}$, match onto the WET operators $\mathcal{O}_{7^{(')}}$ and $\mathcal{O}_{8^{(')}}$ at tree level and at one-loop level. Compared to the SM contribution, effects of these operators on $\mathcal{C}_7$ and $\mathcal{C}_8$ are chirally enhanced by $m_W/m_b$. This enhancement makes $\mathcal{C}_7$ and $\mathcal{C}_8$ and in general also $\mathcal{C}_{7'}$ and $\mathcal{C}_{8'}$
very sensitive probes of SMEFT dipole operators.
In MFV all operator contributions with right-handed bottom quarks are suppressed by $y_b$. This suppression lifts the chiral enhancement in
the matching relation, but still leaves a sizeable effect of $C_{dX}$ in $\mathcal{C}_7$ and $\mathcal{C}_8$. Thanks to the chiral
enhancement, $b\to s$ observables are sensitive to $y_b$-suppressed flavor structures at tree level, unlike top observables, where no such
enhancement occurs. Effects of $C_{dX}$ in $\mathcal{C}_{7'}$ and $\mathcal{C}_{8'}$ in MFV are suppressed by $y_s$, and we neglect them in
our analysis. In Tab.~\ref{tab:chiral-enhancement} in App.~\ref{app:uv-chi}, we list all SMEFT-to-WET matching relations with a non-trivial
dependence on the quark masses and Yukawa couplings.

Operators with right-handed up-type quarks only appear in the SMEFT-to-WET matching at the one-loop level.

Operators with left-handed quarks, $O_{\phi q}^{(1)}$ and $O_{\phi q}^{(3)}$, match onto $\mathcal{O}_{9}$ and $\mathcal{O}_{10}$ at tree level and onto $\mathcal{O}_{7-10}$ at one-loop level. The corresponding Wilson coefficients modify the neutral and charged weak currents with left-handed quarks as
\begin{align}\label{eq:weak-couplings}
    u_L\bar u_L Z:\ C_{\phi q}^{(1)} - C_{\phi q}^{(3)},\qquad d_L\bar d_L Z:\ C_{\phi q}^{(1)} + C_{\phi q}^{(3)},\qquad 
    q_L\bar{q}_L' W:\ C_{\phi q}^{(3)}\,.
\end{align}
The various contributions of $C_{\phi q}^{(1)}$ and $C_{\phi q}^{(3)}$ to $\mathcal{C}_{9}$ and $\mathcal{C}_{10}$ are
 \begin{align}\label{eq:c1c3-flavor}
     \text{tree level}: & \quad C_{\phi q}^{(1),kk} + C_{\phi q}^{(3),kk}\\\nonumber
     tW \text{ loop}: & \quad C_{\phi q}^{(1),33} - C_{\phi q}^{(3),33},\ C_{\phi q}^{(3),3k}\\\nonumber
     bZ,sZ \text{ loop}: & \quad C_{\phi q}^{(1),kk} + C_{\phi q}^{(3),kk}\,,
 \end{align}
where the sum over $k\in \{1,2,3\}$ is implicit. In $\mathcal{C}_7$ and $\mathcal{C}_8$, $C_{\phi q}^{(1)}$ and $C_{\phi q}^{(3)}$ contribute only at loop level. In Eq.~\eqref{eq:c1c3-flavor} only two of the combinations are independent. In our numerical analysis we choose $C_{\phi q}^{(-)} = C_{\phi q}^{(1)} - C_{\phi q}^{(3)}$ and $C^{(3)}_{\phi q}$ as degrees of freedom. Accordingly, we define the relevant parameter combinations in MFV as
\begin{align}\label{eq:phiq-dof}
   A_{\phi q}^{(-)} = A_{\phi q}^{(1)} -  A_{\phi q}^{(3)}\,,\qquad a_{\phi q}^{(-)} =  a_{\phi q}^{(1)} - a_{\phi q}^{(3)}\,,\qquad b_{\phi q}^{(-)} =  b_{\phi q}^{(1)} - b_{\phi q}^{(3)}\,.
\end{align}
In MFV, $O_{\phi q}^{(1)}$ and $O_{\phi q}^{(3)}$ contribute to tree-level and one-loop matching with different flavor structures. For illustration we expand the SMEFT Lagrangian for $O_{\phi q}^{(3)}$ in terms of the relevant quark mass eigenstates 
\begin{align}\label{eq:phiq3}
\mathcal{L} & = C_{\phi q}^{(3),kl} O_{\phi q}^{(3),kl}\\\nonumber
& \supset \frac{e v^2}{2 s_w c_w}Z_{\mu}\left(- C_{\phi q}^{(3),kl} V_{ks}^\ast V_{lb} (\overline{s}_L\gamma^\mu b_L) + C_{\phi q}^{(3),kk} (\overline{u}_L^k\gamma^\mu u_L^k)\right)\\\nonumber
& \quad + \frac{e v^2}{\sqrt{2}s_w}W_{\mu}^+\,C_{\phi q}^{(3),3k} V_{kl}(\overline{t}_L \gamma^\mu d_L^l) + h.c. \\\nonumber
& = \frac{e v^2}{2 s_w c_w}Z_{\mu}\left(- b_{\phi q}^{(3)} y_t^2 V_{tb}V_{ts}^\ast (\overline{s}_L\gamma_\mu b_L) + a_{\phi q}^{(3)}(\overline{u}_L^k\gamma_\mu u_L^k) + b_{\phi q}^{(3)} y_t^2 (\overline{t}_L\gamma_\mu t_L)\right)\\\nonumber
& \quad + \frac{e v^2}{\sqrt{2}s_w}W_{\mu}^+\,(a_{\phi q}^{(3)} + b_{\phi q}^{(3)} y_t^2)\left(V_{tb}(\overline{t}_L \gamma^\mu b_L) + V_{ts}(\overline{t}_L \gamma^\mu s_L)\right) + h.c. + \mathcal{O}(y_b^2)\,,
\end{align}
where the sum over $k,l$ is implicit.
 In the last equation, the first term contributes to $\mathcal{C}_9$ and $\mathcal{C}_{10}$ at tree level directly through the flavor-breaking coupling $b_{\phi q}^{(3)}$. The second and third terms modify the coupling of up-type quarks to the $Z$ boson; they contribute to $\mathcal{C}_{7-10}$ indirectly at one-loop level. Similarly, the couplings in the last line modify the weak charged currents with top quarks and enter $b\to s$ transitions at one-loop level.
 
 In Eq.~\eqref{eq:phiq3}, the flavor-universal coupling $a_{\phi q}^{(3)}$ contributes only to flavor-diagonal neutral currents and to charged currents. At tree level, it cannot induce $b\to s$ FCNCs because of CKM unitarity. At one-loop level, flavor-universal contributions are subject to a partial GIM cancellation in the sum over $u,c,t$ loops. Therefore $b\to s$ processes are most sensitive to $b_{\phi q}^{(3)}$, while $a_{\phi q}^{(3)}$ enters only in the combination $A_{\phi q}^{(3)} = a_{\phi q}^{(3)} + b_{\phi q}^{(3)} y_t^2$ through GIM-breaking contributions from top loops.

\textbf{Four-quark operator contributions}~Four-quark operators contribute to the matching at the one-loop level. In general, $O_{qq}^{(1)}$ and $O_{qq}^{(3)}$ contribute to interactions with up- and down-type quarks in one of three types:
\begin{align}
(\bar{u}_L u_L)(\bar{u}_L u_L),\ (\bar{d}_L d_L)(\bar{d}_L d_L): & \quad C_{qq}^{(1)} + C_{qq}^{(3)}\\\nonumber
(\bar{u}_L u_L)(\bar{d}_L d_L): & \quad C_{qq}^{(1)} - C_{qq}^{(3)}\\\nonumber
(\bar{u}_L d_L)(\bar{d}_L u_L): & \quad C_{qq}^{(3)}\,.
\end{align}
The one-loop matching contributions of SMEFT operators with left-handed tops to $b\to s$ WET coefficients only involve the latter two types of insertions. For convenience, we define the combinations of MFV parameters $(w=1,3)$
\begin{align}\label{eqn:A-At-definitions}
    A_{qq}^{(w)} & = \aa_{qq}^{(w)} + \ba_{qq}^{(w)} y_t^2\,, & \aqqm & = A_{qq}^{(1)} - \aqqth\\\nonumber
    \widetilde{A}_{qq}^{(w)} & = \aat_{qq}^{(w)} + \bat_{qq}^{(w)} y_t^2\,, & \atqqm & =  \widetilde{A}_{qq}^{(1)} - \atqqth\\\nonumber
    B_{qq}^{(w)} & = \ba_{qq}^{(w)} + \bat_{qq}^{(w)} \,, & B_{qq}^{(-)} & =  B_{qq}^{(1)} - B_{qq}^{(3)}\,.
\end{align}
The WET coefficients $\mathcal{C}_9$ and $\mathcal{C}_{10}$ are sensitive to $O_{qq}^{(1)}$, $O_{qq}^{(3)}$ and $O_{qu}^{(1)}$ via
\begin{align}\label{eqn:4q-matching}
F_{qq}^{(-)} & \equiv \atqqm + B_{qq}^{(-)}y_t^2 & = & \
(V_{tb}V_{ts}^\ast)^{-1} \sum_k\left( C_{qq}^{(1),33kk} - C_{qq}^{(3),33kk}\right) V_{ks}^\ast V_{kb}\\\nonumber
F_{qq}^{(3)} & \equiv \aqqth + B_{qq}^{(3)}y_t^2 & = & \ (V_{tb}V_{ts}^\ast)^{-1} \sum_k C_{qq}^{(3),3kk3} V_{ks}^\ast V_{kb}\\\nonumber
F_{qu}^{(1)} & \equiv \big(\aat_{qu}^{(1)} + \ba_{qu}^{(1)}\big)y_t^2 & = & \ (V_{tb}V_{ts}^\ast)^{-1} \sum_{k} C_{qu}^{(1),kk33} V_{ks}^\ast V_{kb}\,,
\end{align}
up to corrections of $\mathcal{O}(y_t^4,y_b^2)$.
The corresponding Feynman diagrams for these three contributions are shown in Figs.~\ref{fig:loop-matching:c} and \ref{fig:loop-matching:d}.
We have only considered contributions of the top loop. Up and charm contributions can affect the relative impact of flavor-universal coefficients $\aa,\aat$ compared to flavor-breaking coefficients $\ba,\bat$ in $F_{qq}^{(-)}$ and $F_{qq}^{(3)}$, but rely on so-far unknown QCD effects. In our numerical analysis, we include only the top contributions and neglect up and charm contributions.

 The WET coefficients $\mathcal{C}_7$ and $\mathcal{C}_8$ receive contributions from $O_{quqd}^{(1)}$ and $O_{quqd}^{(8)}$ via
\begin{align}\label{eq:smeft-to-wet-four-quark-matching}
\mathcal{C}_{7}: \quad & \left(F_{quqd}^{(1)} + \frac{4}{3}F_{quqd}^{(8)}\right)\,y_by_t = (V_{tb}V_{ts}^\ast)^{-1} \sum_{k} \left(C_{quqd}^{(1),k333}  + \frac{4}{3}C_{quqd}^{(8),k333}\right) V_{ks}^\ast V_{kb}\\\nonumber
\mathcal{C}_{8}: \quad & \left(F_{quqd}^{(1)} - \frac{1}{6}F_{quqd}^{(8)}\right)\,y_by_t = (V_{tb}V_{ts}^\ast)^{-1} \sum_{k} \left(C_{quqd}^{(1),k333}  - \frac{1}{6}C_{quqd}^{(8),k333}\right) V_{ks}^\ast V_{kb}\,,
\end{align}
where the relevant directions in flavor space are $(c=1,8)$
\begin{align}
F_{quqd}^{(c)} & \equiv \aa_{quqd}^{(c)}+\big(\ab_{quqd}^{(c)}+\ba_{quqd}^{(c)}+\bat_{quqd}^{(c)}\big)y_t^2\,,
\end{align}
up to corrections of $\mathcal{O}(y_t^4,y_b^2)$. Shown are only contributions from the top loop; light-quark contributions are Yukawa-suppressed. Both operator contributions are chirally enhanced over the Standard Model, see Tab.~\ref{tab:chiral-enhancement} in App.~\ref{app:uv-chi}. As for two-quark operators, the $y_b$ suppression in MFV lifts the chiral enhancement, but still allows for sizeable effects of four-quark operators in $b\to s$ transitions.

Some of the one-loop matching relations for two- and four-quark operators are UV-sensitive, that is, the matching conditions depend logarithmically on the matching scale $\mu$. In Tab.~\ref{tab:uv-sensitivity} in App.~\ref{app:uv-chi}, we summarize all matching conditions that feature this logarithmic dependence. In our numerical analysis, we use $\mu=m_Z$, so that the  $\log(m_{t,W}/m_Z)$ dependence is mild.

For our predictions of $\mathcal{B}(\bsg)$ and $\mathcal{B}(\bsmm)$, we run the set of SMEFT Wilson coefficients of the operators defined in Eqs.~\eqref{eq:2q-operators} and~\eqref{eq:4q-operators} from the scale $\mu = m_t$ to the scale $\mu = m_Z$, where we match them onto the WET operators in Eq.~\eqref{eq:wet-SM+flipped}. Some of the operators mix strongly under the RG evolution, and this mixing can induce large effects in the SMEFT-to-WET matching. These effects are most relevant for SMEFT operators that match onto WET operators at one-loop level, but mix into operators that match onto WET at tree level. Notably the four-quark operators $O_{qq}^{(1/3)}$ and $O_{qu}^{(1)}$ mix into $O_{\phi q}^{(1/3)}$, and the latter induce $b\to s$ transitions at tree level. In Sec.~\ref{sec:fit-results} and App.~\ref{app:polynomials} we discuss the effects of operator mixing numerically. Operator mixing can also generate new SMEFT Wilson coefficients outside of the set of operators we consider. We include these contributions in the SMEFT-to-WET matching, but have checked that they have a small effect on $\mathcal{C}_{7-10}$.

Below $\mu = m_Z$, we evolve the Wilson coefficients in the WET down to the bottom mass scale $\mu = m_b$ via the WET renormalization group. For the entire procedure, we use the \wilson software~\cite{Aebischer:2018bkb},
which provides a numerical implementation of the relevant one-loop matching and running in SMEFT and WET.\footnote{
    We use a modified version of \wilson that includes a bug fix to the one-loop matching conditions for the photon and gluon $b\to s$ dipole
    operators. The modification is expected to be part of \wilson version 2.0.1 and later versions.}

In the Standard Model, the matching conditions of the SM interactions onto $\mathcal{O}_7$, $\mathcal{O}_8$ and $\mathcal{O}_{10}$ are known to NNLO in $\alpha_s$ and partially to NLO in $\alpha_e$~\cite{Bobeth:1999mk,Bobeth:2003at,Huber:2005ig,Bobeth:2013uxa}.
The anomalous dimensions that drive the RG evolution are known to three loops~\cite{Huber:2005ig}. With this information, the SM values of the Wilson coefficients at the bottom scale are known to
next-to-next-to-leading-logarithmic (NNLL) accuracy in QCD.

Beyond the Standard Model, first steps toward a complete set of SMEFT-to-WET matching conditions at dimension six were taken in Refs.~\cite{Aebischer:2017gaw,Jenkins:2017dyc},
which discuss matching at tree level and partially at loop level. The one-loop matching of SMEFT onto the full basis of dimension-six
WET operators was recently published~\cite{Dekens:2019ept} and is the basis for the matching implemented in \wilson. The RG evolution of the BSM contributions relies on the anomalous dimensions
of the full basis of BSM WET operators at dimension six, presently only known at the one-loop level. This restricts the RG evolution to leading-logarithmic (LL) accuracy in QCD.

In our analysis, we adopt a hybrid approach that implements the best available accuracy. We use the SM results to NNLL accuracy and include the BSM terms at LL accuracy.
This approach can and \emph{should} be revisited, once the full set of anomalous dimensions of dimension-six WET operators becomes available at the two-loop level.

Our numerical results for the SMEFT-to-WET matching relations in MFV are compiled in Tab.~\ref{tab:smeft-to-wet-num-Matching}.

\begin{table}[tp]\begin{center}
\renewcommand{\arraystretch}{1.05}
\setlength{\tabcolsep}{2mm}
\begin{tabular}{l|c|c|c|c}
\toprule
  & $\mathcal{C}_7$ &  $\mathcal{C}_8$  & $\mathcal{C}_9$ & $\mathcal{C}_{10}$ \\
\midrule[1pt]
SM & -0.337 & -0.183 & \phantom{-}4.27 & -4.17\\
\midrule
$\left(a_{\phi q}^{(-)},\,b_{\phi q}^{(-)}\right)$ & (0,\,-0.008) & (0,\,0.025) & (-0.01,\,-2.07) & (0.1,\,24.73)\\ 
$\left(a_{\phi q}^{(3)},\,b_{\phi q}^{(3)}\right)$ & (-0.034,\,0.061) & (-0.017,\,0.091) & (0.25,\,-4.18) & (-0.82,\,48.67)\\ 
$\left(a_{\phi u},\,b_{\phi u}\right)$ & (0,\,0) & (0,\,0) & (0.01,\,0.01) & (-0.1,\,-0.1)\\ 
$A_{\phi ud}$ & -0.033 & -0.015 & 0 & 0\\ 
$A_{u B}$ & -0.188 & 0 & 0.148 & 0\\ 
$A_{u W}$ & 0 & 0.024 & 0.115 & -0.440\\ 
$A_{u G}$ & 0 & -0.055 & 0 & 0\\ 
$\left(a_{d B},\,b_{d B}\right)$ & (-0.056,\,19.814) & (0,\,-0.005) & (0,\,0) & (0,\,0)\\ 
$\left(a_{d W},\,b_{d W}\right)$ & (0.059,\,-10.796) & (0.118,\,0.064) & (0,\,0) & (0,\,0)\\ 
$\left(a_{d G},\,b_{d G}\right)$ & (0,\,0) & (-0.016,\,5.816) & (0,\,0) & (0,\,0)\\ 
$F_{qq}^{(-)}$ & 0 & 0 & -0.1 & 0.59\\ 
$F_{qq}^{(3)}$ & 0 & 0 & -0.12 & 0.7\\ 
$F_{qu}^{(1)}$ & 0 & 0 & -0.01 & -0.59\\ 
$F_{quqd}^{(1)}$ & -0.019 & -0.028 & 0 & 0\\ 
$F_{quqd}^{(8)}$ & -0.025 & 0.005 & 0 & 0\\
 \bottomrule[1pt]
\end{tabular}
\end{center}
\caption{Matching contributions of SMEFT coefficients $C(m_Z)$ to WET coefficients $\mathcal{C}(m_Z)$ in MFV with up-alignment at one-loop level in the electroweak theory. The scale of new physics has been set to $\Lambda = 1\tev$. Contributions to $\mathcal{C}_{9'}$ and $\mathcal{C}_{10'}$ are not generated by the operators considered in this work.  \label{tab:smeft-to-wet-num-Matching}}
\end{table}

 At the matching scale $\mu = m_Z$, we find a high sensitivity of $\mathcal{C}_9$ and $\mathcal{C}_{10}$ to the flavor parameters $b_{\phi q}^{(-)}$
and $b_{\phi q}^{(3)}$, which contribute at tree level, see Eq.~\eqref{eq:phiq3}.
The coefficient $\mathcal{C}_9$ is also sensitive to loop-induced contributions of $a_{\phi q}^{(3)}$,
$A_{uB}$ and $A_{uW}$, as well as $\Fqqm$ and $\Fqqth$, but roughly one order
of magnitude less than to $b_{\phi q}^{(-)}$ and $b_{\phi q}^{(3)}$. At a similar level, $\mathcal{C}_{10}$ is sensitive to $a_{\phi q}^{(3)}$, $A_{uW}$, $\Fqqm$, $\Fqqth$ and $F_{qu}^{(1)}$,
roughly two orders of magnitude less than to $b_{\phi q}^{(-)}$ and $b_{\phi q}^{(3)}$.
The electromagnetic WET coefficient $\mathcal{C}_7$ is dominantly sensitive to
$b_{dB}$ and $b_{dW}$, due to chiral enhancements, and receives contributions from $A_{uB}$ 
that are smaller by two orders of magnitude.
Similarly, the chromomagnetic WET coefficient $\mathcal{C}_8$ is mostly sensitive to 
$b_{dG}$, again due to chiral enhancement, and sensitive to $a_{dW}$ by two orders of magnitude less.

The relation between the Wilson coefficients in flavor and top observables can be affected significantly by the RG evolution from the matching scale $m_Z$ down to $m_b$, where the WET coefficients enter the flavor observables. For the dominant contributions to \bsmm and \bsg, however, we find only moderate RG effects. For details we refer the interested reader to App.~\ref{app:polynomials}.

In summary, we find that \bsmm and \bsg are very sensitive to the flavor parameters
\begin{equation*}
    b_{\phi q}^{(-)}, b_{\phi q}^{(3)}, b_{d B}, b_{d W} \text{ and } b_{d G}\,,
\end{equation*}
which contribute to $b\to s$ transitions at tree level. We find 
a smaller sensitivity to
\begin{equation*}
    a_{\phi q}^{(-)}, a_{\phi q}^{(3)}, A_{u W}, A_{u B},
    F_{qq}^{(-)}, F_{qq}^{(3)}, \text{ and } F_{q u}^{(1)}\,,
\end{equation*}
which first contribute at the one-loop level.

\section{Flavor effects in top observables}\label{sec:top}
\noindent Top quark physics at the LHC provide a perfect environment for precision probes of SMEFT couplings involving third-generation quarks.
Precise predictions and measurements of cross sections and kinematic distributions allow us to probe SMEFT couplings in detail. In this work we build on a recent global analysis of top data from the LHC Runs I and II~\cite{Brivio:2019ius}. This analysis was aimed at identifying the gauge structure and quark chirality of effective operators. Here we use the same dataset and observables, but focus on the flavor structure of operators in top observables. In Sec.~\ref{sec:results} we combine top and flavor observables in parameter fits that connect SMEFT effects at the electroweak scale and the bottom mass scale.

In top physics, as for most LHC observables, cross sections and event bins in kinematic distributions depend on the SMEFT coefficients as
\begin{align}
    \sigma = \sigma_{\rm SM} + \sum_a \frac{C_a}{\Lambda^2}\sigma_a + \sum_{a,b} \frac{C_a C_b}{\Lambda^4} \sigma_{ab}\,.
\end{align}
Here $\sigma_{\rm SM}$ is the SM prediction, $\sigma_{a}$ denote contributions from one operator insertion, and $\sigma_{ab}$ are contributions from two interfering amplitudes with one operator insertion each. The sum is over all relevant operators from Eqs.~\eqref{eq:2q-operators} and \eqref{eq:4q-operators}. Below we summarize the main SMEFT contributions to top-antitop production and electroweak top processes. For details, we refer the reader to Ref.~\cite{Brivio:2019ius}.

The focus of our analysis is on the flavor pattern of operator contributions to top observables. To derive the flavor structure of a UV theory from top and bottom observables, we need to test two key features:
\begin{itemize}
    \item[1)] flavor universality;
    \item[2)] flavor alignment, \emph{i.e.}, the orientation of flavor breaking sources in flavor space.
\end{itemize}
 In top observables, the flavor universality of effective interactions can be tested by combining observables that involve either light-quark couplings or top-quark couplings or both of them. If new sources of flavor breaking are present, their alignment with the Yukawa couplings can be probed in FCNCs with top quarks~\cite{Durieux:2014xla}. In this work we focus on tests of flavor universality and leave flavor alignment for future work.

In MFV, flavor universality is strongly broken by the top Yukawa coupling, with significant effects in top and bottom observables. As the Yukawa couplings are the only source of flavor symmetry breaking, our choice of up-alignment is not physical and leads to the same effects as down-alignment. Moreover, in MFV FCNCs are suppressed by small Yukawa couplings and do not lead to sizeable effects in top or bottom observables. In our discussion we neglect contributions to observables that are suppressed by small quark masses or by CKM mixing. We also do not discuss SMEFT effects at NLO in QCD, but include them in our predictions of top-antitop and single top production in our numerical analysis.

\subsection{Two-quark operators}
Most SMEFT operators with two quarks are best probed in electroweak top processes. One exception is $O_{uG}$, which modifies the top-gluon coupling in hadronic $t\bar t$ production. In Tab.~\ref{tab:dof-top-mfv} we summarize the contributions of two-quark operators to top processes at the LHC. Shown are the flavor structures of the Wilson coefficients in MFV with up-alignment, which can be probed by the various channels. All operator contributions interfere with the SM amplitude at tree level. Contributions of $O_{dX}$ and $O_{\phi ud}$ to top observables are suppressed by $m_b^2/m_W^2$ and cannot be probed.

\begin{table}[tp]
    \begin{center}
        \renewcommand{\arraystretch}{1.2}
        \setlength{\tabcolsep}{1.5mm}
        \resizebox{0.75\textwidth}{!}{%
                \begin{tabular}{c|cccccc|c}
                    \toprule
                     &
                    $t\bar{t}$ & single top & $tW$ & $tZ$ & $t\bar{t}Z$ & $t\bar{t}W$ & $\#$ \\
                    \midrule
                    $C_{\phi q}^{(1)}$ & 
                    $-$ & $-$ & $-$ & $a_{\phi q}^{(1)},A_{\phi q}^{(1)}$ & $a_{\phi q}^{(1)},A_{\phi q}^{(1)}$ & $-$ & 2 \\
                    $C_{\phi q}^{(3)}$ & 
                    $-$ & $a_{\phi q}^{(3)},A_{\phi q}^{(3)}$ & $A_{\phi q}^{(3)}$ & $a_{\phi q}^{(3)},A_{\phi q}^{(3)}$ & $a_{\phi q}^{(3)},A_{\phi q}^{(3)}$ & $a_{\phi q}^{(3)}$ & 2\\
                    $C_{\phi u}$ & 
                    $-$ & $-$ & $-$ & $A_{\phi u}$ & $a_{\phi u},A_{\phi u}$ & $-$ & 2 \\
                   $C_{\phi d}$ & $-$ & $-$ & $-$ & $-$ & $a_{\phi d}$ & $-$ & 1\\
                    $C_{uB}$ & 
                    $-$ & $-$ & $-$ & $A_{uB}\,y_t$ & $A_{uB}\,y_t$ & $-$ & 1 \\
                    $C_{uW}$ &
                    $-$ & $A_{uW}\,y_t$ & $A_{uW}\,y_t$ & 
                    $A_{uW}\,y_t$ & $A_{uW}\,y_t$ & $-$ & 1 \\
                    $C_{uG}$ & 
                    $A_{uG}\,y_t$ & $-$ & $A_{uG}\,y_t$ & $-$ & $A_{uG}\,y_t$ & $A_{uG}\,y_t$ & 1\\
                    \bottomrule[1pt]
                \end{tabular}
            }
    \end{center}
    \caption{Two-quark operator contributions to top-quark production in MFV. Shown are the flavor structures for all SMEFT contributions that interfere with the SM amplitude at LO QCD ($t\bar t$, $t\bar t Z$, $t\bar t W$) and at LO EW (single top, $tW$, $tZ$). Contributions to observables of $\mathcal{O}(y_b^2)$ or smaller and CKM-suppressed contributions are neglected. In the last column we count the number of flavor degrees of freedom that can be probed.}
    \label{tab:dof-top-mfv}
\end{table}

In MFV, top observables probe the combination of flavor parameters $A = a + b y_t^2$ through the top coupling, see Eq.~\eqref{eq:3rd-generation}. For $O_{\phi q}^{(1)}$, $O_{\phi q}^{(3)}$ and $O_{\phi u}$, most electroweak top observables also probe the flavor-universal parameter $a$ alone through modifications of light-quark couplings. The sensitivity to both $a$ and $A$ is important, because it allows us to test the degree of flavor universality in a UV theory with top observables alone.

\begin{figure}[t!]
    \centering
    \includegraphics[page=8,scale=1.1]{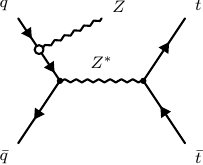}\qquad
    \includegraphics[page=9,scale=1.1]{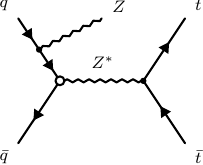}\qquad
    \includegraphics[page=10,scale=1.1]{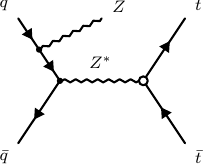}
    \caption{Examples of Feynman diagrams for electroweak contributions to $t\bar t Z$ production in SMEFT. The circles indicate operator insertions in the amplitude.}
    \label{fig:feynman-ew}
\end{figure}

For concreteness, we focus on $O_{\phi q}^{(1)}$ and $O_{\phi q}^{(3)}$, which affect the neutral and charged weak currents with left-handed quarks, see Eq.~\eqref{eq:weak-couplings}. In MFV and other scenarios with mostly flavor-diagonal couplings, the leading contributions to electroweak top production are
\begin{align}\label{eq:ew-top-smeft}
    \text{single top}: & \quad C_{\phi q}^{(3),11},\ C_{\phi q}^{(3),33}\\\nonumber
    t\bar t W: & \quad C_{\phi q}^{(3),11},\ \Big[C_{\phi q}^{(3),33},\ C_{\phi q}^{(1),11} \pm C_{\phi q}^{(3),11},\ C_{\phi q}^{(1),33} - C_{\phi q}^{(3),33} \Big]\\\nonumber
    tW: & \quad C_{\phi q}^{(3),33}\\\nonumber
    t\bar t Z: & \quad C_{\phi q}^{(1),11} \pm C_{\phi q}^{(3),11},\ C_{\phi q}^{(1),33} - C_{\phi q}^{(3),33}\\\nonumber
    tZ: & \quad C_{\phi q}^{(1),11} \pm C_{\phi q}^{(3),11},\ C_{\phi q}^{(1),33} - C_{\phi q}^{(3),33},\ C_{\phi q}^{(3),11},\ C_{\phi q}^{(3),33}.
\end{align}
In the Standard Model, $t\bar t Z$ and $t\bar t W$  production are dominated by QCD-induced processes like $q\bar q \to g^\ast \to t\bar t Z$ or $q\bar q' \to g^\ast W \to t\bar t W$. In SMEFT, electroweak contributions from partonic processes like $q\bar q \to Z^\ast \to t\bar t Z$ or $q\bar q' \to W^\ast \to t\bar t W$ are important, see Fig.~\ref{fig:feynman-ew}. In $t\bar t W$ production they even dominate the sensitivity to $O_{\phi q}^{(3)}$. In Eq.~\eqref{eq:ew-top-smeft} we show in brackets the extra electroweak contributions that can be probed in $t\bar t W$ production. In Sec.~\ref{sec:results}, we will analyze them numerically.

The various coefficients in Eq.~\eqref{eq:ew-top-smeft} are distinguished by the gauge and flavor structures of the Wilson coefficients. Regarding the gauge structure, single top, $tW$ and $t\bar t W$ production probe $C_{\phi q}^{(3)}$ through the $q\bar{q}'W$ coupling. Associated $t\bar t Z$ production is sensitive to ${C_{\phi q}^{(1)} - C_{\phi q}^{(3)}}$ and $C_{\phi q}^{(1)} + C_{\phi q}^{(3)}$ through the $u\bar{u}Z$ and $d\bar d Z$ couplings, and $tZ$ production probes all three directions in SMEFT space~\cite{Degrande:2018fog}. As in flavor observables, Eq.~\eqref{eq:phiq-dof}, we choose ${C_{\phi q}^{(-)} = C_{\phi q}^{(1)} - C_{\phi q}^{(3)}}$ and $C_{\phi q}^{(3)}$ as independent degrees of freedom. In Sec.~\ref{sec:results}, we will show how to resolve them in a combined analysis of top and top+flavor observables.

Regarding flavor, single top, $tZ$ and $t\bar{t}Z$ production probe both $C_{\phi q}^{(w),11}$ and $C_{\phi q}^{(w),33}$ --- and accordingly $a_{\phi q}^{(w)}$ and $A_{\phi q}^{(w)}$ in MFV --- through couplings with light quarks and top quarks. These contributions shift the weak gauge couplings of left-handed quarks by a flavor-dependent constant. This not only affects the total cross section, but also the final-state kinematics. Flavor universality can thus be tested in a combined analysis of electroweak top observables, including both total cross sections and kinematic distributions for an optimal sensitivity. The top polarization in electroweak top production is also sensitive to flavor universality breaking, because SMEFT contributions change the chirality of the top quarks in the final state.

In Sec.~\ref{sec:results}, we will explore the sensitivity of electroweak top processes to the flavor structure of $C_{\phi q}^{(1)}$ and $C_{\phi q}^{(3)}$ numerically. A similar analysis can be carried out for $O_{\phi u}$, which modifies the $Z$ couplings to right-handed up-type quarks.

\subsection{Four-quark operators}\label{sec:4q-in-top}
In top-antitop production, the relevant contributions of four-quark operators can be expressed in terms of vector and axial-vector interactions of two light quarks and two heavy quarks
\begin{align}
    C_{VV}^{(1),q_i} & \big(\bar{t} \gamma^\mu t\big)\big( \bar{q}_i \gamma_\mu q_i\big) & C_{VV}^{(8),q_i} & \big(\bar{t} \gamma^\mu T^A t\big)\big( \bar{q}_i \gamma_\mu T^A q_i\big) \\\nonumber
    C_{AA}^{(1),q_i} & \big(\bar{t} \gamma^\mu\gamma_5 t\big)\big( \bar{q}_i \gamma_\mu\gamma_5 q_i\big) & C_{AA}^{(8),q_i} & \big(\bar{t} \gamma^\mu\gamma_5 T^A t\big)\big( \bar{q}_i \gamma_\mu\gamma_5 T^A q_i\big) \\\nonumber
    C_{AV}^{(1),q_i} & \big(\bar{t} \gamma^\mu \gamma_5 t\big)\big( \bar{q}_i \gamma_\mu q_i\big) & C_{AV}^{(8),q_i} & \big(\bar{t} \gamma^\mu \gamma_5 T^A t\big)\big( \bar{q}_i \gamma_\mu T^A q_i\big) \\\nonumber
    C_{VA}^{(1),q_i} & \big(\bar{t} \gamma^\mu t\big)\big( \bar{q}_i \gamma_\mu \gamma_5 q_i\big) & C_{VA}^{(8),q_i} & \big(\bar{t} \gamma^\mu T^A t\big)\big( \bar{q}_i \gamma_\mu \gamma_5 T^A q_i\big),
\end{align}
where $q_i = u_i,d_i$ with $i\in\{1,2\}$ are quark mass eigenstates of the incoming partons. These coefficients are linear combinations of $C^{33ii}$ and $C^{3ii3}$ from various operators in the Warsaw basis. In App.~\ref{app:4q-in-ttb-warsaw}, we list these combinations explicitly. Here we focus on the operators $O_{qq}^{(1)}$ and $O_{qq}^{(3)}$, which generate interesting effects in both top and flavor observables. A similar analysis can be performed for operators with right-handed quarks, namely $O_{qu}^{(1)}$, $O_{qu}^{(8)}$, $O_{qd}^{(1)}$, $O_{qd}^{(8)}$, $O_{ud}^{(1)}$, $O_{ud}^{(8)}$ and $O_{uu}$.

In MFV the color-singlet and color-octet contributions of $O_{qq}^{(1)}$ and $O_{qq}^{(3)}$ to top-antitop production are (cf. Eq.~\eqref{eqn:A-At-definitions})
\begin{align}\label{eq:4q-up-and-down}
C_{VV}^{(1),u} & = C_{AA}^{(1),u} = - C_{AV}^{(1),u} = - C_{VA}^{(1),u} & = & \ \frac{1}{4}\Big(A_{qq}^{(-)} + 2\aqqth\Big) + \frac{1}{12}\Big(\widetilde{A}_{qq}^{(-)} + 2\atqqth\Big)\\\nonumber
C_{VV}^{(8),u} & = C_{AA}^{(8),u} = - C_{AV}^{(8),u} = - C_{VA}^{(8),u} & = & \ \frac{1}{2}\Big(\widetilde{A}_{qq}^{(-)} + 2\atqqth\Big)\\\nonumber
C_{VV}^{(1),d} & = C_{AA}^{(1),d} = - C_{AV}^{(1),d} = - C_{VA}^{(1),d} & = & \ \frac{1}{4} A_{qq}^{(-)} + \frac{1}{6}\atqqth\\\nonumber
C_{VV}^{(8),d} & = C_{AA}^{(8),d} = - C_{AV}^{(8),d} = - C_{VA}^{(8),d} & = & \ \atqqth\,.
\end{align}
We distinguish between contributions from up and down partons, labelled as $u$ and $d$, respectively. We have neglected contributions of $\mathcal{O}(y_b^2)$ and CKM- or parton-luminosity-suppressed contributions, which only give tiny effects in top-antitop production. Color-singlet coefficients are sensitive to both flavor contractions $A$ and $\widetilde{A}$, corresponding to $C^{33ii}$ and $C^{3ii3}$. Color-octet coefficients are only sensitive to $\widetilde{A}$ via $C^{3ii3}$.

Operator contributions with up and down quarks contribute according to the corresponding parton luminosity. At low energies, top-antitop production effectively probes the distribution of valence quarks inside the protons and is thus sensitive to the combinations
\begin{align}\label{eq:ttb-eff}
    C_{VV}^{(8),\rm eff} & = 2 C_{VV}^{(8),u} + C_{VV}^{(8),d} = \widetilde{A}_{qq}^{(-)} + 3\atqqth\\\nonumber
    C_{VV}^{(1),\rm eff} & = 2 C_{VV}^{(1),u} + C_{VV}^{(1),d} = \frac{1}{6}\left(\widetilde{A}_{qq}^{(-)} + 3\atqqth\right) + \frac{3}{4} A_{qq}^{(-)} + \aqqth\,,
\end{align}
and similarly for $C_{AA},C_{VA},C_{AV}$, see Eq.~\eqref{eq:4q-up-and-down}. Top-antitop symmetric observables probe $C_{VV}^{(8),\rm eff}$ through interference with the QCD amplitude, while $C_{VV}^{(1),\rm eff}$ only contributes at $\mathcal{O}(\Lambda^{-4})$. This allows us to distinguish between the two flavor contractions $A$ and $\widetilde{A}$ by combining symmetric and antisymmetric top-antitop observables~\cite{Rosello:2015sck,Brivio:2019ius}. Kinematic distributions with boosted tops and/or an additional jet in the final state are also sensitive to the relative impact of color-singlet and color-octet coefficients~\cite{Berge:2012rc,Brivio:2019ius,Basan:2020btr}.

Associated $t\bar t Z$ production is sensitive to the same four-quark operators as $t\bar t$ production. However, due to the emission of the $Z$ boson their relative impact on $t\bar t Z$ observables is different from $t\bar t$ observables. This allows us to probe the structure of four-quark coefficients in new directions, thus further resolving the parameter space.

Associated $t\bar t W$ production probes only four-quark operators with at least one left-handed quark, which couples to the $W$ boson. For a detailed analysis of four-quark contributions to $t\bar t Z$ and $t\bar t W$ production in SMEFT, we refer the reader to Ref.~\cite{Brivio:2019ius}.

In single top and $tZ$ production, the dominant partonic processes are $b\bar{u} \to t\bar{d}(Z)$ and $b\bar{d} \to t\bar{u}(Z)$. Only four-quark interactions with a weak triplet configuration contribute. We distinguish between the color-singlet and color-octet contributions
\begin{align}
    C_{Qq}^{(3,1)}\big(\bar{Q}_3\gamma_\mu\tau^I Q_3\big)\big(\bar{Q}_i\gamma^\mu\tau^I Q_i\big),\qquad     C_{Qq}^{(3,8)}\big(\bar{Q}_3\gamma_\mu\tau^I T^A Q_3\big)\big(\bar{Q}_i\gamma^\mu\tau^I T^A Q_i\big),
\end{align}
with $i\in \{1,2\}$. As in top-antitop production, we have neglected CKM-suppressed contributions from other quark flavors. The color-singlet interaction contributes at $\mathcal{O}(\Lambda^{-2})$ through interference with the SM amplitude; the color-octet interaction enters at $\mathcal{O}(\Lambda^{-4})$.

In MFV, the two combinations read~\cite{AguilarSaavedra:2018nen}
\begin{align}\label{eq:single-top-mfv}
    C_{Qq}^{(3,1)} & = \frac{1}{6}\left(C_{qq}^{(1),3ii3} - C_{qq}^{(3),3ii3}\right) + C_{qq}^{(3),33ii} & = & \ \frac{1}{6} \widetilde{A}_{qq}^{(-)} + \aqqth\\\nonumber
    C_{Qq}^{(3,8)} & = C_{qq}^{(1),3ii3} - C_{qq}^{(3),3ii3} & = & \ \widetilde{A}_{qq}^{(-)}.
\end{align}
By comparing with Eq.~\eqref{eq:ttb-eff}, we see that electroweak top production and top-antitop production probe complementary directions of $C_{qq}^{(1)}$ and $C_{qq}^{(3)}$ in flavor space.

The combinations from Eq.~\eqref{eq:single-top-mfv} also contribute to single top and $tZ$ production via ${d_i\bar{u} \to t\bar{d}(Z)}$ processes with light quarks in the initial state. Compared with $b\bar{u} \to t\bar{d}(Z)$, they are enhanced by the down-quark parton luminosity as $d_i \bar{u}/b\bar{u}$. However, in MFV the Wilson coefficients are suppressed by $|V_{ti}|^2/|V_{tb}|^2$, which cannot be compensated by the partonic enhancement. We therefore neglect these contributions in our analysis. In s-channel single top production, $d\bar{u}\to t\bar{d}$ is the dominant partonic process, both in the Standard Model and in SMEFT.

Previous SMEFT analyses of LHC observables in the top sector have assumed a $U(2)_q\times U(2)_u \times U(2)_d$ flavor symmetry among quarks of the first two generations, as outlined in Ref.~\cite{AguilarSaavedra:2018nen}. The underlying assumption is that top observables are blind to the flavor of light quarks which share the same gauge quantum numbers. In MFV, the leading contributions to top observables are the same as in a $U(2)$-symmetric scenario, assuming that the only sources of flavor breaking are $y_t$ and $y_b$.

\section{Flavor tests with top and bottom observables}\label{sec:results}
Equipped with our analytical results for the top and bottom observables in MFV and their connection within the SMEFT, we carry out
a joint fit to data from the LHC and $b$-factory experiments. In Sec.~\ref{sec:statistics} we describe the fitting framework and the
relevant inputs. In Sec.~\ref{sec:fit-results} we present
the fit results for our selection of top and bottom observables, which showcase a path towards resolving the flavor structure
of the SMEFT coefficients. We consider our analysis as a proof of concept. Using our framework, future analyses can --- and should --- be extended by further observables to gain a more complete picture of possible structures of UV physics.

\subsection{Statistical analysis}\label{sec:statistics}
For our numerical analysis we use the statistics tool \texttt{sfitter}~\cite{Lafaye:2004cn,Lafaye:2007vs,Lafaye:2009vr}, which was mainly developed to perform global fits in high-energy physics, including the Higgs sector~\cite{Klute:2012pu,Corbett:2015ksa,Butter:2016cvz,Biekotter:2018rhp} and the top sector~\cite{Brivio:2019ius}.
 In our fit we build on the global top fit~\cite{Brivio:2019ius} and include the flavor observables $\mathcal{B}(\bsmm)$ and $\mathcal{B}(\bsg)$ without modifications to the statistical treatment in the \texttt{sfitter} setup. Including further flavor observables, chiefly in $B\to K^{(*)}\mu^+\mu^-$, would require substantial
modifications to account for the large number of hadronic nuisance parameters in bottom observables. To overcome this technical bottleneck, in future fits the WET Wilson coefficients should be used as the direct interface between SMEFT fits of high-energy observables and the bounds obtained from low-energy observables. We advocate for the flavor community to provide
a global likelihood of the WET Wilson coefficients from low-energy observables that takes into account all relevant hadronic nuisance
parameters through marginalization or profiling.

In our statistical analysis each term in the likelihood function follows the Rfit scheme~\cite{Hocker:2001xe}, which treats theoretical uncertainties of observables and systematic uncertainties of measurements that are not data-driven as a flat ``core'' around the central value of a fit parameter. At the edges of the core two half-gaussian ``tails'' attach, whose fall-off is governed
by the experimental statistical and data-driven systematic uncertainties. Correlations between experimental systematic uncertainties in the top sector are taken into account as in Ref.~\cite{Brivio:2019ius}.

In the following we briefly discuss the construction of the individual terms of the likelihood.

\paragraph{Flavor observables}
 The likelihood for the $B_s\to \mu^+\mu^-$ branching ratio comprises measurements by the
ATLAS~\cite{Aaboud:2018mst}, CMS~\cite{Sirunyan:2019xdu} and LHCb collaborations~\cite{Aaij:2017vad}.
Each of the three measurements arises from a combined analyses of the branching fractions of $\bsmm$
and $B_d\to \mu^+\mu^-$ decays, the latter being irrelevant to our analysis. We use the average of the three
two-dimensional likelihoods published in Ref.~\cite{LHCb-CONF-2020-002}, which accounts for correlated systematic
uncertainties shared among the three analyses. By marginalizing over $B_d\to \mu^+\mu^-$, we project onto the $\bsmm$ branching fraction and symmetrize the uncertainties around the central value, using the larger of the two uncertainties. In this way we obtain
\begin{equation}
    \mathcal{B}(B_s\to \mu^+\mu^-)_\text{exp} = (2.69 \pm 0.37) \times 10^{-9}\,.
\end{equation}
The experimental uncertainty governs the tails of the Rfit-scheme likelihood, while
the core is spanned by two sources of theory uncertainty around the experimental central value:
the CKM matrix elements ($\pm 1.5\%$) and the $B_s$ decay constant ($\pm 1.2\%$), which are added linearly.

The likelihood for the $\bsg$ branching ratio is based on the measurements at the BaBar~\cite{Aubert:2007my,Lees:2012ym,Lees:2012wg}, Belle~\cite{Limosani:2009qg,Saito:2014das}
and CLEO-2~\cite{Chen:2001fja} experiments. It represents the branching ratio of an
admixture of the two $B$-meson isospin modes produced on the $\Upsilon(4S)$ resonance at the
$b$ factory experiments. The individual measurements of $\mathcal{B}(\bsg)$ are carried
out for a variety of cuts $E_\gamma \geq E_{\gamma,\text{min}}$ on the photon energy
in the $B$-meson rest frame. We use the world average that encompasses each measurement extrapolated to a cut ${E_{\gamma,\text{min}} = 1.9\GeV}$~\cite{Amhis:2019ckw,Zyla:2020zbs},
\begin{equation}
    \mathcal{B}(B\to X_s\gamma)_\text{exp} = \left(3.49 \pm 0.19\right) \times 10^{-4}\,.
\end{equation}
The experimental uncertainty governs the tails of the Rfit-scheme likelihood.
The core is spanned by several sources of theory uncertainties around the experimental central value.
We use a theory uncertainty of $\pm 5\%$, which corresponds to the uncertainty obtained in Ref.~\cite{Misiak:2020vlo}.

\paragraph{Top observables}
Our selection of top observables is --- up to a few modifications that we point out below --- identical to the set in Ref.~\cite{Brivio:2019ius}. The predictions for top-antitop and single-top production are calculated in SMEFT at NLO QCD. The likelihood for the top sector comprises a large set of observables:
\begin{itemize}
    \item the $t\bar{t}$ production cross section in $87$ measurements and the $t\bar{t}$ production charge asymmetry in five measurements;
    \item the $t\bar{t}Z$ and $t\bar{t}W$ production cross section in four measurements in total;
    \item the $t$-channel and $s$-channel single top production cross sections in $13$ measurements;
    \item the $tZ$ and $tW$ production cross section in seven measurements in total;
    \item and the $W$ helicity fractions in top decay in eight measurements.
\end{itemize}
Recent global top fits focus on SMEFT operators with top couplings and discard operator contributions with
light quarks only~\cite{Hartland:2019bjb,Brivio:2019ius}. Unlike these analyses, our fit accounts for light-quark couplings where they are numerically relevant. In particular, we include contributions
of $C_{\phi q}^{(3),ii}$ with $i\in\lbrace 1, 2 \rbrace$ in single top production
and of $C_{\phi q}^{(-),ii}$, $C_{\phi q}^{(3),ii}$ with $i\in \lbrace 1, 2\rbrace$ in $tZ$, $t\bar t Z$, and $t\bar t W$ production. In these cases our theory prediction of the SMEFT effects is at LO in QCD.

We also include LO electroweak SMEFT contributions to $t\bar t Z$ and $t\bar t W$ production observables, which had not been considered in previous analyses. We find that electroweak SMEFT contributions are numerically relevant. This is evident from Eq.~(\ref{eq:phiq-top}), where we highlight the electroweak contributions in boldface. Our fits show that electroweak SMEFT contributions
are crucial for an accurate interpretation of the data, see Fig.~\ref{fig:phiq-ew}.


\subsection{Fit results and discussion}\label{sec:fit-results}
Based on our analysis of SMEFT effects in bottom and top observables in Secs.~\ref{sec:flavor} and \ref{sec:top}, 
we now compare these predictions with data and derive bounds on the flavor structure of a UV theory in Minimal Flavor Violation.

We classify the top and bottom observables included in our analysis according to their sensitivity to the SMEFT operators
listed in Eqs.~\eqref{eq:2q-operators} and \eqref{eq:4q-operators}:
\begin{align}
    \text{bottom:} & \quad O_{\phi d},\ O_{\phi ud},\ O_{dB},\ O_{dW},\ O_{dG}\\\nonumber
    \text{top:} & \quad O_{\phi u},\ O_{uB},\ O_{uW},\ O_{uG},\ O_{uu},\ O_{qu}^{(8)},\ O_{qd}^{(1)},\ O_{qd}^{(8)},\ O_{ud}^{(1)},\ O_{ud}^{(8)}\\\nonumber
    \text{top \& bottom:} & \quad O_{\phi q}^{(1)},\ O_{\phi q}^{(3)},\ O_{qq}^{(1)},\ O_{qq}^{(3)},\ O_{qu}^{(1)}.
\end{align}
The operators $O_{quqd}^{(1)}$, $O_{quqd}^{(8)}$ and $O_{dd}$ are not probed by any of the observables we consider.

The operators in the first row are probed in bottom observables only. These are all operators with scalar or tensor currents with right-handed down quarks. Their contributions to top observables are suppressed by the bottom quark mass or by $y_b$ in MFV.

The operators in the second row are best probed in top observables. Two-quark operators with right-handed tops also contribute to flavor observables, but with a loop suppression.

The operators in the third row are equally well probed by both top and bottom observables. All of these operators involve weak doublets of left-handed up and down quarks. Due to the built-in $SU(2)_L$ invariance in SMEFT, effective top and bottom couplings are related through weak interactions. This relation leads to non-trivial correlations of SMEFT effects in top and bottom observables. In our numerical analysis we will therefore focus on these five operators.

In what follows we show that by combining top and bottom observables we gain information about the flavor structure of SMEFT coefficients that cannot be obtained from either top or bottom observables in isolation.
To this end, we perform dedicated tests of flavor universality for the Wilson coefficients $C_{\phi q}^{(1)},\, C_{\phi q}^{(3)}$
and $C_{qq}^{(1)},\, C_{qq}^{(3)}$ in MFV. All SMEFT coefficients and flavor parameters are defined at the reference scale
$\mu_0 = m_t$.
By default we present our numerical bounds on the SMEFT flavor parameters in two-dimensional contour plots. The contour lines correspond to $\Delta\chi^2=2.30$ and $\Delta\chi^2=5.99$ for the profiled likelihoods,
\emph{i.e.}, the projections of the full $n$-dimensional likelihoods onto a two-dimensional surface by means of profiling. In a fully Gaussian case, these contours correspond to the 68\% and 95\% confidence levels.

\paragraph{Two-quark operators} In MFV the Wilson coefficients $C_{\phi q}^{(1)}$ and $C_{\phi q}^{(3)}$ are determined by the four flavor parameters
\begin{align}\label{eq:phiq-param}
a_{\phi q}^{(-)},\ b_{\phi q}^{(-)},\ a_{\phi q}^{(3)},\ b_{\phi q}^{(3)}\,.
\end{align}
Electroweak processes are sensitive to either all or a subset of these parameters. The relevant top observables at the LHC with $13$ TeV collision energy can be written as
\begin{align}\label{eq:phiq-top}
    \sigma_{t}\ [\text{pb}] & = 126 + 15.1\big[a_{\phi q}^{(3)} + A_{\phi q}^{(3)}\big] + 0.5\big[a_{\phi q}^{(3)} + A_{\phi q}^{(3)}\big]^2\\\nonumber
    \sigma_{tW}\ [\text{pb}] & = 75.3 + 9.1 A_{\phi q}^{(3)} + 0.27 \big(A_{\phi q}^{(3)}\big)^2\\\nonumber
    \sigma_{tZ}\ [\text{pb}] & = 0.78 + 0.17\big[a_{\phi q}^{(3)} + A_{\phi q}^{(3)}\big] + 0.01 a_{\phi q}^{(-)} + 0.10 \big(a_{\phi q}^{(3)}\big)^2 + 0.02 \big(A_{\phi q}^{(3)}\big)^2\\\nonumber
    \sigma_{t\bar t Z}\ [\text{pb}] & = 0.679 + 0.023 a_{\phi q}^{(3)}  - 0.070 A_{\phi q}^{(-)} + {\bf 0.008} \big(a_{\phi q}^{(-)}\big)^2 + {\bf 0.004}\big[2a_{\phi q}^{(3)} + a_{\phi q}^{(-)}\big]^2\\\nonumber
    \sigma_{t\bar t W}\ [\text{pb}] & = 0.446 + (0.054 + {\bf 0.008})a_{\phi q}^{(3)} + {\bf 0.062} \big(a_{\phi q}^{(3)}\big)^2,
\end{align}
and the bottom observables are given by
\begin{align}\label{eq:phiq-flavor}
    \mathcal{B}(\bsg)\times 10^4 & = 3.26 + 0.36\,a_{\phi q}^{(3)} - 0.76\,b_{\phi q}^{(3)}\\\nonumber
    \mathcal{B}(\bsmm)\times 10^9 & = 3.57 - 41.0\big[2 b_{\phi q}^{(3)} + b_{\phi q}^{(-)}\big]
    + 117.8\big[2b_{\phi q}^{(3)}+ b_{\phi q}^{(-)}\big]^2,
\end{align}
up to percent corrections or less for $a,b \sim 1$. In Tab.~\ref{tab:2q-num} in App.~\ref{app:polynomials}, we list the full numerical expressions, including electroweak corrections to $t\bar t Z$ and $t\bar t W$ production. Contributions of $a_{\phi q}^{(3)},a_{\phi q}^{(-)}$ and $A_{\phi q}^{(3)},A_{\phi q}^{(-)}$ are due to light-quark and top-quark couplings, respectively. While top observables probe flavor-violating effects in the combination $A = a+b y_t^2$, \bsg and \bsmm also involve the flavor-breaking parameter $b$ alone. In Eq.~\eqref{eq:phiq-top} we have highlighted sizeable electroweak contributions in boldface.

\begin{figure}[t]
    \centering
   \includegraphics[width=0.33\textwidth]{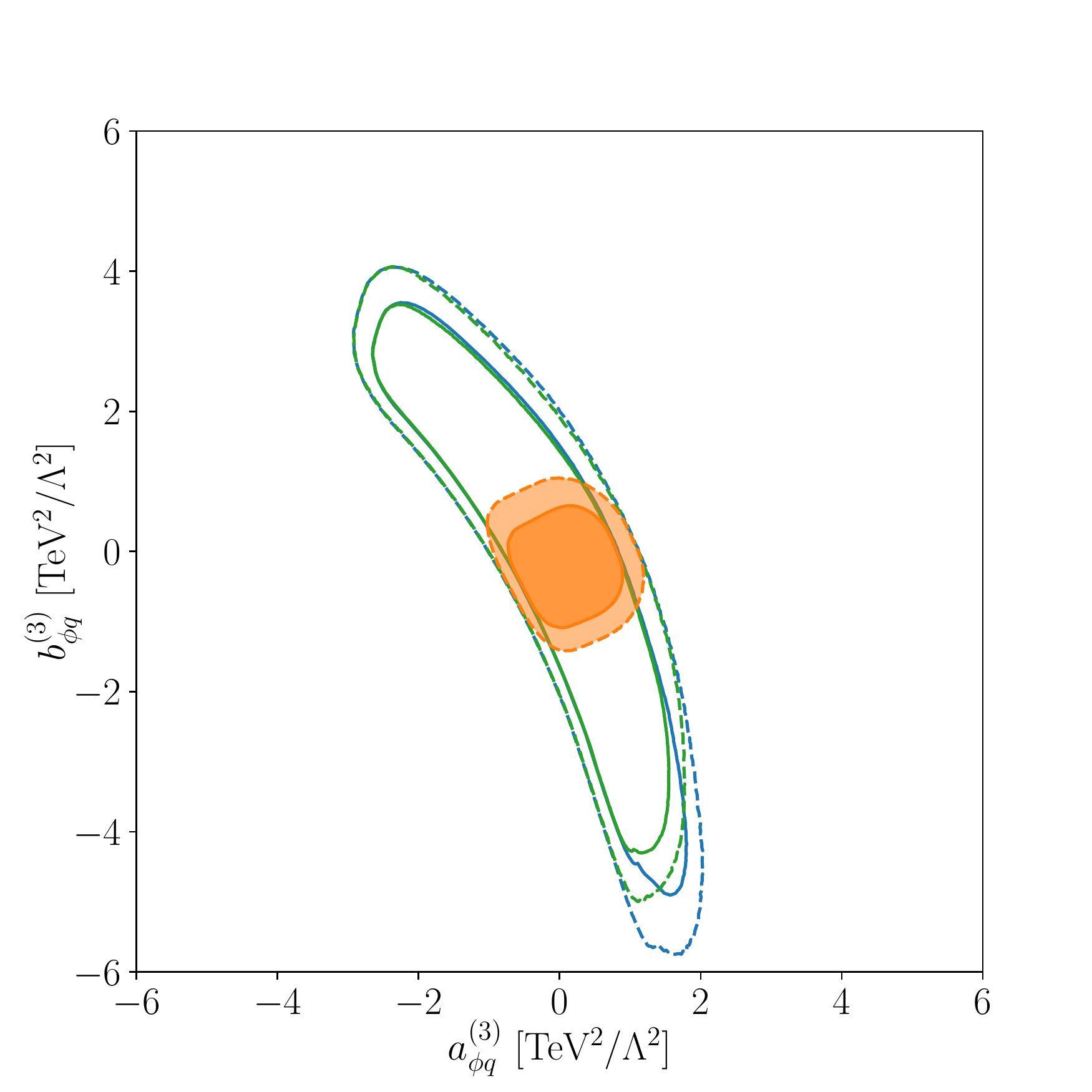}\hspace*{0.18cm}
   \includegraphics[width=0.33\textwidth]{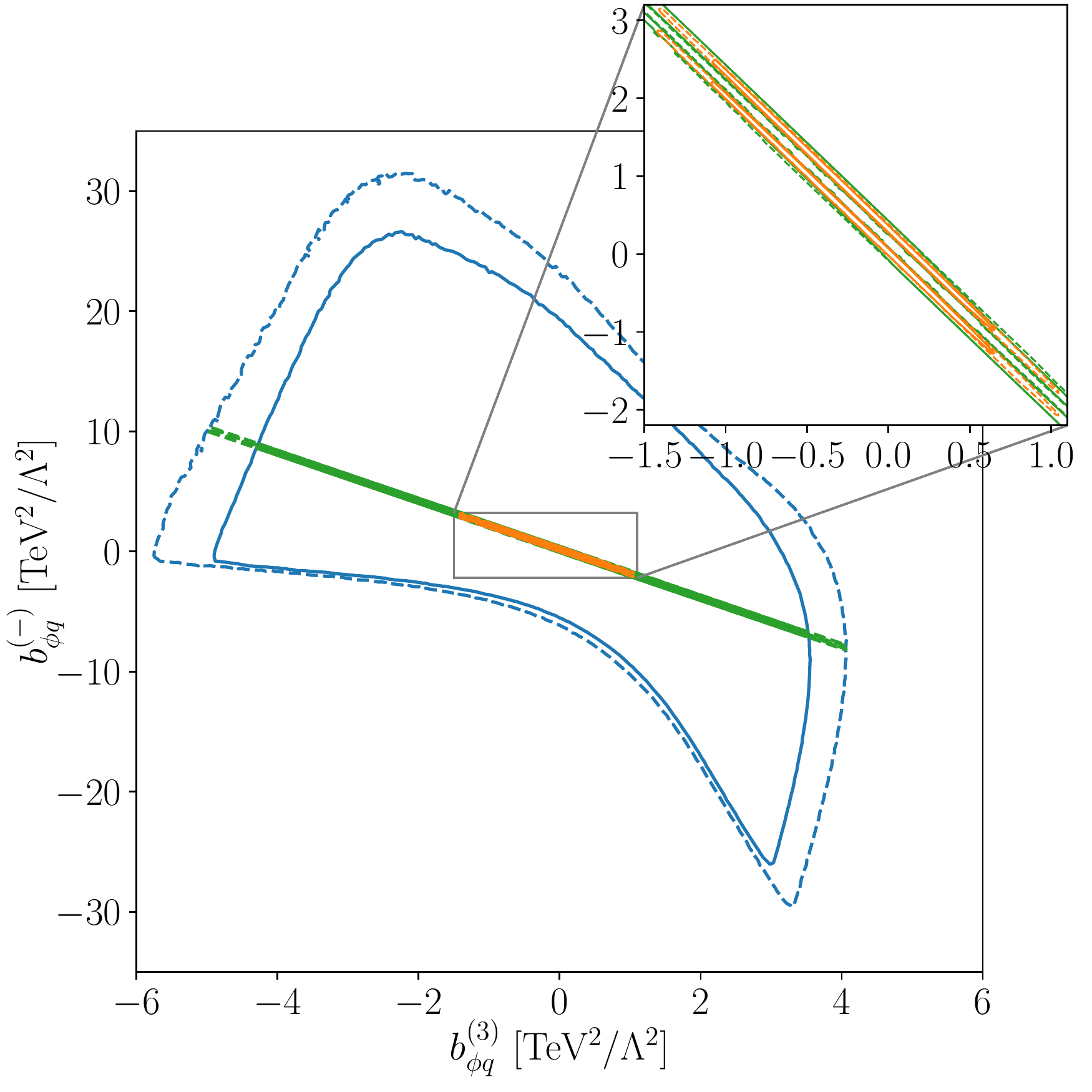}\hspace*{0.18cm}
   \includegraphics[width=0.33\textwidth]{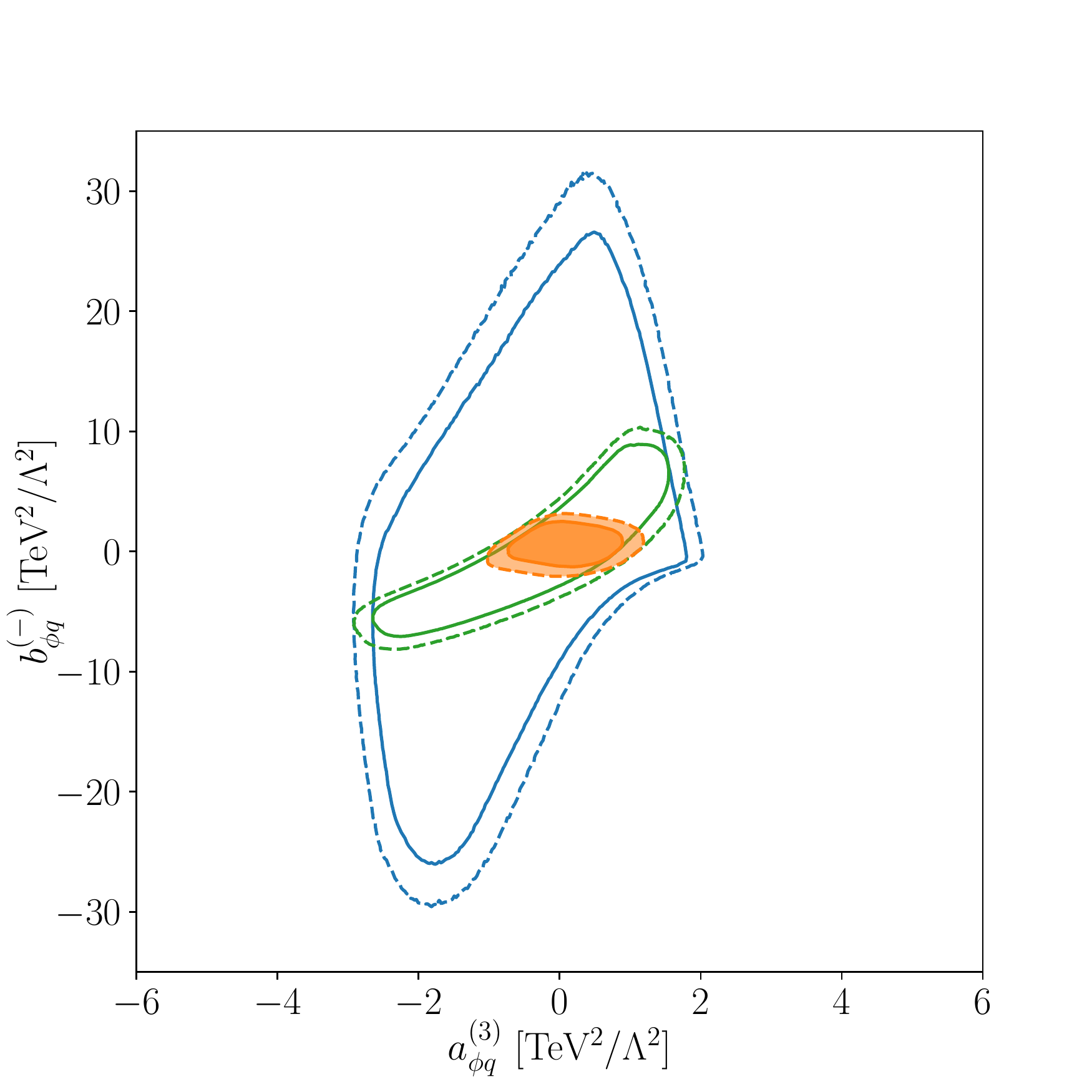}
    \caption{Flavor structure of two-quark Wilson coefficients $C_{\phi q}^{(-)}(m_t) = C_{\phi q}^{(1)} - C_{\phi q}^{(3)}$ and $C_{\phi q}^{(3)}(m_t)$ in MFV. Shown are the $\Delta \chi^2=2.30$ (solid) and $\Delta \chi^2=5.99$ (dashed) contours obtained from a four-parameter fit of $\{a_{\phi q}^{(-)},b_{\phi q}^{(-)},a_{\phi q}^{(3)},b_{\phi q}^{(3)}\}$ to top (blue), top \& \bsmm (green), and top \& \bsmm \& \bsg (orange) data. Left: flavor universality test of weak triplet interactions $a_{\phi q}^{(3)},b_{\phi q}^{(3)}$. Center: flavor breaking in charged currents ($b_{\phi q}^{(3)}$) versus neutral up-quark currents ($b_{\phi q}^{(-)}$). Right: interplay of charged currents ($a_{\phi q}^{(3)}$) and neutral up-quark currents ($b_{\phi q}^{(-)}$).}
    \label{fig:phiq}
\end{figure}

In Fig.~\ref{fig:phiq} we display the results of a four-parameter fit to the top and flavor observables described in Sec.~\ref{sec:statistics}: top observables (blue), top \& \bsmm (green), and top \& \bsmm \& \bsg (orange). Shown are the $\Delta \chi^2=2.30$ (solid) and $\Delta \chi^2=5.99$ (dashed) contours for selected pairs of flavor parameters, obtained by profiling over the remaining two parameters. A global fit of electroweak top observables sets bounds in all directions of the parameter space spanned by Eq.~\eqref{eq:phiq-param}. Including \bsg and \bsmm in the fit significantly enhances the resolution of the gauge and flavor structure for the considered operators.

The left panel shows bounds on flavor universality breaking for $C_{\phi q}^{(3)}$. In a flavor-universal UV theory, $b_{\phi q}^{(3)} = 0$ and all SMEFT contributions lie along the horizontal line $(a_{\phi q}^{(3)},0)$. Single top production constrains the direction $(2,1)$, cf. Eq.~\eqref{eq:phiq-top}. The orthogonal direction $(1,-2)$ is bounded by contributions of $(a_{\phi q}^{(3)})^2$ and $(A_{\phi q}^{(3)})^2$ in $tZ$, $tW$ and $t\bar t W$ production. In combination, these top observables (blue) are sensitive to flavor universality breaking through the interplay of effective couplings with light quarks and top quarks. Adding {\bsmm} to the fit (green) does not add further bounds on $b_{\phi q}^{(3)}$, because $b_{\phi q}^{(-)}$ is only loosely bounded by $\sigma_{t\bar t Z}$ and compensates $b_{\phi q}^{(3)}$ contributions in a combined fit. Measurements of $\mathcal{B}(\bsg)$ probe the direction $(1,-2.1)$, cf. Eq.~\eqref{eq:phiq-flavor}, which accidentally is almost aligned with the least bounded direction in top observables. Adding \bsg to the fit (orange) results in a much stronger bound on flavor universality breaking than from top observables alone. Remarkably, the bound from \bsg is due to a one-loop effect, as $C_{\phi q}^{(1)}$ and $C_{\phi q}^{(3)}$ do not contribute to $\mathcal{C}_7$ and $\mathcal{C}_8$ at tree level, see Tab.~\ref{tab:smeft-to-wet-loop}.

In the central panel, we illustrate how to distinguish between flavor universality breaking in charged and neutral weak currents. Among the various top processes, only $t\bar t Z$ production is sensitive to flavor breaking in neutral currents through $A_{\phi q}^{(-)}$, see Eq.~\eqref{eq:phiq-top}. The bounds in the $(b_{\phi q}^{(3)},b_{\phi q}^{(-)})$ plane arise from a non-trivial interplay of different top observables, because the two parameters can only be probed together with $a_{\phi q}^{(3)}$ and $a_{\phi q}^{(-)}$. Resolving the $(b_{\phi q}^{(3)},b_{\phi q}^{(-)})$ parameter space is only possible in a combined analysis of at least four electroweak top processes. Among the flavor observables, $\mathcal{B}(\bsmm)$ plays a big role in pinning down the amount of flavor breaking in the electroweak sector. The process is highly sensitive to the direction $(2,1)$, because the flavor breaking induces down-quark FCNCs at tree level, see Eq.~\eqref{eq:phiq-flavor}. The two narrow ellipses (green) are due to the shift of the quadratic contributions $(2b_{\phi q}^{(3)} + b_{\phi q}^{(-)})^2$ in \bsmm by a linear term $2b_{\phi q}^{(3)} + b_{\phi q}^{(-)}$, which leads to two distinct maxima in the likelihood function. Moreover, when including \bsmm the fit disfavors the Standard Model $(0,0)$ at the 95\% CL, due to the current discrepancy in $\mathcal{B}(\bsmm)$. Adding \bsg (orange) sets an additional bound on $b_{\phi q}^{(3)}$. The combined analysis leaves room for new physics in the direction $(1,-2)$. This includes the possibility of simultaneous flavor universality breaking in charged currents $(b_{\phi q}^{(3)})$ and neutral currents among up-type quarks $(b_{\phi q}^{(-)})$.

In the right panel of Fig.~\ref{fig:phiq}, we focus on the interplay of charged currents $a_{\phi q}^{(3)}$ and neutral flavor-breaking currents $b_{\phi q}^{(-)}$ in top and flavor observables. Both parameters enter the observables in combination with $b_{\phi q}^{(3)}$ and/or $a_{\phi q}^{(-)}$, see Eqs.~\eqref{eq:phiq-top} and \eqref{eq:phiq-flavor}. To resolve the $(a_{\phi q}^{(3)},b_{\phi q}^{(-)})$ parameter space, a combined analysis of top and flavor data is therefore indispensable. Top observables alone (blue) only set a loose bound on $b_{\phi q}^{(-)}$ through $t\bar t Z$ production. Adding $\bsmm$ to the fit (green) enhances the sensitivity to $b_{\phi q}^{(-)}$ and introduces a correlation with $a_{\phi q}^{(3)}$, which is bound in top observables. Finally, adding \bsg (orange) leads to an even stronger bound on $b_{\phi q}^{(-)}$, even though \bsg is not directly sensitive to neutral vector currents. The enhanced sensitivity to $a_{\phi q}^{(3)}$ and $b_{\phi q}^{(-)}$ in the four-parameter fit when including flavor observables is largely due to the interplay of \bsg, \bsmm and top observables. This example illustrates that in a combined fit flavor-universal effects can have an impact on the sensitivity to flavor-breaking effects --- and vice versa.

\begin{figure}[t!]
    \centering
   \includegraphics[width=0.45\textwidth]{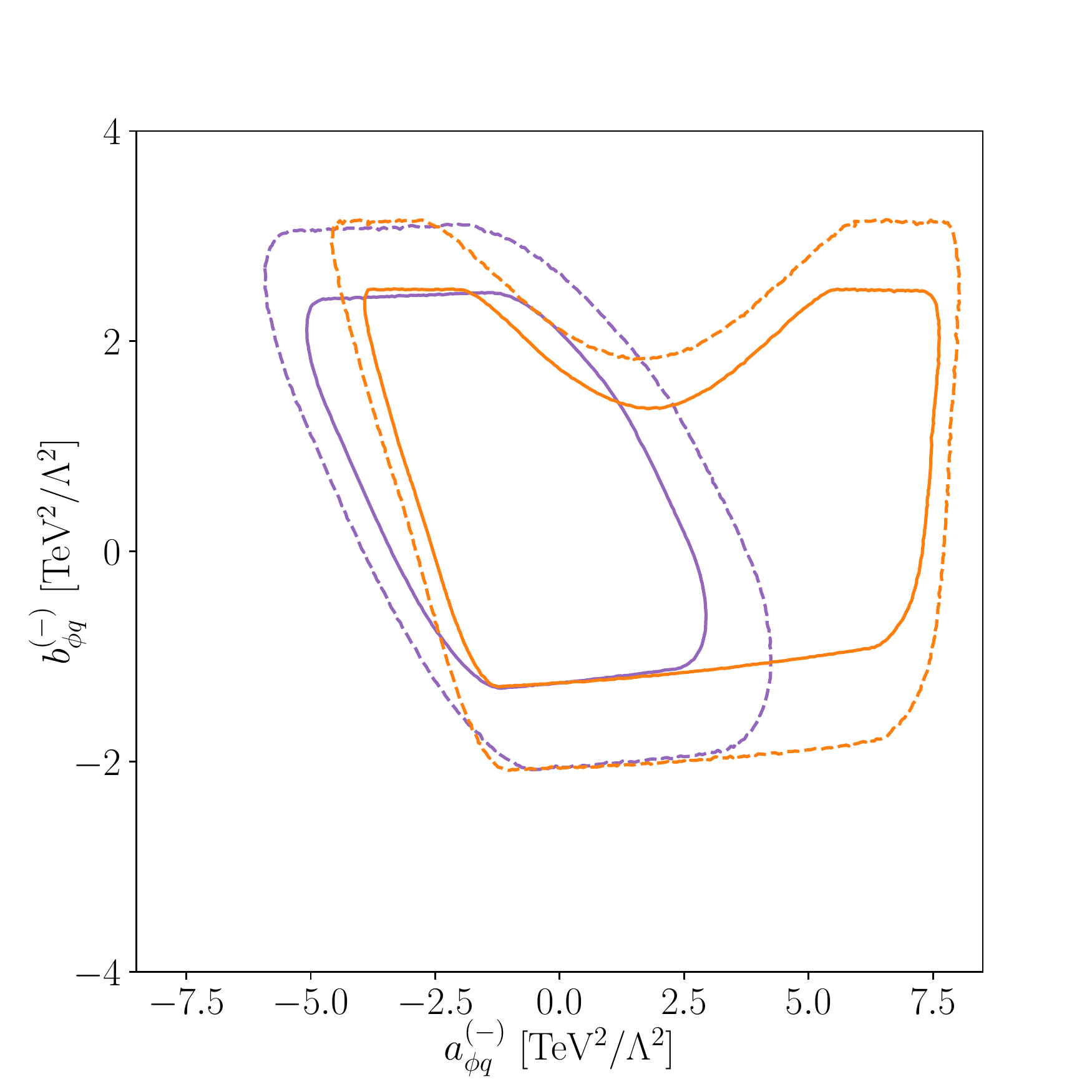}\hspace*{1cm}
   \includegraphics[width=0.45\textwidth]{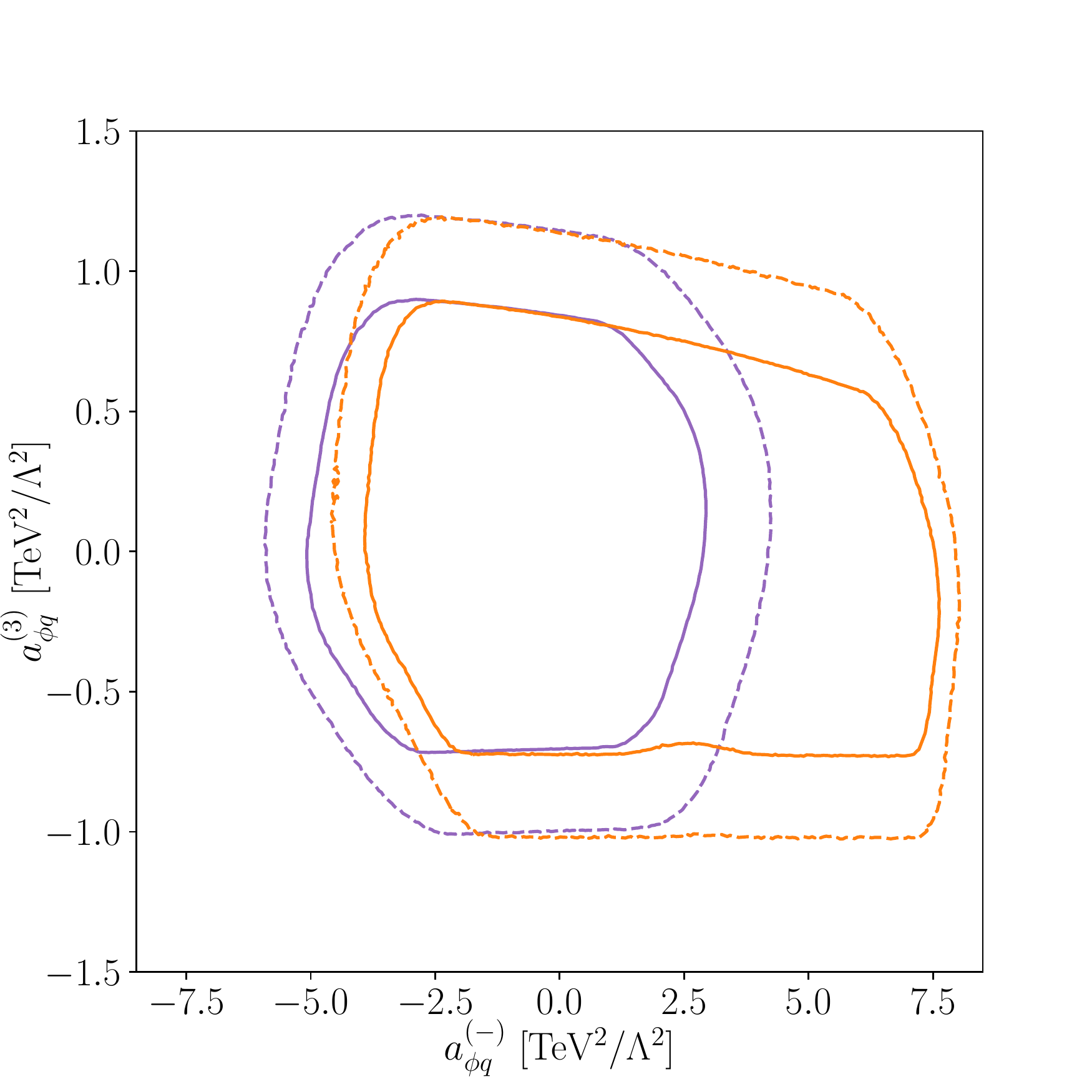}
    \caption{Impact of electroweak contributions of $C_{\phi q}^{(-)}(m_t) = C_{\phi q}^{(1)} - C_{\phi q}^{(3)}$ and $C_{\phi q}^{(3)}(m_t)$ to $t\bar t Z$ and $t\bar t W$ production in MFV. Shown are the $\Delta \chi^2=2.30$ (solid) and ${\Delta \chi^2=5.99}$ (dashed) contours obtained from four-parameter fits of $\{a_{\phi q}^{(-)},b_{\phi q}^{(-)},a_{\phi q}^{(3)},b_{\phi q}^{(3)}\}$ to top observables with (orange) and without (purple) electroweak contributions. \bsmm and \bsg are also included in the fits. Left: flavor universality test of weak neutral currents with up-quarks ($a_{\phi q}^{(-)},b_{\phi q}^{(-)}$). Right: flavor-universal neutral currents ($a_{\phi q}^{(-)}$) versus charged currents ($a_{\phi q}^{(3)}$).}
    \label{fig:phiq-ew}
\end{figure}

We close our discussion with an important comment on the impact of electroweak contributions in $t\bar t Z$ production, which so far have not been considered in the literature. Partonic processes like $q\bar q \to Z^\ast \to t\bar t Z$ affect the sensitivity to $a_{\phi q}^{(3)}$ and $a_{\phi q}^{(-)}$, see Eq.~\eqref{eq:phiq-top}. While $a_{\phi q}^{(3)}$ by itself is well constrained by $\sigma_{t}$, $\sigma_{tW}$ and $\sigma_{t\bar t}$, the bound on $a_{\phi q}^{(-)}$ relies significantly on parameter correlations in $\sigma_{t\bar t Z}$, \bsg and \bsmm. In Fig.~\ref{fig:phiq-ew}, we show the bounds on $C_{\phi q}^{(1)}$ and $C_{\phi q}^{(3)}$ in MFV, as obtained from a combined top-flavor fit including (orange) and discarding (purple) electroweak contributions to $t\bar t Z$ and $t\bar t W$ production.
 From Eq.~\eqref{eq:phiq-flavor} it is apparent that \bsmm favors $2 b_{\phi q}^{(3)} + b_{\phi q}^{(-)} \neq 0$ to
 account for the large deviation from the SM prediction in this measurement.
 In a combined fit, a sizeable contribution $b_{\phi q}^{(-)} > 0$ to $\sigma_{t\bar t Z}$ entails an upper bound on $a_{\phi q}^{(-)}$, due to the negative interference of $A_{\phi q}^{(-)} = a_{\phi q}^{(-)} + b_{\phi q}^{(-)} y_t^2$ (purple contours). This bound is relaxed by electroweak contributions, which re-open the parameter space for $a_{\phi q}^{(-)} > 0$ (orange contours) through an interplay of (negative) linear and (positive) quadratic contributions of $a_{\phi q}^{(-)}$ to $\sigma_{t\bar t Z}$. As $t\bar t Z$ production plays a key role to determine the flavor structure of effective operators with weak gauge bosons, electroweak contributions should be included in SMEFT parameter fits.

\paragraph{Four-quark operators} Besides the two-quark operators $O_{\phi q}^{(1)}$ and $O_{\phi q}^{(3)}$, three four-quark operators $O_{qq}^{(1)}$, $O_{qq}^{(3)}$ and $O_{qu}^{(1)}$ contribute to both top observables and \bsmm. Notice that \bsg is not sensitive to any of these parameters and does not play a role in our analysis here. In our attempt to pin down the flavor structure of SMEFT coefficients, we focus mostly on $O_{qq}^{(1)}$, $O_{qq}^{(3)}$, \emph{i.e.}, on operators with left-handed quarks only. In MFV, the relevant degrees of freedom that can be probed in top observables are (see Eq.~\eqref{eqn:A-At-definitions})
\begin{align}
\aqqm, \atqqm, \aqqth, \atqqth.
\end{align}
These four parameters can be resolved in a combined fit of top-antitop production and electroweak top production, cf. Eqs.~\eqref{eq:ttb-eff} and \eqref{eq:single-top-mfv}. In addition, \bsmm is sensitive to the combination
\begin{align}
     B_{qq} & = \sin\theta\, B_{qq}^{(-)}(m_t) + \cos\theta\, B_{qq}^{(3)}(m_t),\qquad \sin\theta = 0.09
\end{align}
through the Wilson coefficient (see Eq.~\eqref{eqn:4q-matching})
\begin{align}\label{eq:c10-mfv}
    \mathcal{C}_{10}(m_b) = 0.29 \Big(A_{qq}^{(3)}(m_t) + B_{qq}^{(3)}(m_t) y_t^2\Big) + 0.03 \Big(\atqqm(m_t) + B_{qq}^{(-)}(m_t) y_t^2\Big).
\end{align}
As mentioned in Sec.~\ref{sec:matching}, the sensitivity of $\mathcal{C}_{10}$ to $F_{qq}^{(3)}(\mu) = A_{qq}^{(3)}(\mu) + B_{qq}^{(3)}(\mu) y_t^2$ and $F_{qq}^{(-)}(\mu) = \atqqm(\mu) + B_{qq}^{(-)}(\mu) y_t^2$  depends on the choice of the scale $\mu$, due to operator mixing in the SMEFT RG evolution. By comparing the matching relations in Tab.~\ref{tab:smeft-to-wet-num-Matching} with Tabs.~\ref{tab:smeft-to-wet-num} and \ref{tab:smeft-to-wet-num-Matching-runningWET}, we see that $\mathcal{C}_{10}$ has a good sensitivity to $F_{qq}^{(3)}(m_Z)$ and $F_{qq}^{(-)}(m_Z)$, while the sensitivity to $F_{qq}^{(3)}(m_t)$ and $F_{qq}^{(-)}(m_t)$ is much lower. As a consequence, the relative impact of top and flavor observables in the SMEFT fit is sensitive to the energy scale probed in top observables.

\begin{figure}[t!]
    \centering
    \includegraphics[width=0.5\textwidth]{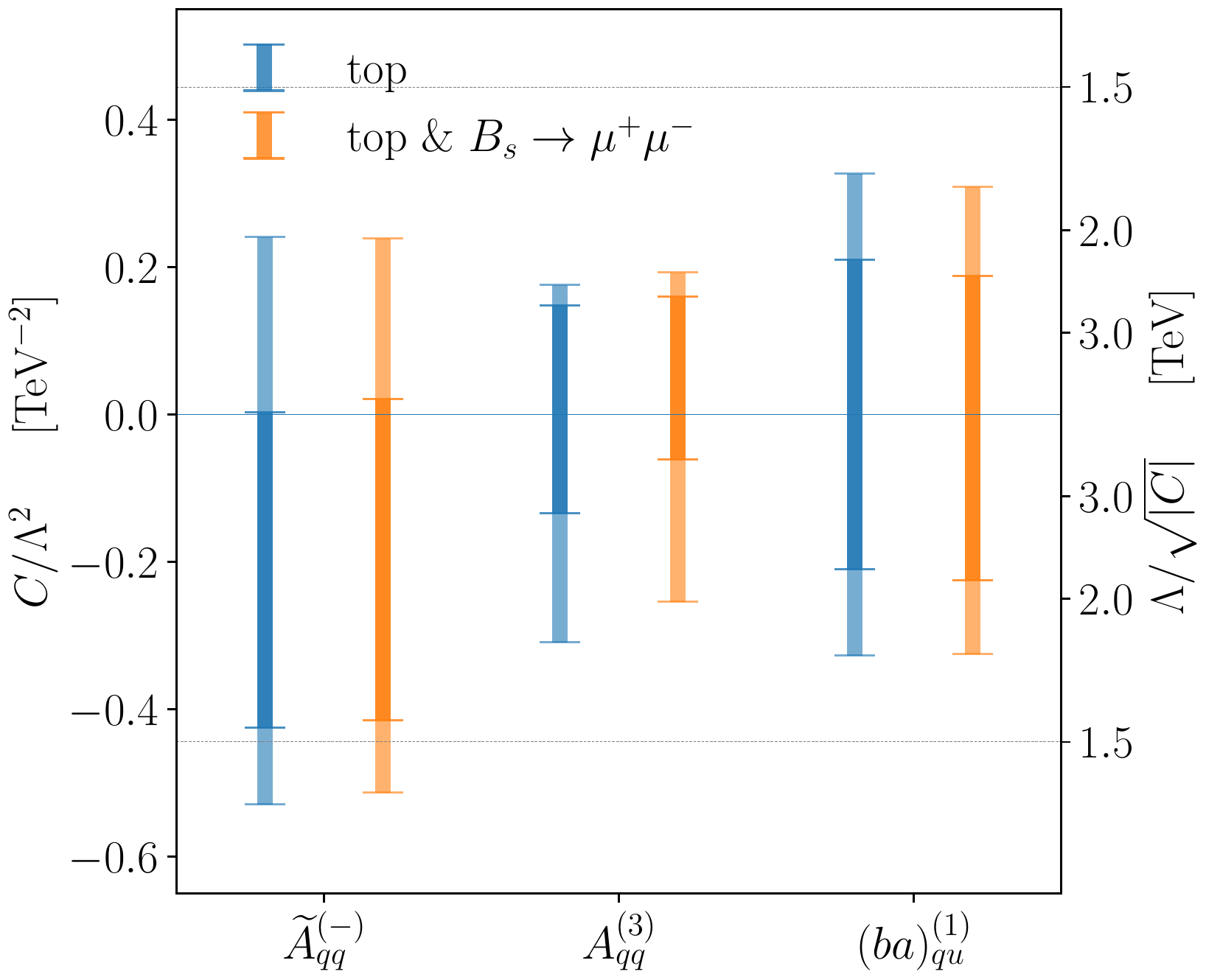}
    \caption{Bounds on the four-quark Wilson coefficients $C_{qq}^{(-)}(m_t) = C_{qq}^{(1)} - C_{qq}^{(3)}$, $C_{qq}^{(3)}(m_t)$, and $C_{qu}^{(1)}(m_t)$ in MFV. Shown are the results from one-parameter fits of $\atqqm$, $\aqqth$ and $\ba_{qu}^{(1)}$, see Eqs.~\eqref{eqn:A-At-definitions} and \eqref{eqn:4q-matching}, to top observables (blue) and top \& \bsmm (orange) at 68\% (dark shade) and 95\% CL (light shade).\label{fig:qq_1d}}
\end{figure}

Before analyzing the flavor structure of the four-quark coefficients, it is interesting to compare the sensitivity of top and flavor observables to individual flavor parameters. To this end, we have performed separate fits to $\atqqm$, $\aqqth$ and $\ba_{qu}^{(1)}$, \emph{i.e.}, to those parameters that contribute to both top observables and \bsmm. In Fig.~\ref{fig:qq_1d}, we show the bounds obtained from one-parameter fits of top observables (blue) and top \& \bsmm (orange).
%
 Top observables are sensitive to all three parameters through top-antitop production (see Eq.~\eqref{eq:ttb-eff}) and single top production (see Eq.~\eqref{eq:single-top-mfv}). Adding \bsmm does add sensitivity through loop-induced effects in $\mathcal{C}_{10}$, see Eq.~\eqref{eq:c10-mfv}, but has only a mild impact on the bounds obtained in a combined fit. Notice, however, that the impact of \bsmm depends on the assumed scale $\mu = m_t$ for the Wilson coefficients, due to SMEFT operator mixing.

While the magnitude and gauge structure of four-quark operator coefficients are well constrained by top observables alone~\cite{Brivio:2019ius}, a combined analysis with flavor observables like \bsmm is indispensable to disentangle the flavor structure of these coefficients. In what follows we will test the flavor structure of $C_{qq}^{(-)}$ and $C_{qq}^{(3)}$ with respect to flavor universality and flavor contractions $A$ versus $\widetilde{A}$.

\paragraph{Flavor universality} In MFV, top observables always probe the combinations $A_{qq}=\aa + \ba y_t^2$ and $\widetilde{A}_{qq}=\aat + \bat y_t^2$ through couplings with two light quarks and two third-generation quarks, see Eq.~\eqref{eqn:A-At-definitions}. As the contributions $\aa$ and $\ba$ cannot be distinguished, top observables alone are not effective in probing flavor universality. In turn, \bsmm is sensitive to the flavor-breaking combination $B_{qq}$ from Eq.~\eqref{eq:c10-mfv}, calling for a combined fit of top \& \bsmm to test flavor universality in four-quark coefficients.

\begin{figure}[t!]
    \centering
       \includegraphics[width=0.34\textwidth]{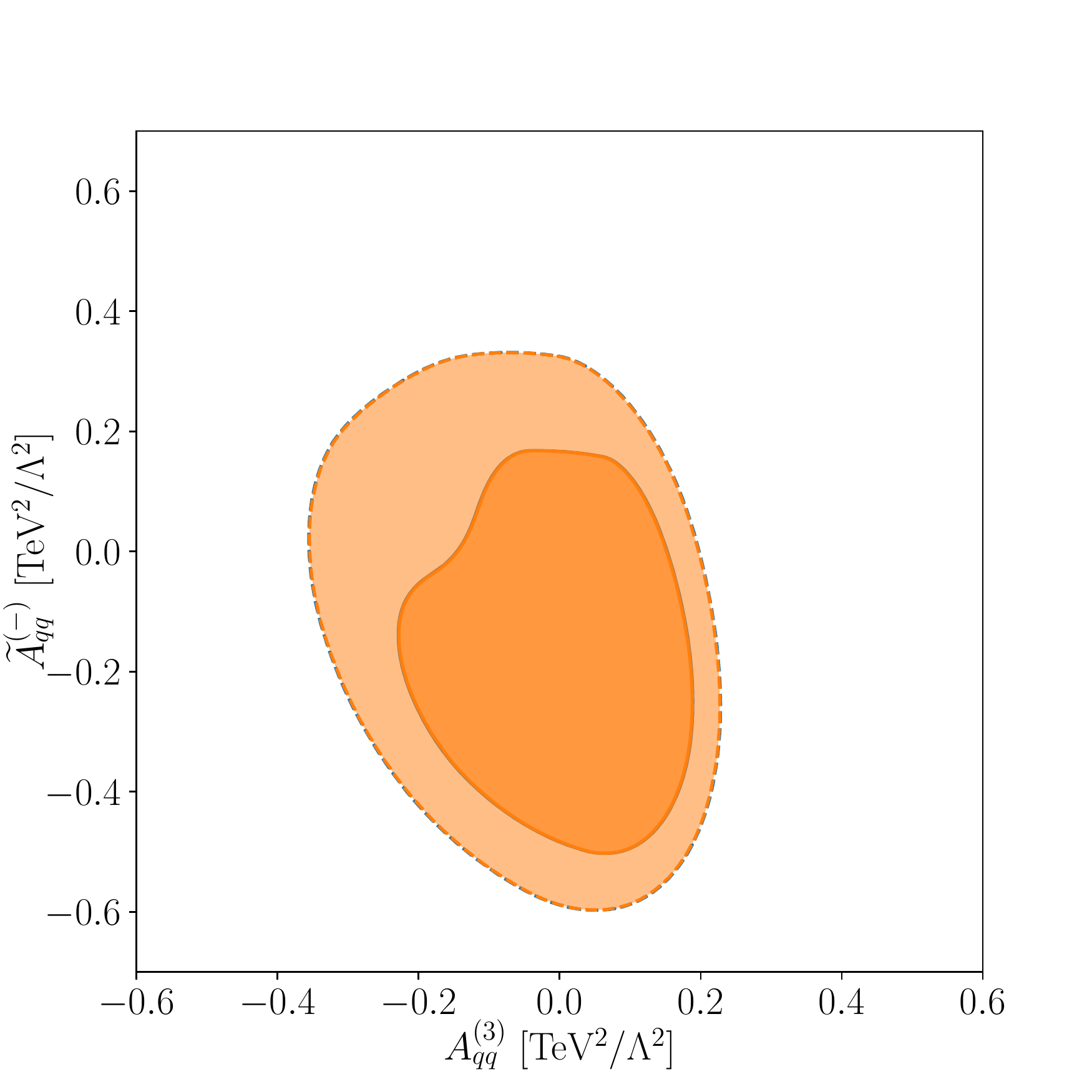}\hspace*{-0.05cm}
   \includegraphics[width=0.34\textwidth]{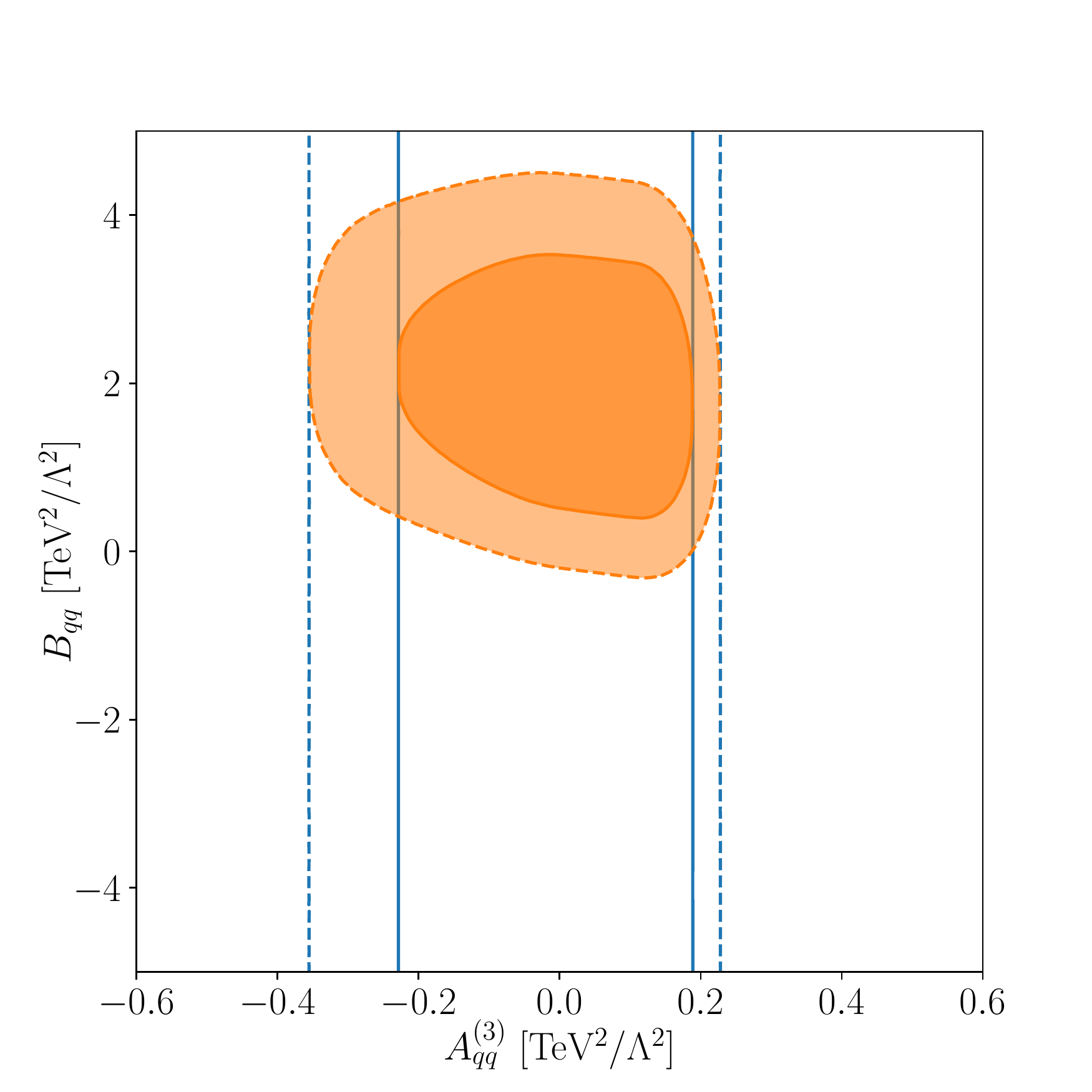}\hspace*{-0.05cm}
   \includegraphics[width=0.34\textwidth]{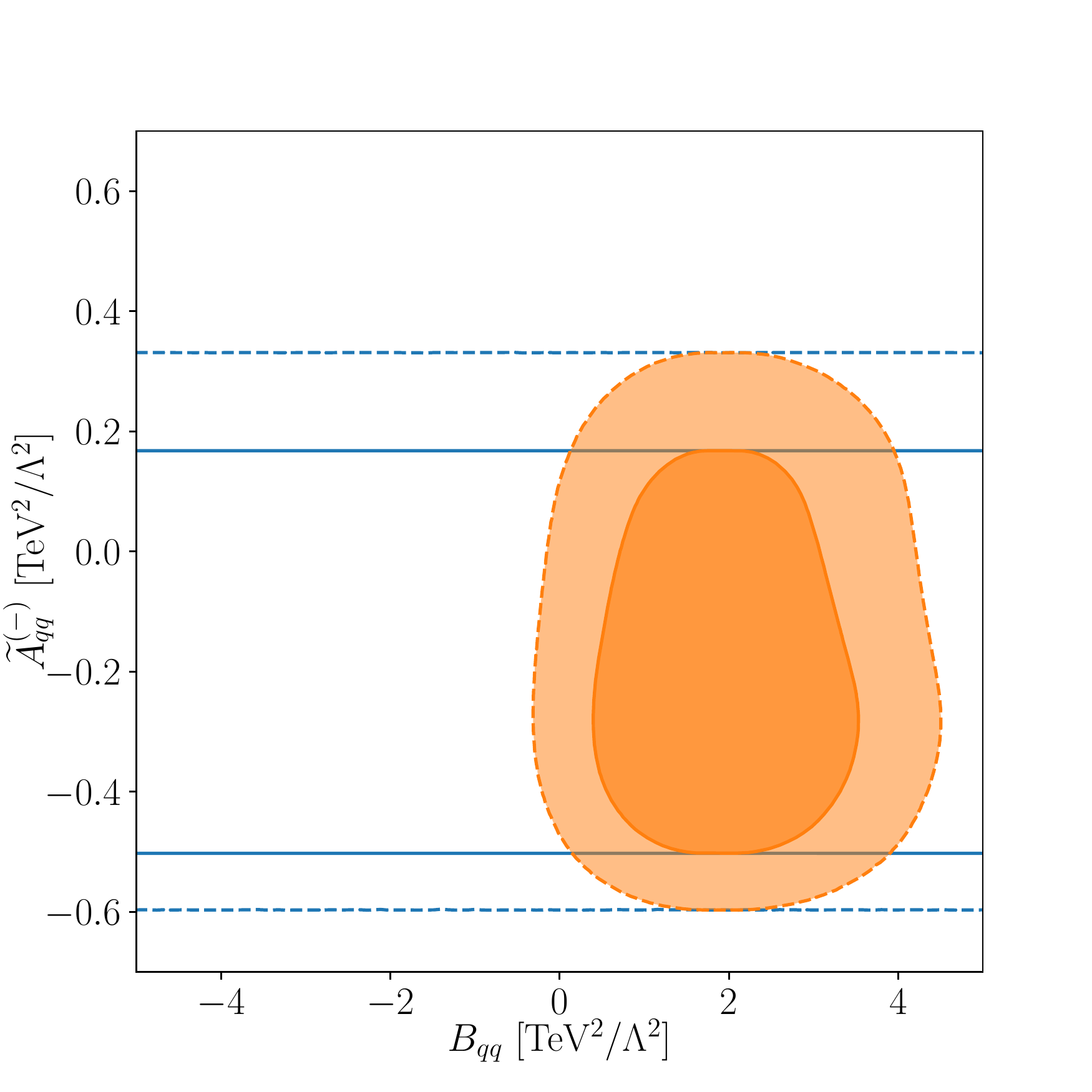}
    \caption{Flavor universality test of the four-quark Wilson coefficients ${C_{qq}^{(-)}(m_t) = C_{qq}^{(1)} - C_{qq}^{(3)}}$ and $C_{qq}^{(3)}(m_t)$ in MFV. Shown are the bounds obtained from a three-parameter fit of $\{\aqqth,\atqqm,B_{qq}\}$ to top (blue) and top \& \bsmm (orange) data at 68\% and 95\% CL.}
    \label{fig:qq_3d}
\end{figure}

In Fig.~\ref{fig:qq_3d}, we show the bounds obtained from a three-parameter fit of $\{\aqqth,\atqqm,B_{qq}\}$ to top observables alone (blue) and top \& \bsmm (orange).
%
 As suggested by the single-parameter fits, the parameter space of flavor-universal contributions $\aqqth$ and $\atqqm$ (shown in the left panel) is well constrained by top observables alone. Moreover, \bsmm does not contribute to the bounds, due to its additional sensitivity to $B_{qq}$. When projecting the likelihood onto the $(\aqqth,\atqqm)$ plane, the profiling over $B_{qq}$ compensates any effect of $\aqqth$ and $\atqqm$ in \bsmm.

The impact of flavor universality breaking is apparent in the $(\aqqth,B_{qq})$ and $(B_{qq},\atqqm)$ planes (central and right panels). Top observables alone leave a blind direction along $B_{qq}$ (blue). When adding \bsmm (orange),
the fact that $\aqqth$ and $\atqqm$ are already constrained from top measurements
leads to a bound on $B_{qq}$, thus breaking the blind direction. Notice that the currently observed discrepancy in $\mathcal{B}(\bsmm)$ pulls the best-fit region towards positive $B_{qq}$. This means that an explanation of the discrepancy in terms of four-quark operator contributions requires flavor universality breaking in the underlying UV theory.

Probing flavor universality breaking at the LHC requires adding observables that probe four-quark operators through light quarks or through top quarks only. For instance, adding four-top production~\cite{Zhang:2017mls,Banelli:2020iau} and/or dijet production~\cite{Alte:2017pme} to the top fit would allow us to distinguish between flavor-universal and flavor-breaking four-quark interactions. These are interesting directions for future work, especially in the context of the current discrepancies in $b\to s\ell^+\ell^-$ observables.

 \paragraph{Flavor contractions} As we discussed in Sec.~\ref{sec:mfv}, four-quark operators allow for two possible contractions of the flavor indices, indicated by $A$ and $\widetilde{A}$. The Wilson coefficients of color-singlet operators $(\overline{Q}^k\gamma^\mu Q^l)(\overline{Q}^l\gamma_\mu Q^k)$ with a flavor structure $(\widetilde{\mathcal{A}}_Q)_{kk}(\widetilde{\mathcal{A}}_Q)_{ll}$ are related to the coefficients of color-octet operators $(\overline{Q}^k\gamma^\mu T^A Q^k)(\overline{Q}^l\gamma_\mu T^A Q^l)$ with flavor structure $(\mathcal{A}_Q)_{kk}(\mathcal{A}_Q)_{ll}$ through a Fierz transformation, see App.~\ref{app:4q-in-ttb-warsaw} and Ref.~\cite{AguilarSaavedra:2018nen}. Probing for flavor contractions of Wilson coefficients in the Warsaw basis thus means resolving the color structure of four-quark interactions.

Top-antitop production is sensitive to both color-singlet and color-octet four-quark couplings through a variety of kinematic distributions and charge asymmetry observables. In MFV, this translates to a sensitivity to both flavor contractions $\aqqth$ and $\atqqth$, as well as $\aqqm$ and $\atqqm$, see Sec.~\ref{sec:4q-in-top}. In turn, electroweak top production and \bsmm are only sensitive to $\aqqth$ and $\atqqm$. As we will see, a combined fit allows us to resolve all four directions and to pin down the flavor contractions $A$ and $\widetilde{A}$.

\begin{figure}[t!]
    \centering
    \includegraphics[width=0.45\textwidth]{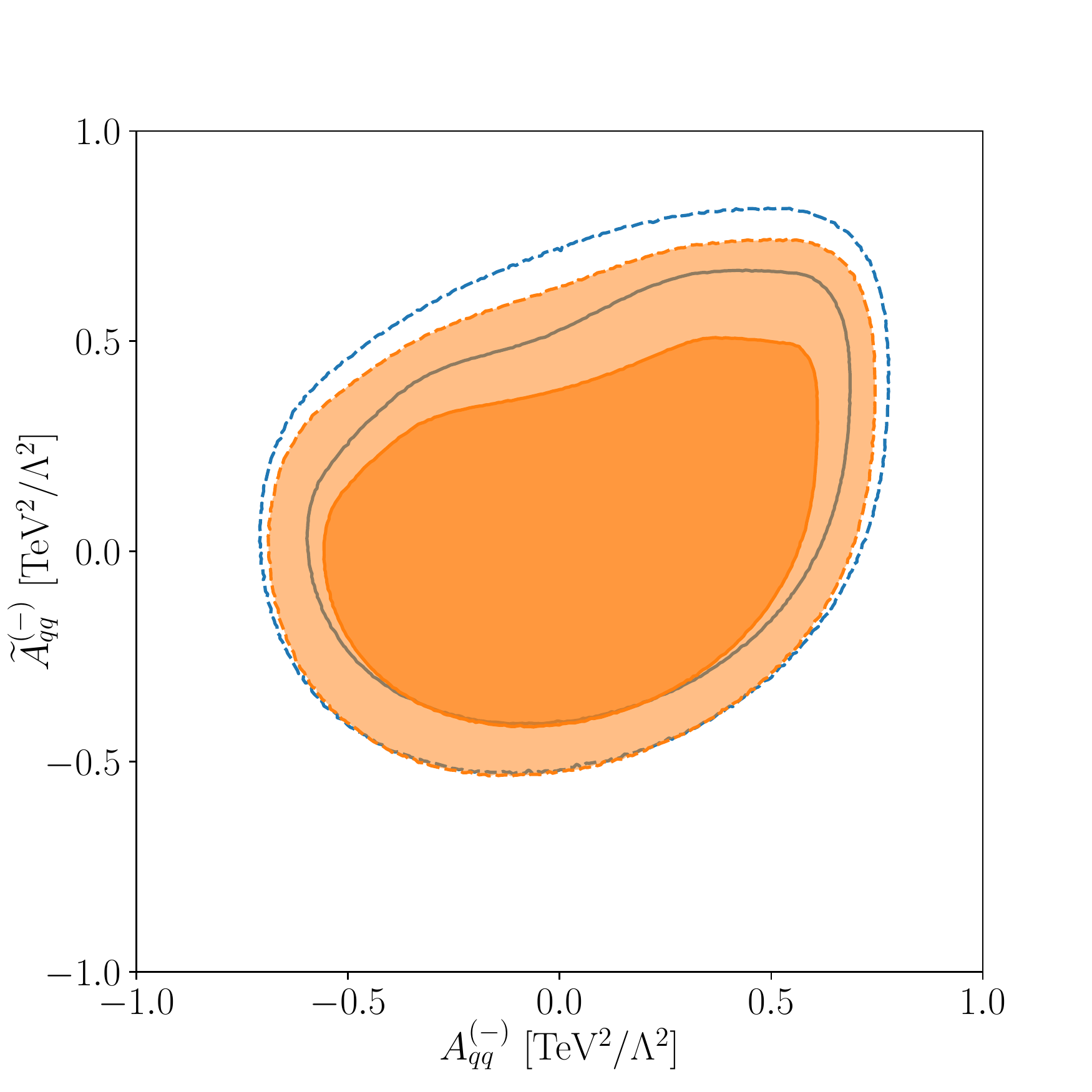}\hspace*{1cm}
    \includegraphics[width=0.45\textwidth]{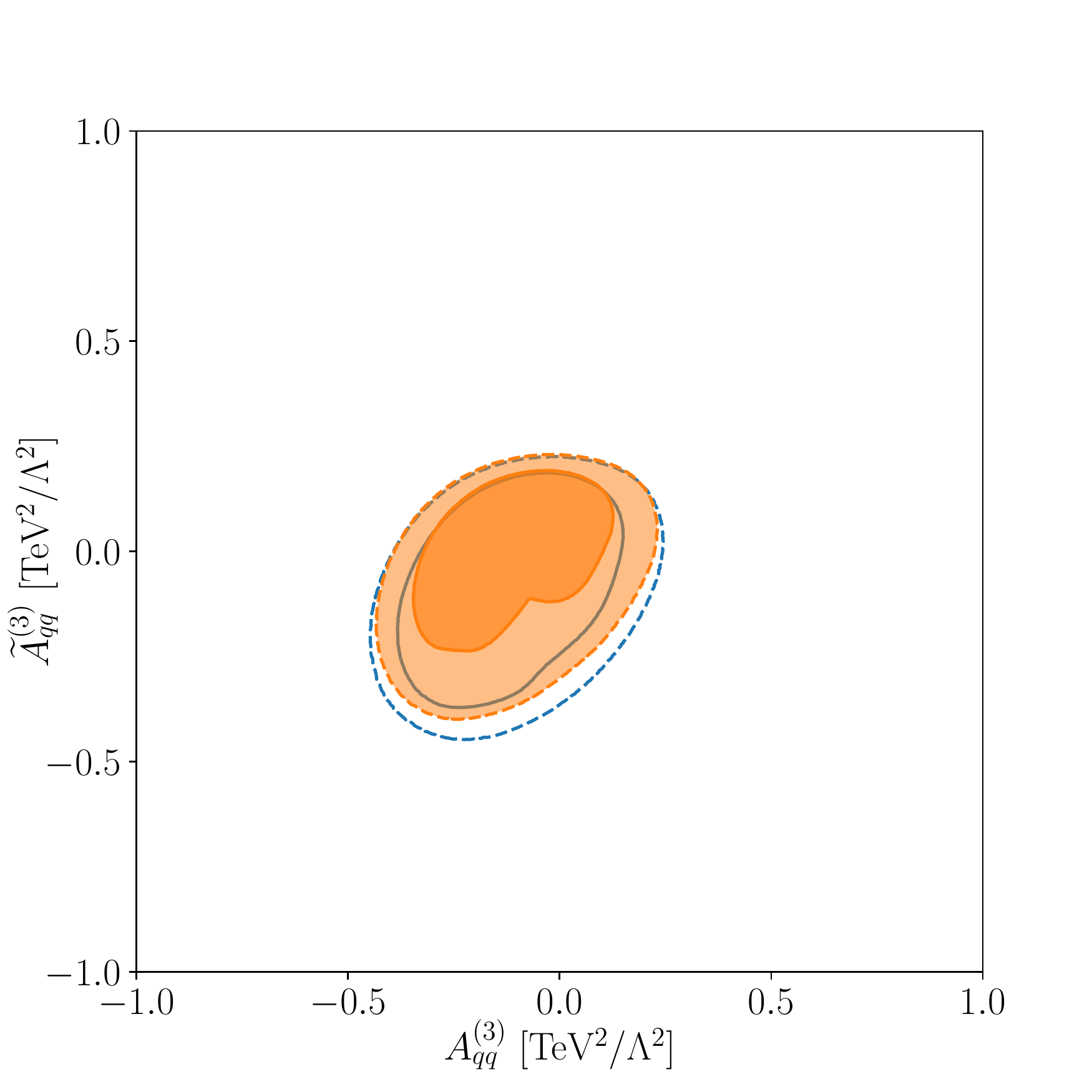}
    \caption{Flavor contractions in the four-quark coefficients $C_{qq}^{(-)}(m_t) = C_{qq}^{(1)} - C_{qq}^{(3)}$ (left) and $C_{qq}^{(3)}(m_t)$ (right) in MFV. Shown are the bounds obtained from a four-parameter fit of $\{\aqqm,\atqqm,\aqqth,\atqqth\}$ to top (blue) and top \& \bsmm (orange) data at 68\% and 95\% CL.}
    \label{fig:qq_4d}
\end{figure}

In Fig.~\ref{fig:qq_4d}, we show the bounds on flavor contractions obtained from a four-parameter fit of $\{\aqqm,\atqqm,\aqqth,\atqqth\}$ to top versus top \& \bsmm observables.
 As expected, top observables alone (blue) constrain the two flavor contractions $A$ and $\widetilde{A}$ in both $C_{qq}^{(-)}$ (left panel) and $C_{qq}^{(3)}$ (right panel). A slight correlation is apparent, disfavoring flavor contractions of opposite sign. Unlike in the three-parameter fit, adding \bsmm (orange) does modify these results, because all four parameters are bounded in top observables, leaving no blind direction in the fit. As before, the impact of \bsmm in the fit is very moderate, due to the relatively low sensitivity of $\mathcal{C}_{10}$ to loop contributions of $C_{qq}^{(-)}(m_t)$ and $C_{qq}^{(3)}(m_t)$.

We conclude our analysis with a few comments on possible extensions. To make the interplay of operator contributions to top and flavor observables apparent, we have limited ourselves to a small set of flavor parameters in each fit. When including more operators, and accordingly more parameters, blind directions generally occur in the global fit,
even if they don't necessarily manifest themselves in single parameter fits.
Such blind directions can be resolved by adding observables to the fit that are sensitive to these directions and do not introduce new parameters.

In the low-energy sector, adding $b\to s\ell^+\ell^-$ observables introduces a sensitivity to $\mathcal{C}_9$. This would allow us to probe all directions in WET that are relevant in $b\to s$ transitions in MFV, see Tab.~\ref{tab:smeft-to-wet-num-Matching}. Another interesting addition would be $B_s$ meson mixing~\cite{Aebischer:2020dsw}, which is sensitive to flavor structures with only third-generation quarks. Due to the strong correlation of the hadronic inputs for \bsmm and $\Delta m_s$, adding $B_s$ mixing require a combined analysis with \bsmm. Once more, we stress that a global analysis of SMEFT coefficients in flavor data and the availability of a 'flavor likelihood function' would be very valuable in combined fits with high-energy data.

At high energies, Higgs and electroweak physics are widely considered as probes of electroweak symmetry breaking, not as flavor tests. However, the sensitivity of Higgs and electroweak observables to the third quark generation~\cite{Biekotter:2018rhp,Ellis:2018gqa} makes them valuable constituents in combined fit with top and flavor observables. Indeed, recent analyses including Higgs and electroweak data show a different sensitivity to SMEFT operators in MFV and flavor-universal scenarios~\cite{Aoude:2020dwv,Ellis:2020unq}. This suggests that Higgs and electroweak observables add valuable information on flavor symmetry breaking to a combined fit.

Throughout our analysis, we have assumed Minimal Flavor Violation as the flavor structure of the Wilson coefficients. Other phenomenologically viable scenarios are flavor universality among the first and second generations, corresponding to a $U(2)^3$ flavor symmetry; and flavor universality, corresponding to a $U(3)^3$ symmetry in the quark sector. Imposing a $U(3)^3$ symmetry reduces MFV to the subset of flavor-universal interactions of the form $(\overline{Q}Q)$, $(\overline{U}U)$ and $(\overline{D}D)$. The absence of flavor violation eliminates all effects of $b$, $\ab$, $\ba$ \emph{etc.}~in flavor observables. In top observables, light-quark couplings $a$ and top couplings $A$ are identical and effects of $O_{uX}$ are absent. Relaxing the symmetry to $U(2)^3$ leaves the top sector essentially unchanged compared to MFV, but eliminates chirally enhanced effects of $O_{dX}^{i3}$ in flavor observables. The different flavor scenarios can thus be distinguished by probing for these effects. Yukawa misalignment is best probed with FCNCs, which are strongly suppressed in MFV and absent in $U(2)^3$- or $U(3)^3$-symmetric scenarios.


\section{Conclusions and outlook}\label{sec:conclusions}
\noindent In this work we have explored what we can learn about the flavor structure of a UV theory with a combined fit of top and bottom observables in the SMEFT-WET framework. Assuming that new physics couples to quarks, we have classified all contributing SMEFT operators by their flavor structure in Minimal Flavor Violation. Assuming instead a $U(2)^3$ flavor symmetry among quarks of the first and second generation would lead to similar results; a $U(3)^3$ symmetry means flavor universality and corresponds to a subset of the degrees of freedom in MFV.

In our analysis we have combined the flavor observables \bsmm and \bsg with top observables at the LHC, including a large set of observables in top-antitop production, electroweak top production and $t\bar tZ$, $t\bar t W$ production. The flavor observables are uncorrelated and sensitive to different directions in SMEFT space. Including more bottom observables in the fit is complicated by correlations and by hadronic uncertainties, which introduce many nuisance parameters. A global statistical analysis of flavor observables, presented in terms of a likelihood function of WET coefficients, would greatly improve the sensitivity in combined fits with high-energy observables. In this respect, meson mixing and $b\to s\ell^+\ell^-$ observables are particularly interesting to probe new directions in the SMEFT space.

Among the flavor observables, $\bsg$ is very sensitive to SMEFT dipole operators with right-handed bottom quarks, due to a chiral enhancement in $\mathcal{C}_7$ compared to the SM contribution. In turn, \bsmm is very sensitive to flavor breaking in operators with left-chiral quark currents through tree-level contributions to $\mathcal{C}_{10}$. Four-quark operators enter \bsg and \bsmm only in loops through local top-quark contributions and non-local charm contributions. For a reliable interpretation of the fit results, the full set of non-local hadronic form factors is needed, but currently only known for the SM contributions.

Top observables probe dipole operators with right-handed top quarks and the large set of four-quark operators with tops. In combining top with bottom observables, we observe a large impact of $t\bar t Z$ production.
In particular, electroweak SMEFT contributions to the total cross section are sizeable for two-quark operators and induce sensitivity to directions in the parameter space that can otherwise only be probed in bottom observables.
We have computed and analyzed electroweak SMEFT contributions to $t\bar t Z$ and $t\bar t W$ production for the first time.

The combination of top and bottom observables proves most powerful for operators with left-chiral quark currents, which contribute to both sectors. Top observables alone have a moderate sensitivity to the flavor structure of SMEFT coefficients through the interplay of light-quark and top-quark couplings. Adding bottom observables to the fit not only improves the resolution of the SMEFT parameter space, but also allows to detect potential new sources of flavor violation through correlations with top observables. In MFV the combined top-bottom fit strongly constrains flavor breaking among left-handed quarks, suggesting that UV physics that modifies electroweak interactions should couple mostly flavor-universally. This important result largely relies on a non-trivial interplay of flavor-universal and flavor-breaking effects in \bsg.

Four-quark operators are well constrained by top observables. However, flavor universality violation cannot be tested in top observables alone, because hadronic top production probes only operators with two light quarks and two top quarks. In turn, observables in \bsmm, $b\to s\ell^+\ell^-$, meson mixing, dijet production and/or four-top production probe different quark flavors and resolve the flavor structure in a combined analysis with top observables. In particular, we find that a combined fit of top and \bsmm observables shows a good sensitivity to flavor breaking in MFV. However, the sensitivity of $b\to s$ observables to four-quark operators is affected by the strong operator mixing under the renormalization group and should be interpreted with care.

In the course of our work, we became aware of many new research topics that will improve combined SMEFT fits with low- and high-energy observables. The most notable topics are a reliable description of non-local effects in rare $B$ decay observables within
the SMEFT framework, the connection of SMEFT effects in observables at different energy scales through the renormalization group, and the use of Higgs and electroweak observables in the flavor context. The flavor universality tests presented here give just a glimpse of possible UV physics; much more can be learned in combining observables across the scales.


\section*{Acknowledgments}
\noindent We thank Ilaria Brivio, Oscar Cata, Thorsten Feldmann and Thomas Mannel for helpful discussions and David Straub and Peter Stangl for their advice with \wilson. The research of SB and SW has been supported by the DFG (German Research Foundation) under grant no. 396021762$-$TRR 257. RS acknowledges support of the DFG through the research training group {\it Particle physics beyond the Standard Model} (GRK 1940).
The work of DvD is supported by the DFG within the Emmy Noether Programme under grant DY 130/1-1
and the DFG Collaborative Research Center 110 {\it Symmetries and the Emergence of Structure in QCD}. The authors acknowledge support by the state of Baden-W\"urttemberg through bwHPC
and the German Research Foundation (DFG) through grant no INST 39/963-1 FUGG (bwForCluster NEMO).


\newpage

\appendix

\section{Predictions of top and bottom observables}
\label{app:polynomials}

In this appendix we provide details about the top and bottom observables within the WET \& the SMEFT as included in our fit.

\subsection{Bottom observables and hadronic uncertainties}
 $\boldsymbol{\bsmm}$\textbf{:}~In the Standard Model, the branching ratio of $\bsmm$ is dominated by
$\mathcal{O}_{10}$~\cite{Bobeth:2013uxa}. The SM Wilson coefficient $\mathcal{C}_{10}^\text{SM}$ has been calculated to NNLO in $\alpha_s$~\cite{Hermann:2013kca} and to NLO in the electroweak coupling~\cite{Bobeth:2013tba}.
In Eq.~\eqref{eq:bsmm-br} we have inserted $\mathcal{C}_{10}^\text{SM}$ explicitly, so that $\mathcal{C}_{10}$ parametrizes only genuine BSM contributions.
The constant term therefore corresponds to the SM prediction and is compatible with the results of Ref.~\cite{Bobeth:2013uxa} within uncertainties.

Within the WET, the decay is fully described by a single hadronic matrix element, parametrized in terms of the $B_s$ decay constant $f_{B_s}$. Lattice QCD calculations of $f_{B_s}$ have a total uncertainty of $0.6\%$~\cite{Bazavov:2017lyh}. A further source of uncertainty for $\mathcal{B}(\bsmm)$ is the CKM element $V_{ts}$, which is obtained from a recent update of a global CKM fit~\cite{Bona:2006ah}.\footnote{
The numerical value is taken from the summer 2018 update, available \href{https://utfit.org}{online}.} Altogether, $\mathcal{B}(\bsmm)$ is precisely predicted in and beyond the Standard Model, and it is therefore a clean probe of SMEFT contributions to $\mathcal{C}_{10^{(')}}$.

$\boldsymbol{\bsg}$\textbf{:}~In the Standard Model, the branching ratio of $\bsg$ is dominated by $\mathcal{O}_7$~\cite{Ali:1991fy}.
Further SM contributions arise from $\mathcal{O}_8$
and the four-quark operators $\mathcal{O}_1$ and $\mathcal{O}_2$ through NLO QCD corrections and operator mixing under
the renormalization group evolution.

The hadronic uncertainty in $\mathcal{B}(\bsg)$ is governed by a series of hadronic matrix elements from both local and non-local
contributions. In the Standard Model, these are extracted from moments of the kinematic distribution of $B\to X_{c,u}\ell\bar\nu$ decays
and the isospin asymmetry of $\bsg$ decays,
see Refs.~\cite{Gunawardana:2019gep,Misiak:2020vlo} for a recent reanalysis. In our analysis, we assume that the hadronic uncertainties for the BSM contributions
are comparable to the SM uncertainty and assign an overall uncertainty to the branching ratio predictions.
Ideally, this setup should be replaced with a comprehensive multivariate statistical analysis of flavor observables for the full basis of WET operators.

Non-local contributions from four-quark operators induce constant shifts to $\mathcal{C}_7$, as discussed in Sec.~\ref{sec:wet}.
We include the known
shift induced by SM contributions of $\mathcal{O}_1$ and $\mathcal{O}_2$, as defined for instance in Ref.~\cite{Bobeth:2017vxj}.
However, we neglect CKM-suppressed shifts within the Standard Model and BSM shifts due to four-quark WET operators;
calculating the latter would take us far beyond the scope of this analysis.
Due to the non-local nature, the shifts of $\mathcal{C}_7$ and $\mathcal{C}_{7'}$ by four-quark operators are
process-dependent. However, they can always be expressed as a sum of four-quark coefficients, weighted with the respective non-local hadronic
matrix elements at $q^2 = 0$. Currently only the matrix elements for the SM contributions are known. Once a complete BSM analysis becomes
available, our fit results can be reinterpreted to include BSM contributions from four-quark operators by shifting $\mathcal{C}_7$ and $\mathcal{C}_{7'}$ accordingly.

$\boldsymbol{B\to K^{(*)}\mu^+\mu^-}$\textbf{:}~In the Standard Model, the branching ratios and the full angular distributions
are dominated by the semileptonic operators $\mathcal{O}_{9}$ and $\mathcal{O}_{10}$. However, this picture is strongly dependent
on the dilepton mass square $q^2$, and other operators dominate in parts of the spectrum. For $q^2 \simeq 0$, the so-called photon pole dominates
in $B\to K^*\mu^+\mu^-$ decays. For $q^2$ close to the mass squared of either of the $J/\psi$ or $\psi(2S)$ resonance,
the non-local contributions of the four-quark operators dominate.
The latter effect has been investigated within the scope of the SM in Ref.~\cite{%
Dimou:2012un,Lyon:2013gba,%
Beneke:2001at,Beneke:2004dp,%
Khodjamirian:2010vf,Khodjamirian:2012rm,%
Bobeth:2017vxj,Gubernari:2020eft%
}. However, hadronic matrix elements of BSM four-quark operators are still unknown.
We therefore see no way to include $B\to K^{(\ast)}\mu^+\mu^-$ decays or observables that are dominated
by $\mathcal{O}_{9^{(')}}$ in our analysis in a consistent manner,
since, unlike for $\mathcal{C}_{7^{(')}}$, we would not be able to reinterpret our fit results for
$\mathcal{C}_{9^{(')}}$ due to the momentum dependence of the non-local shifts.
We emphasize that this problem is not specific to our analysis, but also affects all other SMEFT analyses
in which BSM effects potentially enter the WET four-quark operators.
Hence, a complete analysis of all non-local hadronic matrix elements for $b\to s$ transitions that arise in
SMEFT is highly requested, especially in light of the currently observed tensions in $B\to K^{(\ast)}\mu^+\mu^-$
observables. We discuss the potential impact of including $\mathcal{C}_9$ in a global SMEFT analysis in Sec.~\ref{sec:results}.

\subsection{Connection between bottom and top observables}

\textbf{Matching and running}~We provide some details on the 
matching and running of the SMEFT and WET Wilson coefficients, as outlined in
Section~\ref{sec:matching}.
The RG evolution of the SMEFT coefficients from the high scale $\mu_0$
to the WET matching scale
induces operators beyond the set
considered in this work. In the SMEFT-to-WET matching, we include these RG-induced operators.
Exploiting the linear relation between the WET coefficients at $\mu = m_b$ and the SMEFT coefficients at $\mu_0 = m_t$, we express the WET coefficients as polynomials of SMEFT coefficients,
\begin{equation}
    \mathcal{C}_a(m_b) = \big(\mathcal{R}^\text{WET}(m_b,m_Z)\big)_{ab} \, \big(\mathcal{M}(m_Z)\big)_{bc} \, \big(\mathcal{R}^\text{SMEFT}(m_Z,m_t)\big)_{cd} \, C_d(m_t),
\end{equation}
where the sum over repeated indices is implied.
The RG evolution in WET and in SMEFT is contained in the matrices $\mathcal{R}^\text{WET}$ and $\mathcal{R}^\text{SMEFT}$, and $\mathcal{M}$ comprises the SMEFT-to-WET matching relations. The full evolution $\mathcal{R}^\text{WET}\mathcal{M}\mathcal{R}^\text{SMEFT}$, which is used in our numerical analysis, is displayed for MFV parameters in
Tab.~\ref{tab:smeft-to-wet-num}.
For completeness, we also display the WET running plus matching $\mathcal{R}^\text{WET}\mathcal{M}$
in Tab.~\ref{tab:smeft-to-wet-num-Matching-runningWET}
and the matching $\mathcal{M}(m_Z)$
in Tab.~\ref{tab:smeft-to-wet-num-Matching}. We only display the submatrices pertinent to the set of operators in
Eqs.~\eqref{eq:2q-operators} and \eqref{eq:4q-operators}.

 This approach allows us to consistently combine flavor observables with global top fits, where the Wilson coefficients in the fit are defined at $\mu_0 = m_t$.
However, it makes the interpretation of the fit results in terms of a UV theory more difficult, since the set of SMEFT Wilson coefficients in top observables
is not self-contained under the RG evolution. To match the fit results to a UV completion at a scale $\Lambda > m_t$, the full set of SMEFT coefficients $C_a(m_t)$ has to be RG-evolved up to the BSM scale $\Lambda$.

\begin{figure}[t!]
    \centering
    \includegraphics[width=0.5\textwidth]{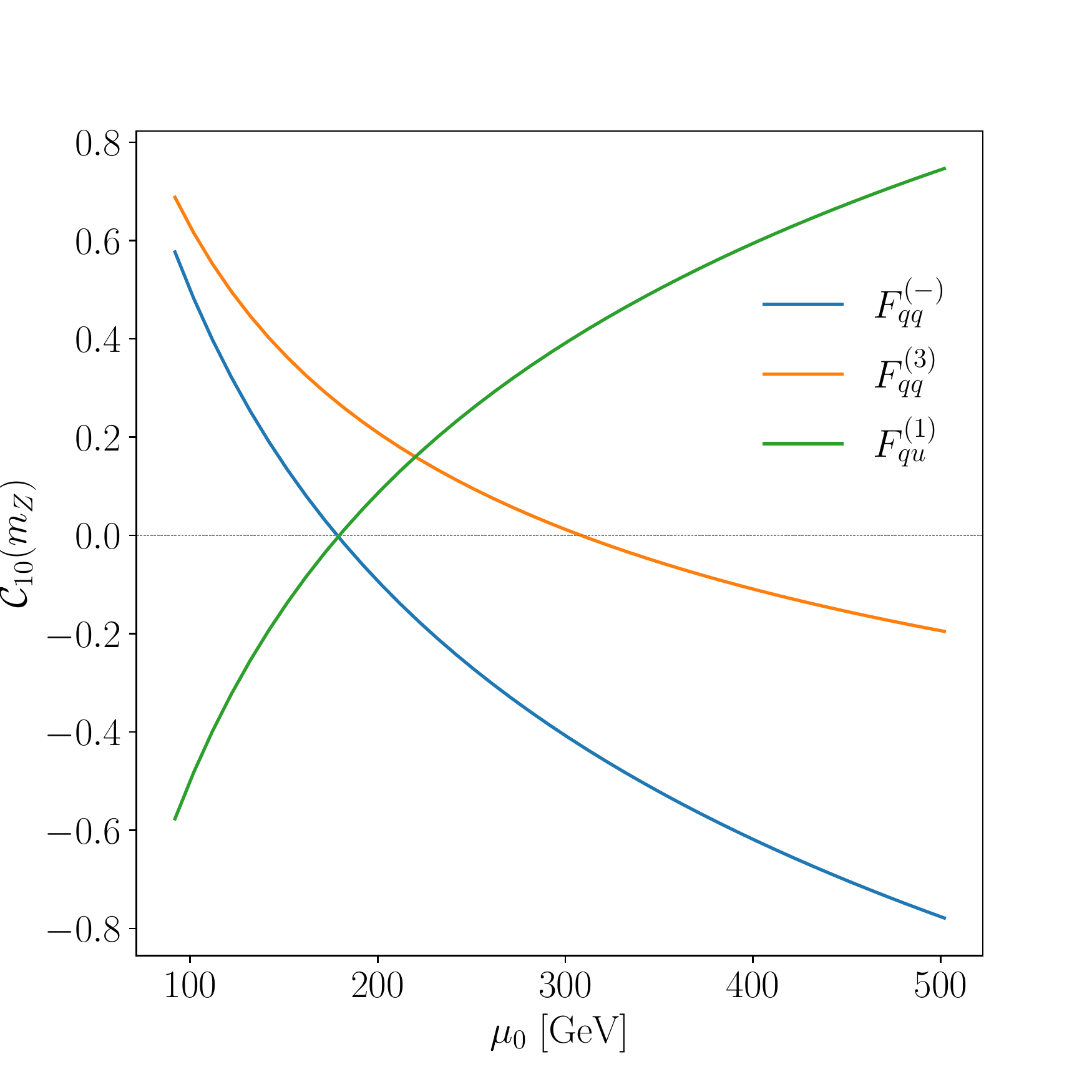}\hspace*{1cm}
    \caption{SMEFT contributions of four-quark coefficients $F_i(\mu_0)$ in MFV to $\mathcal{C}_{10}(m_Z)$ as a function of the high scale $\mu_0$. The MFV parameters have been set to $F_i(\mu_0) = 1$, evolved down to $\mu=m_Z$, and matched onto $\mathcal{C}_{10}$. The cancellation of $F_{qq}^{(-)}$ and $F_{qu}^{(1)}$ contributions close to
    $\mu_0 = m_t$ is accidental.}
    \label{fig:SMEFT-running}
\end{figure}

\textbf{Sensitivity to SMEFT parameters}
~Based on Tab.~\ref{tab:smeft-to-wet-num-Matching}, Tab.~\ref{tab:smeft-to-wet-num}, and Tab.~\ref{tab:smeft-to-wet-num-Matching-runningWET}, we can assess the sensitivity of the individual WET coefficients
to the flavor parameters in MFV. It is useful to compare the matching relations from Tab.~\ref{tab:smeft-to-wet-num-Matching}
with the WET running plus matching in Tab.~\ref{tab:smeft-to-wet-num-Matching-runningWET}: The RG evolution in WET enhances the sensitivity to some coefficients compared to the matching relations.
However, we emphasize that the effects described below are based on the
WET RG evolution to LL accuracy and should be revisited once the NLL evolution becomes available.
Compared to $\mathcal{C}_{10}(m_Z)$, we find the sensitivity of $\mathcal{C}_{10}(m_b)$ to the flavor parameters to be largely unchanged. The sensitivity of $\mathcal{C}_9(m_b)$ is also very similar compared to $\mathcal{C}_9(m_Z)$; only the sensitivity to $b_{\phi q}^{(-)}$ and $b_{\phi q}^{(3)}$ is reduced by about $10\%$.
The largest effect occurs in the dipole coefficients $\mathcal{C}_{7}(m_b)$ and $\mathcal{C}_{8}(m_b)$, where the running of the $b$-quark mass reduces the sensitivity to the chirally-enhanced SMEFT coefficients.
An additional effect arises due to operator mixing: the coefficient $\mathcal{C}_7(m_b)$ develops a sizeable sensitivity to $b_{d G}$, due to the mixing of $\mathcal{C}_8$ into $\mathcal{C}_7$.
A counter effect is the suppression or outright cancellation of SMEFT contributions by the WET running. We
observe such an effect in the contribution of $F_{quqd}^{(1)}$ and $F_{quqd}^{(8)}$ to $\mathcal{C}_7(m_b)$, which reduces the sensitivity by more than one order of magnitude and renders $\mathcal{C}_7(m_b)$ essentially insensitive to these SMEFT coefficients. We conclude that the RG evolution in the WET does not overly affect the dominant contributions to \bsmm and \bsg.

On the contrary, operator mixing in the SMEFT RG evolution can have sizeable effects. We find that the mixing renders
the contributions of the four-quark coefficients $F_{qq}^{(-)}$, $F_{qq}^{(3)}$ and $F_{qu}^{(1)}$ to $\mathcal{C}_{10}$ especially sensitive to the scale $\mu_0$.
This effect is illustrated in Fig.~\ref{fig:SMEFT-running}, where we show the contributions of the four-quark flavor parameters $F_i(\mu_0)$ to $\mathcal{C}_{10}(m_Z)$ as a function
of $\mu_0$.

The surprisingly strong variation with $\mu_0$
is due to the mixing of $O_{qq}^{(1)}$, $O_{qu}^{(1)}$ and $O_{qq}^{(3)}$ into $O_{\phi q}^{(1)}$ and $O_{\phi q}^{(3)}$
which match onto $\mathcal{C}_{10}$ at tree level.

The following numerical example illustrates the importance of this effect: Starting with $F_{qq}^{(-)}(\mu_0 = 200\,\text{GeV})=1$, we compare three cases:
\begin{itemize}
    \item We match $F_{qq}^{(-)}(m_Z)$ onto $\mathcal{C}_{10}(m_Z)$ and find $\mathcal{C}_{10}(m_Z)=0.57$.
    \item We match $F_{qq}^{(-)}(m_Z)$, $C_{\phi q}^{(1),33}(m_Z)$ and $C_{\phi q}^{(3),33}(m_Z)$ and find $\mathcal{C}_{10}(m_Z)=-0.11$.
    \item We match the full SMEFT and find $\mathcal{C}_{10}(m_Z)=-0.09$.
\end{itemize}
We conclude that the sensitivity of $\mathcal{C}_{10}$ to the scale $\mu_0$ indeed arises from the mixing of the four-quark coefficients into $C_{\phi q}^{(1)}$ and $C_{\phi q}^{(3)}$. Notice, however, that the relation between $\mathcal{C}(m_Z)$ and $C(\mu_0)$ relies on the SMEFT RG evolution at leading order. At the next order, the RG evolution will presumably stabilize and might modify the relation between the coefficients in WET and in SMEFT.

\begin{table}[tp]\begin{center}
\renewcommand{\arraystretch}{1.05}
\setlength{\tabcolsep}{1.5mm}
\begin{tabular}{l|c|c|c|c}
\toprule
  & $\mathcal{C}_7$ & $\mathcal{C}_8$  & $\mathcal{C}_9$ & $\mathcal{C}_{10}$ \\
\midrule[1pt]
SM & -0.3373 & -0.1829 & 4.2734 & -4.1661\\
\midrule
$\left(a_{\phi q}^{(-)},\,b_{\phi q}^{(-)}\right)$ & (0.001,\,-0.002) & (0,\,0.016) & (-0.19,\,-1.98) & (0,\,23.94)\\ 
$\left(a_{\phi q}^{(3)},\,b_{\phi q}^{(3)}\right)$ & (-0.023,\,0.047) & (-0.011,\,0.061) & (0.01,\,-3.91) & (-0.26,\,48.08)\\ 
$\left(a_{\phi u},\,b_{\phi u}\right)$ & (0,\,0) & (0,\,0) & (0,\,0) & (0,\,0)\\ 
$A_{\phi ud}$ & -0.022 & -0.010 & 0 & 0\\ 
$A_{u B}$ & -0.082 & 0 & 0.14 & 0\\ 
$A_{u W}$ & -0.015 & 0.016 & 0.111 & -0.423\\ 
$A_{u G}$ & -0.003 & -0.027 & -0.002 & 0.004\\ 
$\left(a_{d B},\,b_{d B}\right)$ & (-0.036,\,12.273) & (0,\,0.070) & (0,\,0) & (0,\,0)\\  
$\left(a_{d W},\,b_{d W}\right)$ & (0.163,\,-6.617) & (0.079,\,-0.069) & (0,\,0) & (0,\,0)\\ 
$\left(a_{d G},\,b_{d G}\right)$ & (-0.002,\,0.663) & (-0.009,\,4.024) & (0,\,0) & (0,\,0)\\
$F_{qq}^{(-)}$ & 0 & 0 & -0.04 & 0.03\\ 
$F_{qq}^{(3)}$ & 0 & 0 & -0.13 & 0.29\\ 
$F_{qu}^{(1)}$ & 0 & 0 & 0.05 & -0.03\\ 
$F_{quqd}^{(1)}$ & 0 & -0.002 & 0 & 0\\ 
$F_{quqd}^{(8)}$ & -0.001 & 0 & 0 & 0\\
 \bottomrule[1pt]
\end{tabular}
\end{center}
\caption{Contributions of SMEFT coefficients $C(m_t)$ to WET coefficients $\mathcal{C}(m_b)$ in MFV with up-alignment. The SMEFT coefficients are matched onto WET at $\mu = m_Z$ at one-loop level in the electroweak theory and then RG-evolved down to $\mu = m_b$ at LO in QCD. The scale of new physics has been set to $\Lambda = 1\tev$. Contributions to $\mathcal{C}_{9'}$ and $\mathcal{C}_{10'}$ are not generated by the operators considered in this work.}
\label{tab:smeft-to-wet-num}
\end{table}

\begin{table}[tp]\begin{center}
\renewcommand{\arraystretch}{1.05}
\setlength{\tabcolsep}{1.5mm}
\begin{tabular}{c|c|c|c|c}
\toprule
  & $\mathcal{C}_7$  & $\mathcal{C}_8$  & $\mathcal{C}_9$ & $\mathcal{C}_{10}$ \\
\midrule[1pt]
SM &  -0.337 
& -0.183 
& 4.27 
& -4.17
\\
\midrule
$\left(a_{\phi q}^{(-)},\,b_{\phi q}^{(-)}\right)$ & (0,\,-0.003) & (0,\,0.017) & (-0.01,\,-1.86) & (0.1,\,24.72)\\
$\left(a_{\phi q}^{(3)},\,b_{\phi q}^{(3)}\right)$ & (-0.024,\,0.048) & (-0.012,\,0.061) & (0.25,\,-3.76) & (-0.82,\,48.64)\\ 
$\left(a_{\phi u},\,b_{\phi u}\right)$ & (0,\,0) & (0,\,0) & (0.01,\,0.01) & (-0.1,\,-0.1)\\ 
$A_{\phi ud}$ & -0.023 & -0.010 & 0 & 0\\
$A_{u B}$ & -0.120 & -0.001 & 0.15 & 0\\
$A_{u W}$ & 0.002 & 0.016 & 0.11 & -0.44\\
$A_{u G}$ & -0.005 & -0.037 & 0 & 0\\
$\left(a_{d B},\,b_{d B}\right)$ & (-0.035,\,12.595) & (0,\,0.068) & (0,\,0) & (0,\,0)\\ 
$\left(a_{d W},\,b_{d W}\right)$ & (0.049,\,-6.856) & (0.080,\,0.004) & (0,\,0) & (0,\,0)\\ 
$\left(a_{d G},\,b_{d G}\right)$ & (-0.002,\,0.571) & (-0.011,\,3.918) & (0,\,0) & (0,\,0)\\ 
$F_{qq}^{(-)}$ & 0 & 0 & -0.09 & 0.58\\ 
$F_{qq}^{(3)}$ & 0 & 0 & -0.11 & 0.69\\ 
$F_{qu}^{(1)}$ & 0 & 0 & -0.01 & -0.59\\ 
$F_{quqd}^{(1)}$ & -0.015 & -0.019 & 0 & 0\\ 
$F_{quqd}^{(8)}$ & -0.015 & 0.003 & 0 & 0\\
 \bottomrule[1pt]
\end{tabular}
\end{center}
\caption{Contributions of SMEFT coefficients $C(m_Z)$ to WET coefficients $\mathcal{C}(m_b)$ in MFV with up-alignment. The SMEFT coefficients are matched onto WET at $\mu = m_Z$ at one-loop level in the electroweak theory and then RG-evolved down to $\mu = m_b$ at LO in QCD. The scale of new physics has been set to $\Lambda = 1\tev$. Contributions to $\mathcal{C}_{9'}$ and $\mathcal{C}_{10'}$ are not generated by the operators considered in this work.}
\label{tab:smeft-to-wet-num-Matching-runningWET}
\end{table}

\subsection{Top observables} In Tab.~\ref{tab:2q-num}, we summarize the contributions of the SMEFT coefficients $C_{\phi q}^{(-)}$ and $C_{\phi q}^{(3)}$ to selected top observables, as well as \bsmm and \bsg. Shown are the contributions of each flavor parameter in MFV at $\mathcal{O}(\Lambda^{-2})$ and all numerically relevant combinations of parameters at $\mathcal{O}(\Lambda^{-4})$.

\begin{table}[t!]
\begin{center}
\renewcommand{\arraystretch}{1.05}
 \setlength{\tabcolsep}{1.5mm}
 \resizebox{\textwidth}{!}{%
\begin{tabular}{l|cc|cc|c|c|c|c|c}
\toprule
                                   & \multicolumn{2}{c|}{$\sigma_{t\bar{t}Z}$ [pb]} & \multicolumn{2}{c|}{$\sigma_{t\bar{t}W}$ [pb]} & $\sigma_{tZ}$ [pb] & $\sigma_{tW}$ [pb] & $\sigma_{t}$ [pb] & $\mathcal{B}(\bsmm)$ & $\mathcal{B}(\bsg)$ \\
                                   & QCD                  & EW                 & QCD                 & EW                  &               &               &              &      $[10^{-9}]$                &        $[10^{-4}]$             \\
\midrule[1pt]
SM                                 & \phantom{-}0.669                & 0.011              & 0.443               & 0.004               & 0.782         & 75.3\phantom{-}          & 126\phantom{-}          & 3.57                 & 3.26                \\
\midrule
$a_{\phi q}^{(3)}$                 & \phantom{-}0.022                & x                  & 0.054               & 0.008               & 0.308         & 9.2           & 31           & 0.45                 & 0.36                \\
$b_{\phi q}^{(3)}$                 & x                    & x                  & --                  & x                   & 0.169         & 9.0           & 15           & -82.32\phantom{--}               & -0.76\phantom{-}               \\
$a_{\phi q}^{(-)}$                 & -0.069               & x                  & --                  & x                   & 0.014         & --            & --           & x                    & x                   \\
$b_{\phi q}^{(-)}$                 & -0.070                & x                  & --                  & x                   & 0.017         & --            & --           & 41.0\phantom{-}                & x                   \\
\midrule
$\big(a_{\phi q}^{(3)}\big)^2$  & x                    & 0.016              & x                   & 0.062               & 0.123         & 0.3           & 2            & x                    & x                   \\
$\big(b_{\phi q}^{(3)}\big)^2$  & x                    & x                  & --                  & x                   & 0.021         & 0.3           & x            & 475.03\phantom{--}               & 0.04                \\
$\big(a_{\phi q}^{(-)}\big)^2$  & x                    & 0.012              & --                  & x                   & x             & --            & --           & x                    & x                   \\
$\big(b_{\phi q}^{(-)}\big)^2$  & x                    & x                  & --                  & x                   & x             & --            & --           & 117.82\phantom{--}               & x                   \\
$a_{\phi q}^{(3)}b_{\phi q}^{(3)}$ & x                    & x                  & --                  & x                   & 0.048         & 0.5           & 2            & -5.15                & -0.04\phantom{-}               \\
$a_{\phi q}^{(3)}b_{\phi q}^{(-)}$ & x                    & x                  & --                  & x                   & x             & --            & --           & -2.57                & x                   \\
$a_{\phi q}^{(3)}a_{\phi q}^{(-)}$ & x                    & 0.016              & --                  & x                   & x             & --            & --           & x                    & x                   \\
$b_{\phi q}^{(3)}b_{\phi q}^{(-)}$ & x                    & x                  & --                  & x                   & x             & --            & --           & 473.15\phantom{--}               & x        \\
\bottomrule[1pt]
\end{tabular}
}
\caption{SMEFT contributions to top and flavor observables in MFV. The top observables are predictions for the LHC at 13 TeV. Contributions to $\sigma_{ t\bar{t}Z}$, $\sigma_{t\bar{t}W}$, $\sigma_{tZ}$ and $t$-channel single top production $\sigma_{t}$ are given at LO QCD / EW; contributions to $\sigma_{tW}$ are given at NLO QCD. The SMEFT contributions to the flavor observables rely on one-loop SMEFT-to-WET matching in the electroweak theory and RG evolution at LO in QCD. The symbol x stands for contributions of less than $1\%$ of the SM prediction and the symbol -- stands for no contribution.
\label{tab:2q-num}}
\end{center}
\end{table}

\section{UV sensitivity and chiral enhancement}\label{app:uv-chi}
In this appendix we list all contributions of SMEFT operators that are chirally enhanced or UV sensitive when matched onto WET. In Tab.~\ref{tab:chiral-enhancement} we show the scaling of the SMEFT coefficients with the relevant quark masses and the corresponding flavor structure in MFV. In Tab.~\ref{tab:uv-sensitivity} we list all one-loop matching relations that are logarithmically sensitive to the matching scale $\mu$.

\begin{table}[t!]\centering
\renewcommand{\arraystretch}{1.2}
\begin{tabular}{l|cccc|c}\toprule
  & $\mathcal{C}_7$ & $\mathcal{C}_{7'}$ & $\mathcal{C}_8$ & $\mathcal{C}_{8'}$ & MFV \\\midrule
    $C_{\phi q}^{(1),\Sigma\Sigma}, C_{\phi q}^{(3),3\Sigma/\Sigma3/\Sigma\Sigma}$ & 1 & $\frac{m_s}{m_b}$ & 1 & $\frac{m_s}{m_b}$ & 1 \\
    $C_{\phi d}^{32}$ & & 1 & & 1 & $y_by_s$ \\
    $C_{\phi d}^{23}$ & $\frac{m_s}{m_b}$ & & $\frac{m_s}{m_b}$ & & $y_by_s$\\
    $C_{\phi ud}^{32}, C_{dW}^{32}$ & & $\frac{m_W}{m_b}$ & & $\frac{m_W}{m_b}$ & $y_s$\\
    $C_{\phi ud}^{33}, C_{dW}^{23}$ & $\frac{m_W}{m_b}$ & & $\frac{m_W}{m_b}$ & & $y_b$\\
    $C_{uB}^{33/\Sigma3}, C_{uW}^{33}$ & 1 & $\frac{m_s}{m_b}$ & & & 1\\
    $C_{dB}^{32}$ &  & $\frac{m_W}{m_b}$ & & & $y_s$\\
    $C_{dB}^{33}$ & $\frac{m_W}{m_b}$ & & & & $y_b$\\
    $C_{dB}^{\Sigma2}, C_{dW}^{\Sigma2}$ & & $\mathbf{\frac{m_W}{m_b}}$ & & $\frac{m_W}{m_b}$ & $y_s$ \\
    $C_{dB}^{\Sigma3}, C_{dW}^{\Sigma3}$ & $\mathbf{\frac{m_W}{m_b}}$ & & $\frac{m_W}{m_b}$ & & $y_b$\\
    $C_{uW}^{\Sigma3}$ & 1 & $\frac{m_s}{m_b}$ & 1 & $\frac{m_s}{m_b}$ & 1\\
    $C_{uG}^{33/\Sigma3}$ & & & 1 & $\frac{m_s}{m_b}$ & 1\\
    $C_{dG}^{32}$ & & & & $\frac{m_W}{m_b}$ & $y_s$\\
    $C_{dG}^{33}$ & & & $\frac{m_W}{m_b}$ & & $y_b$\\
    $C_{dG}^{\Sigma2}$ & & & & $\mathbf{\frac{m_W}{m_b}}$ & $y_s$\\
    $C_{dG}^{\Sigma3}$ & & & $\mathbf{\frac{m_W}{m_b}}$ & & $y_b$\\\midrule
    $C_{quqd}^{(1),\Sigma332},C_{quqd}^{(8),\Sigma332}$ & & $\frac{m_W}{m_b}$ & & $\frac{m_W}{m_b}$ & $y_s$\\
    $C_{quqd}^{(1),\Sigma333},C_{quqd}^{(8),\Sigma333}$ & $\frac{m_W}{m_b}$ & & $\frac{m_W}{m_b}$ & & $y_b$\\\bottomrule
\end{tabular}
\caption{Scaling of SMEFT coefficients with the external quark masses in SMEFT-to-WET matching at one-loop level, relative to the SM contribution. Tree-level contributions with chiral enhancement are written in boldface. Sub-dominant contributions in the quark masses have been neglected. In matching contributions with left-handed quarks the sum over flavor indices $k\in\{1,2,3\}$ is denoted as $\sum$. Contributions to $\mathcal{C}_9$, $\mathcal{C}_{9'}$, $\mathcal{C}_{10}$ and $\mathcal{C}_{10'}$ do not feature a chiral enhancement. In the last column we show the scaling of the flavor structure of SMEFT coefficients with the Yukawa couplings in MFV.\label{tab:chiral-enhancement}}
\end{table}

\begin{table}[t!]\centering
\renewcommand{\arraystretch}{1.2}
\begin{tabular}{l|cccccccc}\toprule
   ln = ln$(m_{t,W}/\mu)$ & $\mathcal{C}_7$ & $\mathcal{C}_{7'}$ & $\mathcal{C}_8$ & $\mathcal{C}_{8'}$ & $\mathcal{C}_9$ & $\mathcal{C}_{9'}$ & $\mathcal{C}_{10}$ & $\mathcal{C}_{10'}$\\\midrule
    $C_{\phi q}^{(1),3\Sigma/\Sigma3}$ & & & & & $\ln$ & & $\ln$\\
    $C_{\phi q}^{(1),\Sigma\Sigma}$  & x & x & x & x & $\ln$ & & $\ln$\\
    $C_{\phi q}^{(3),33}$ & & & & & $\ln$ & $\ln$ & $\ln$ & $\ln$   \\
    $C_{\phi q}^{(3),3\Sigma/\Sigma3/\Sigma\Sigma}$ & x & x & x & x & $\ln$ & & $\ln$ \\
    $C_{\phi u}^{33}$ & & & & & $\ln$ & & $\ln$ & \\
    $C_{\phi d}^{23}$ & x & & x & & & $\ln$ & & $\ln$ \\
    $C_{uB}^{\Sigma3}$ & $\ln$ & x & & & x & & & \\
    $C_{dB}^{32}$ &  & $\ln$  \\
    $C_{dB}^{33}$ & $\ln$ \\
    $C_{dB}^{\Sigma2},C_{dW}^{\Sigma2}$ &  & $\ln$ & & $\ln$ \\
    $C_{dB}^{\Sigma3},C_{dW}^{\Sigma3}$ &  $\ln$ & & $\ln$\\
    $C_{uW}^{\Sigma3}$ & $\ln$ & x & x & x & x & & x & \\
    $C_{dW}^{32}$ & & $\ln$ & & x \\
    $C_{dW}^{33}$ & $\ln$ & & x \\
    $C_{uG}^{\Sigma3}$ & & & $\ln$ & x \\
    $C_{dG}^{32},C_{dG}^{\Sigma2}$ &  & & & $\ln$\\
    $C_{dG}^{33},C_{dG}^{\Sigma3}$ & & & $\ln$ \\\midrule
    $C_{qq}^{(1),\Sigma\Sigma33}, C_{qq}^{(3),\Sigma\Sigma33/\Sigma33\Sigma}$ & & & & & $\ln$ & & $\ln$ \\
    $C_{qu}^{(1),\Sigma\Sigma33}$ & & & & & $\ln$ & & $\ln$  \\
    $C_{qd}^{(1),3323}$ & & & & & & $\ln$ & & $\ln$ \\
    $C_{ud}^{(1),3323}$ & & & & & & $\ln$ & & $\ln$ \\
    $C_{quqd}^{(1),\Sigma332},C_{quqd}^{(8),\Sigma332}$ & & $\ln$ & & $\ln$ \\
    $C_{quqd}^{(1),\Sigma333},C_{quqd}^{(8),\Sigma333}$ & $\ln$ & & $\ln$ \\\bottomrule
\end{tabular}
\caption{UV sensitivity in SMEFT-to-WET matching relations at one-loop level, resulting in a logarithmic dependence on the matching scale $\mu$. The symbol x stands for finite matching relations, $\ln$ stands for a UV-sensitive matching relation, and $\sum$ denotes the sum over flavor indices $k\in\{1,2,3\}$. SMEFT coefficients with finite matching relations to all eight WET coefficients are not shown.\label{tab:uv-sensitivity}}
\end{table}

\section{Four-quark operators in top-antitop production}
\label{app:4q-in-ttb-warsaw}
The contributions of four-quark SMEFT coefficients in the Warsaw basis~\cite{Grzadkowski:2010es} can be classified according to their color and chiral structure. Color-singlet interactions for up-quarks are expressed as~\cite{AguilarSaavedra:2018nen,Brivio:2019ius}
\begin{align}
4 C_{VV}^{(1),u_i} & = \phantom{-}C_{qq}^{(1),33ii} + C_{qq}^{(3),33ii} + \frac{1}{3}\left(C_{qq}^{(1),3ii3} + C_{qq}^{(3),3ii3}\right) + C_{uu}^{33ii} + \frac{1}{3}C_{uu}^{3ii3}\\\nonumber
& \quad + C_{qu}^{(1),ii33} + C_{qu}^{(1),33ii}\\\nonumber
4 C_{AA}^{(1),u_i} & = \phantom{-}C_{qq}^{(1),33ii} + C_{qq}^{(3),33ii} + \frac{1}{3}\left(C_{qq}^{(1),3ii3} + C_{qq}^{(3),3ii3}\right) + C_{uu}^{33ii} + \frac{1}{3}C_{uu}^{3ii3}\\\nonumber
& \quad - C_{qu}^{(1),ii33} - C_{qu}^{(1),33ii}\\\nonumber
4 C_{AV}^{(1),u_i} & = - C_{qq}^{(1),33ii} - C_{qq}^{(3),33ii} - \frac{1}{3}\left(C_{qq}^{(1),3ii3} + C_{qq}^{(3),3ii3}\right) + C_{uu}^{33ii} + \frac{1}{3}C_{uu}^{3ii3}\\\nonumber
& \quad + C_{qu}^{(1),ii33} - C_{qu}^{(1),33ii}\\\nonumber
4 C_{VA}^{(1),u_i} & = - C_{qq}^{(1),33ii} - C_{qq}^{(3),33ii} - \frac{1}{3}\left(C_{qq}^{(1),3ii3} + C_{qq}^{(3),3ii3}\right) + C_{uu}^{33ii} + \frac{1}{3}C_{uu}^{3ii3}\\\nonumber
& \quad - C_{qu}^{(1),ii33} + C_{qu}^{(1),33ii}
\end{align}
and for down-quarks as
\begin{align}
4 C_{VV}^{(1),d_i} & = \phantom{-}C_{qq}^{(1),33ii} - C_{qq}^{(3),33ii} + \frac{2}{3} C_{qq}^{(3),3ii3} + C_{ud}^{(1),33ii} 
+ C_{qu}^{(1),ii33} + C_{qd}^{(1),33ii}\\\nonumber
4 C_{AA}^{(1),d_i} & = \phantom{-}C_{qq}^{(1),33ii} - C_{qq}^{(3),33ii} + \frac{2}{3} C_{qq}^{(3),3ii3} + C_{ud}^{(1),33ii} 
- C_{qu}^{(1),ii33} - C_{qd}^{(1),33ii}\\\nonumber
4 C_{AV}^{(1),d_i} & = - C_{qq}^{(1),33ii} + C_{qq}^{(3),33ii} - \frac{2}{3} C_{qq}^{(3),3ii3} + C_{ud}^{(1),33ii} + C_{qu}^{(1),ii33} - C_{qd}^{(1),33ii}\\\nonumber
4 C_{VA}^{(1),d_i} & = - C_{qq}^{(1),33ii} + C_{qq}^{(3),33ii} - \frac{2}{3} C_{qq}^{(3),3ii3} + C_{ud}^{(1),33ii} - C_{qu}^{(1),ii33} + C_{qd}^{(1),33ii}.
\end{align}
Color-octet interactions are given for up-quarks by
\begin{align}
4 C_{VV}^{(8),u_i} & = \phantom{-}2\left(C_{qq}^{(1),3ii3} + C_{qq}^{(3),3ii3}\right) + 2C_{uu}^{3ii3} + C_{qu}^{(8),ii33} + C_{qu}^{(8),33ii}\\\nonumber
4 C_{AA}^{(8),u_i} & = \phantom{-}2\left(C_{qq}^{(1),3ii3} + C_{qq}^{(3),3ii3}\right) + 2C_{uu}^{3ii3} - C_{qu}^{(8),ii33} - C_{qu}^{(8),33ii}\\\nonumber
4 C_{AV}^{(8),u_i} & = - 2\left(C_{qq}^{(1),3ii3} + C_{qq}^{(3),3ii3}\right) + 2C_{uu}^{3ii3} + C_{qu}^{(8),ii33} - C_{qu}^{(8),33ii}\\\nonumber
4 C_{VA}^{(8),u_i} & = - 2\left(C_{qq}^{(1),3ii3} + C_{qq}^{(3),3ii3}\right) + 2C_{uu}^{3ii3} - C_{qu}^{(8),ii33} + C_{qu}^{(8),33ii}
\end{align}
and for down-quarks by
\begin{align}
4 C_{VV}^{(8),d_i} & = \phantom{-} 4 C_{qq}^{(3),3ii3} + C_{ud}^{(8),33ii} + C_{qu}^{(8),ii33} + C_{qd}^{(8),33ii}\\\nonumber
4 C_{AA}^{(8),d_i} & = \phantom{-} 4 C_{qq}^{(3),3ii3} + C_{ud}^{(8),33ii} - C_{qu}^{(8),ii33} - C_{qd}^{(8),33ii}\\\nonumber
4 C_{AV}^{(8),d_i} & = - 4 C_{qq}^{(3),3ii3} + C_{ud}^{(8),33ii} + C_{qu}^{(8),ii33} - C_{qd}^{(8),33ii}\\\nonumber
4 C_{VA}^{(8),d_i} & = - 4 C_{qq}^{(3),3ii3} + C_{ud}^{(8),33ii} - C_{qu}^{(8),ii33} + C_{qd}^{(8),33ii}.
\end{align}
We have neglected contributions of $C^{3333}$ and other flavor combinations, because they are suppressed by CKM elements or by $b$-quark parton distributions.

\bibliographystyle{JHEP}
\bibliography{top-flavor-v1.bib}

\providecommand{\href}[2]{#2}\begingroup\raggedright\begin{thebibliography}{100}

\bibitem{Buchmuller:1985jz}
W.~Buchmuller and D.~Wyler, \emph{{Effective Lagrangian Analysis of New
  Interactions and Flavor Conservation}},
  \href{https://doi.org/10.1016/0550-3213(86)90262-2}{\emph{Nucl. Phys. B}
  {\bfseries 268} (1986) 621}.

\bibitem{Grzadkowski:2010es}
B.~Grzadkowski, M.~Iskrzynski, M.~Misiak and J.~Rosiek, \emph{{Dimension-Six
  Terms in the Standard Model Lagrangian}},
  \href{https://doi.org/10.1007/JHEP10(2010)085}{\emph{JHEP} {\bfseries 10}
  (2010) 085} [\href{https://arxiv.org/abs/1008.4884}{{\ttfamily 1008.4884}}].

\bibitem{Brivio:2017vri}
I.~Brivio and M.~Trott, \emph{{The Standard Model as an Effective Field
  Theory}}, \href{https://doi.org/10.1016/j.physrep.2018.11.002}{\emph{Phys.
  Rept.} {\bfseries 793} (2019) 1}
  [\href{https://arxiv.org/abs/1706.08945}{{\ttfamily 1706.08945}}].

\bibitem{Dawson:2020oco}
S.~Dawson, S.~Homiller and S.~D. Lane, \emph{{Putting standard model EFT fits
  to work}}, \href{https://doi.org/10.1103/PhysRevD.102.055012}{\emph{Phys.
  Rev. D} {\bfseries 102} (2020) 055012}
  [\href{https://arxiv.org/abs/2007.01296}{{\ttfamily 2007.01296}}].

\bibitem{David:2020pzt}
A.~David and G.~Passarino, \emph{{Use and reuse of SMEFT}},
  \href{https://arxiv.org/abs/2009.00127}{{\ttfamily 2009.00127}}.

\bibitem{Ellis:2020unq}
J.~Ellis, M.~Madigan, K.~Mimasu, V.~Sanz and T.~You, \emph{{Top, Higgs, Diboson
  and Electroweak Fit to the Standard Model Effective Field Theory}},
  \href{https://arxiv.org/abs/2012.02779}{{\ttfamily 2012.02779}}.

\bibitem{Zhang:2020jyn}
C.~Zhang and S.-Y. Zhou, \emph{{Convex Geometry Perspective on the (Standard
  Model) Effective Field Theory Space}},
  \href{https://doi.org/10.1103/PhysRevLett.125.201601}{\emph{Phys. Rev. Lett.}
  {\bfseries 125} (2020) 201601}
  [\href{https://arxiv.org/abs/2005.03047}{{\ttfamily 2005.03047}}].

\bibitem{Chakrabortty:2020mbc}
J.~Chakrabortty, S.~Prakash, S.~U. Rahaman and M.~Spannowsky, \emph{{Uncovering
  the Root of LEFT in SMEFT}},
  \href{https://arxiv.org/abs/2011.00859}{{\ttfamily 2011.00859}}.

\bibitem{Feldmann:2008ja}
T.~Feldmann and T.~Mannel, \emph{{Large Top Mass and Non-Linear Representation
  of Flavour Symmetry}},
  \href{https://doi.org/10.1103/PhysRevLett.100.171601}{\emph{Phys. Rev. Lett.}
  {\bfseries 100} (2008) 171601}
  [\href{https://arxiv.org/abs/0801.1802}{{\ttfamily 0801.1802}}].

\bibitem{Buchalla:1995vs}
G.~Buchalla, A.~J. Buras and M.~E. Lautenbacher, \emph{{Weak decays beyond
  leading logarithms}},
  \href{https://doi.org/10.1103/RevModPhys.68.1125}{\emph{Rev. Mod. Phys.}
  {\bfseries 68} (1996) 1125}
  [\href{https://arxiv.org/abs/hep-ph/9512380}{{\ttfamily hep-ph/9512380}}].

\bibitem{Eilam:1990zc}
G.~Eilam, J.~Hewett and A.~Soni, \emph{{Rare decays of the top quark in the
  standard and two Higgs doublet models}},
  \href{https://doi.org/10.1103/PhysRevD.44.1473}{\emph{Phys. Rev. D}
  {\bfseries 44} (1991) 1473}.

\bibitem{Mele:1998ag}
B.~Mele, S.~Petrarca and A.~Soddu, \emph{{A New evaluation of the $t \to cH$
  decay width in the standard model}},
  \href{https://doi.org/10.1016/S0370-2693(98)00822-3}{\emph{Phys. Lett. B}
  {\bfseries 435} (1998) 401}
  [\href{https://arxiv.org/abs/hep-ph/9805498}{{\ttfamily hep-ph/9805498}}].

\bibitem{Buras:2000dm}
A.~Buras, P.~Gambino, M.~Gorbahn, S.~Jager and L.~Silvestrini, \emph{{Universal
  unitarity triangle and physics beyond the standard model}},
  \href{https://doi.org/10.1016/S0370-2693(01)00061-2}{\emph{Phys. Lett. B}
  {\bfseries 500} (2001) 161}
  [\href{https://arxiv.org/abs/hep-ph/0007085}{{\ttfamily hep-ph/0007085}}].

\bibitem{DAmbrosio:2002vsn}
G.~D'Ambrosio, G.~Giudice, G.~Isidori and A.~Strumia, \emph{{Minimal flavor
  violation: An Effective field theory approach}},
  \href{https://doi.org/10.1016/S0550-3213(02)00836-2}{\emph{Nucl. Phys. B}
  {\bfseries 645} (2002) 155}
  [\href{https://arxiv.org/abs/hep-ph/0207036}{{\ttfamily hep-ph/0207036}}].

\bibitem{Barbieri:2012uh}
R.~Barbieri, D.~Buttazzo, F.~Sala and D.~M. Straub, \emph{{Flavour physics from
  an approximate $U(2)^3$ symmetry}},
  \href{https://doi.org/10.1007/JHEP07(2012)181}{\emph{JHEP} {\bfseries 07}
  (2012) 181} [\href{https://arxiv.org/abs/1203.4218}{{\ttfamily 1203.4218}}].

\bibitem{Efrati:2015eaa}
A.~Efrati, A.~Falkowski and Y.~Soreq, \emph{{Electroweak constraints on
  flavorful effective theories}},
  \href{https://doi.org/10.1007/JHEP07(2015)018}{\emph{JHEP} {\bfseries 07}
  (2015) 018} [\href{https://arxiv.org/abs/1503.07872}{{\ttfamily
  1503.07872}}].

\bibitem{AguilarSaavedra:2018nen}
D.~Barducci et~al., \emph{{Interpreting top-quark LHC measurements in the
  standard-model effective field theory}},
  \href{https://arxiv.org/abs/1802.07237}{{\ttfamily 1802.07237}}.

\bibitem{Faroughy:2020ina}
D.~A. Faroughy, G.~Isidori, F.~Wilsch and K.~Yamamoto, \emph{{Flavour
  symmetries in the SMEFT}},
  \href{https://doi.org/10.1007/JHEP08(2020)166}{\emph{JHEP} {\bfseries 08}
  (2020) 166} [\href{https://arxiv.org/abs/2005.05366}{{\ttfamily
  2005.05366}}].

\bibitem{Hewett:1993em}
J.~L. Hewett and T.~G. Rizzo, \emph{{Using $b \to s\gamma$ to probe top quark
  couplings}}, \href{https://doi.org/10.1103/PhysRevD.49.319}{\emph{Phys. Rev.
  D} {\bfseries 49} (1994) 319}
  [\href{https://arxiv.org/abs/hep-ph/9305223}{{\ttfamily hep-ph/9305223}}].

\bibitem{Grzadkowski:2008mf}
B.~Grzadkowski and M.~Misiak, \emph{{Anomalous Wtb coupling effects in the weak
  radiative B-meson decay}},
  \href{https://doi.org/10.1103/PhysRevD.78.077501}{\emph{Phys. Rev. D}
  {\bfseries 78} (2008) 077501}
  [\href{https://arxiv.org/abs/0802.1413}{{\ttfamily 0802.1413}}].

\bibitem{Kamenik:2011dk}
J.~F. Kamenik, M.~Papucci and A.~Weiler, \emph{{Constraining the dipole moments
  of the top quark}},
  \href{https://doi.org/10.1103/PhysRevD.85.071501}{\emph{Phys. Rev. D}
  {\bfseries 85} (2012) 071501}
  [\href{https://arxiv.org/abs/1107.3143}{{\ttfamily 1107.3143}}].

\bibitem{Drobnak:2011aa}
J.~Drobnak, S.~Fajfer and J.~F. Kamenik, \emph{{Probing anomalous tWb
  interactions with rare B decays}},
  \href{https://doi.org/10.1016/j.nuclphysb.2011.10.004}{\emph{Nucl. Phys. B}
  {\bfseries 855} (2012) 82} [\href{https://arxiv.org/abs/1109.2357}{{\ttfamily
  1109.2357}}].

\bibitem{Brod:2014hsa}
J.~Brod, A.~Greljo, E.~Stamou and P.~Uttayarat, \emph{{Probing anomalous $
  t\overline{t}Z $ interactions with rare meson decays}},
  \href{https://doi.org/10.1007/JHEP02(2015)141}{\emph{JHEP} {\bfseries 02}
  (2015) 141} [\href{https://arxiv.org/abs/1408.0792}{{\ttfamily 1408.0792}}].

\bibitem{Aebischer:2018iyb}
J.~Aebischer, J.~Kumar, P.~Stangl and D.~M. Straub, \emph{{A Global Likelihood
  for Precision Constraints and Flavour Anomalies}},
  \href{https://doi.org/10.1140/epjc/s10052-019-6977-z}{\emph{Eur. Phys. J. C}
  {\bfseries 79} (2019) 509}
  [\href{https://arxiv.org/abs/1810.07698}{{\ttfamily 1810.07698}}].

\bibitem{Silvestrini:2018dos}
L.~Silvestrini and M.~Valli, \emph{{Model-independent Bounds on the Standard
  Model Effective Theory from Flavour Physics}},
  \href{https://doi.org/10.1016/j.physletb.2019.135062}{\emph{Phys. Lett. B}
  {\bfseries 799} (2019) 135062}
  [\href{https://arxiv.org/abs/1812.10913}{{\ttfamily 1812.10913}}].

\bibitem{Aebischer:2020dsw}
J.~Aebischer, C.~Bobeth, A.~J. Buras and J.~Kumar, \emph{{SMEFT ATLAS of
  $\Delta$F = 2 transitions}},
  \href{https://doi.org/10.1007/JHEP12(2020)187}{\emph{JHEP} {\bfseries 12}
  (2020) 187} [\href{https://arxiv.org/abs/2009.07276}{{\ttfamily
  2009.07276}}].

\bibitem{Fox:2007in}
P.~J. Fox, Z.~Ligeti, M.~Papucci, G.~Perez and M.~D. Schwartz,
  \emph{{Deciphering top flavor violation at the LHC with $B$ factories}},
  \href{https://doi.org/10.1103/PhysRevD.78.054008}{\emph{Phys. Rev. D}
  {\bfseries 78} (2008) 054008}
  [\href{https://arxiv.org/abs/0704.1482}{{\ttfamily 0704.1482}}].

\bibitem{Cirigliano:2016nyn}
V.~Cirigliano, W.~Dekens, J.~de~Vries and E.~Mereghetti, \emph{{Constraining
  the top-Higgs sector of the Standard Model Effective Field Theory}},
  \href{https://doi.org/10.1103/PhysRevD.94.034031}{\emph{Phys. Rev. D}
  {\bfseries 94} (2016) 034031}
  [\href{https://arxiv.org/abs/1605.04311}{{\ttfamily 1605.04311}}].

\bibitem{Alioli:2017ces}
S.~Alioli, V.~Cirigliano, W.~Dekens, J.~de~Vries and E.~Mereghetti,
  \emph{{Right-handed charged currents in the era of the Large Hadron
  Collider}}, \href{https://doi.org/10.1007/JHEP05(2017)086}{\emph{JHEP}
  {\bfseries 05} (2017) 086}
  [\href{https://arxiv.org/abs/1703.04751}{{\ttfamily 1703.04751}}].

\bibitem{Biekotter:2018rhp}
A.~Biekoetter, T.~Corbett and T.~Plehn, \emph{{The Gauge-Higgs Legacy of the
  LHC Run II}},
  \href{https://doi.org/10.21468/SciPostPhys.6.6.064}{\emph{SciPost Phys.}
  {\bfseries 6} (2019) 064} [\href{https://arxiv.org/abs/1812.07587}{{\ttfamily
  1812.07587}}].

\bibitem{Bissmann:2019gfc}
{Bi\ss{}mann, Stefan and Erdmann, Johannes and Grunwald, Cornelius and Hiller,
  Gudrun and Kr\"oninger, Kevin}, \emph{{Constraining top-quark couplings
  combining top-quark and $\boldsymbol{B}$ decay observables}},
  \href{https://doi.org/10.1140/epjc/s10052-020-7680-9}{\emph{Eur. Phys. J. C}
  {\bfseries 80} (2020) 136}
  [\href{https://arxiv.org/abs/1909.13632}{{\ttfamily 1909.13632}}].

\bibitem{Falkowski:2019hvp}
A.~Falkowski and D.~Straub, \emph{{Flavourful SMEFT likelihood for Higgs and
  electroweak data}},
  \href{https://doi.org/10.1007/JHEP04(2020)066}{\emph{JHEP} {\bfseries 04}
  (2020) 066} [\href{https://arxiv.org/abs/1911.07866}{{\ttfamily
  1911.07866}}].

\bibitem{Aoude:2020dwv}
R.~Aoude, T.~Hurth, S.~Renner and W.~Shepherd, \emph{{The impact of flavour
  data on global fits of the MFV SMEFT}},
  \href{https://arxiv.org/abs/2003.05432}{{\ttfamily 2003.05432}}.

\bibitem{Bissmann:2020mfi}
S.~Bi\ss{}mann, C.~Grunwald, G.~Hiller and K.~Kr\"oninger, \emph{{Top and
  Beauty synergies in SMEFT-fits at present and future colliders}},
  \href{https://arxiv.org/abs/2012.10456}{{\ttfamily 2012.10456}}.

\bibitem{Zhang:2010dr}
C.~Zhang and S.~Willenbrock, \emph{{Effective-Field-Theory Approach to
  Top-Quark Production and Decay}},
  \href{https://doi.org/10.1103/PhysRevD.83.034006}{\emph{Phys. Rev. D}
  {\bfseries 83} (2011) 034006}
  [\href{https://arxiv.org/abs/1008.3869}{{\ttfamily 1008.3869}}].

\bibitem{Faller:2013gca}
S.~Faller, S.~Gadatsch and T.~Mannel, \emph{{Minimal flavor violation and
  anomalous top decays}},
  \href{https://doi.org/10.1103/PhysRevD.88.035006}{\emph{Phys. Rev. D}
  {\bfseries 88} (2013) 035006}
  [\href{https://arxiv.org/abs/1304.2675}{{\ttfamily 1304.2675}}].

\bibitem{Rontsch:2014cca}
R.~R\"ontsch and M.~Schulze, \emph{{Constraining couplings of top quarks to the
  Z boson in $ t\overline{t} $ + Z production at the LHC}},
  \href{https://doi.org/10.1007/JHEP09(2015)132}{\emph{JHEP} {\bfseries 07}
  (2014) 091} [\href{https://arxiv.org/abs/1404.1005}{{\ttfamily 1404.1005}}].

\bibitem{Rontsch:2015una}
R.~R\"ontsch and M.~Schulze, \emph{{Probing top-Z dipole moments at the LHC and
  ILC}}, \href{https://doi.org/10.1007/JHEP08(2015)044}{\emph{JHEP} {\bfseries
  08} (2015) 044} [\href{https://arxiv.org/abs/1501.05939}{{\ttfamily
  1501.05939}}].

\bibitem{Bylund:2016phk}
O.~Bessidskaia~Bylund, F.~Maltoni, I.~Tsinikos, E.~Vryonidou and C.~Zhang,
  \emph{{Probing top quark neutral couplings in the Standard Model Effective
  Field Theory at NLO in QCD}},
  \href{https://doi.org/10.1007/JHEP05(2016)052}{\emph{JHEP} {\bfseries 05}
  (2016) 052} [\href{https://arxiv.org/abs/1601.08193}{{\ttfamily
  1601.08193}}].

\bibitem{Buckley:2015lku}
A.~Buckley, C.~Englert, J.~Ferrando, D.~J. Miller, L.~Moore, M.~Russell et~al.,
  \emph{{Constraining top quark effective theory in the LHC Run II era}},
  \href{https://doi.org/10.1007/JHEP04(2016)015}{\emph{JHEP} {\bfseries 04}
  (2016) 015} [\href{https://arxiv.org/abs/1512.03360}{{\ttfamily
  1512.03360}}].

\bibitem{Brivio:2019ius}
I.~Brivio, S.~Bruggisser, F.~Maltoni, R.~Moutafis, T.~Plehn, E.~Vryonidou
  et~al., \emph{{O new physics, where art thou? A global search in the top
  sector}}, \href{https://doi.org/10.1007/JHEP02(2020)131}{\emph{JHEP}
  {\bfseries 02} (2020) 131}
  [\href{https://arxiv.org/abs/1910.03606}{{\ttfamily 1910.03606}}].

\bibitem{Hartland:2019bjb}
N.~P. Hartland, F.~Maltoni, E.~R. Nocera, J.~Rojo, E.~Slade, E.~Vryonidou
  et~al., \emph{{A Monte Carlo global analysis of the Standard Model Effective
  Field Theory: the top quark sector}},
  \href{https://doi.org/10.1007/JHEP04(2019)100}{\emph{JHEP} {\bfseries 04}
  (2019) 100} [\href{https://arxiv.org/abs/1901.05965}{{\ttfamily
  1901.05965}}].

\bibitem{Durieux:2019rbz}
G.~Durieux, A.~Irles, V.~Miralles, A.~Pe\~nuelas, R.~P\"oschl, M.~Perell\'o
  et~al., \emph{{The electro-weak couplings of the top and bottom quarks --
  global fit and future prospects}},
  \href{https://doi.org/10.1007/JHEP12(2019)098}{\emph{JHEP} {\bfseries 12}
  (2019) 098} [\href{https://arxiv.org/abs/1907.10619}{{\ttfamily
  1907.10619}}].

\bibitem{Aebischer:2015fzz}
J.~Aebischer, A.~Crivellin, M.~Fael and C.~Greub, \emph{{Matching of gauge
  invariant dimension-six operators for $b\to s$ and $b\to c$ transitions}},
  \href{https://doi.org/10.1007/JHEP05(2016)037}{\emph{JHEP} {\bfseries 05}
  (2016) 037} [\href{https://arxiv.org/abs/1512.02830}{{\ttfamily
  1512.02830}}].

\bibitem{Jenkins:2017jig}
E.~E. Jenkins, A.~V. Manohar and P.~Stoffer, \emph{{Low-Energy Effective Field
  Theory below the Electroweak Scale: Operators and Matching}},
  \href{https://doi.org/10.1007/JHEP03(2018)016}{\emph{JHEP} {\bfseries 03}
  (2018) 016} [\href{https://arxiv.org/abs/1709.04486}{{\ttfamily
  1709.04486}}].

\bibitem{Dekens:2019ept}
W.~Dekens and P.~Stoffer, \emph{{Low-energy effective field theory below the
  electroweak scale: matching at one loop}},
  \href{https://doi.org/10.1007/JHEP10(2019)197}{\emph{JHEP} {\bfseries 10}
  (2019) 197} [\href{https://arxiv.org/abs/1908.05295}{{\ttfamily
  1908.05295}}].

\bibitem{Hurth:2019ula}
T.~Hurth, S.~Renner and W.~Shepherd, \emph{{Matching for FCNC effects in the
  flavour-symmetric SMEFT}},
  \href{https://doi.org/10.1007/JHEP06(2019)029}{\emph{JHEP} {\bfseries 06}
  (2019) 029} [\href{https://arxiv.org/abs/1903.00500}{{\ttfamily
  1903.00500}}].

\bibitem{Aebischer:2017gaw}
J.~Aebischer, M.~Fael, C.~Greub and J.~Virto, \emph{{B physics Beyond the
  Standard Model at One Loop: Complete Renormalization Group Evolution below
  the Electroweak Scale}},
  \href{https://doi.org/10.1007/JHEP09(2017)158}{\emph{JHEP} {\bfseries 09}
  (2017) 158} [\href{https://arxiv.org/abs/1704.06639}{{\ttfamily
  1704.06639}}].

\bibitem{Jenkins:2017dyc}
E.~E. Jenkins, A.~V. Manohar and P.~Stoffer, \emph{{Low-Energy Effective Field
  Theory below the Electroweak Scale: Anomalous Dimensions}},
  \href{https://doi.org/10.1007/JHEP01(2018)084}{\emph{JHEP} {\bfseries 01}
  (2018) 084} [\href{https://arxiv.org/abs/1711.05270}{{\ttfamily
  1711.05270}}].

\bibitem{Kagan:2009bn}
A.~L. Kagan, G.~Perez, T.~Volansky and J.~Zupan, \emph{{General Minimal Flavor
  Violation}}, \href{https://doi.org/10.1103/PhysRevD.80.076002}{\emph{Phys.
  Rev. D} {\bfseries 80} (2009) 076002}
  [\href{https://arxiv.org/abs/0903.1794}{{\ttfamily 0903.1794}}].

\bibitem{EOS}
D.~van Dyk et~al., ``{\textit{EOS --- A HEP program for flavour
  observables}}.''
\newblock \url{https://eos.github.io/}.

\bibitem{Aebischer:2017ugx}
J.~Aebischer et~al., \emph{{WCxf: an exchange format for Wilson coefficients
  beyond the Standard Model}},
  \href{https://doi.org/10.1016/j.cpc.2018.05.022}{\emph{Comput. Phys. Commun.}
  {\bfseries 232} (2018) 71}
  [\href{https://arxiv.org/abs/1712.05298}{{\ttfamily 1712.05298}}].

\bibitem{Beneke:2001at}
M.~Beneke, T.~Feldmann and D.~Seidel, \emph{{Systematic approach to exclusive
  $B \to V l^+ l^-$, $V \gamma$ decays}},
  \href{https://doi.org/10.1016/S0550-3213(01)00366-2}{\emph{Nucl. Phys. B}
  {\bfseries 612} (2001) 25}
  [\href{https://arxiv.org/abs/hep-ph/0106067}{{\ttfamily hep-ph/0106067}}].

\bibitem{Zyla:2020zbs}
{\scshape Particle Data Group} collaboration, \emph{{Review of Particle
  Physics}}, \href{https://doi.org/10.1093/ptep/ptaa104}{\emph{PTEP} {\bfseries
  2020} (2020) 083C01}.

\bibitem{Lyon:2014hpa}
J.~Lyon and R.~Zwicky, \emph{{Resonances gone topsy turvy - the charm of QCD or
  new physics in $b \to s \ell^+ \ell^-$?}},
  \href{https://arxiv.org/abs/1406.0566}{{\ttfamily 1406.0566}}.

\bibitem{Ciuchini:2015qxb}
M.~Ciuchini, M.~Fedele, E.~Franco, S.~Mishima, A.~Paul, L.~Silvestrini et~al.,
  \emph{{$B\to K^* \ell^+ \ell^-$ decays at large recoil in the Standard Model:
  a theoretical reappraisal}},
  \href{https://doi.org/10.1007/JHEP06(2016)116}{\emph{JHEP} {\bfseries 06}
  (2016) 116} [\href{https://arxiv.org/abs/1512.07157}{{\ttfamily
  1512.07157}}].

\bibitem{Capdevila:2017ert}
B.~Capdevila, S.~Descotes-Genon, L.~Hofer and J.~Matias, \emph{{Hadronic
  uncertainties in $B \to K^* \mu^+ \mu^-$: a state-of-the-art analysis}},
  \href{https://doi.org/10.1007/JHEP04(2017)016}{\emph{JHEP} {\bfseries 04}
  (2017) 016} [\href{https://arxiv.org/abs/1701.08672}{{\ttfamily
  1701.08672}}].

\bibitem{Jager:2017gal}
S.~J\"ager, M.~Kirk, A.~Lenz and K.~Leslie, \emph{{Charming new physics in rare
  B-decays and mixing?}},
  \href{https://doi.org/10.1103/PhysRevD.97.015021}{\emph{Phys. Rev. D}
  {\bfseries 97} (2018) 015021}
  [\href{https://arxiv.org/abs/1701.09183}{{\ttfamily 1701.09183}}].

\bibitem{Bobeth:2017vxj}
C.~Bobeth, M.~Chrzaszcz, D.~van Dyk and J.~Virto, \emph{{Long-distance effects
  in $B\rightarrow K^*\ell \ell $ from analyticity}},
  \href{https://doi.org/10.1140/epjc/s10052-018-5918-6}{\emph{Eur. Phys. J. C}
  {\bfseries 78} (2018) 451}
  [\href{https://arxiv.org/abs/1707.07305}{{\ttfamily 1707.07305}}].

\bibitem{Arbey:2018ics}
A.~Arbey, T.~Hurth, F.~Mahmoudi and S.~Neshatpour, \emph{{Hadronic and New
  Physics Contributions to $b \to s$ Transitions}},
  \href{https://doi.org/10.1103/PhysRevD.98.095027}{\emph{Phys. Rev. D}
  {\bfseries 98} (2018) 095027}
  [\href{https://arxiv.org/abs/1806.02791}{{\ttfamily 1806.02791}}].

\bibitem{Gubernari:2020eft}
N.~Gubernari, D.~van Dyk and J.~Virto, \emph{{Non-local matrix elements in
  $B_{(s)}\to \{K^{(*)},\phi\}\ell^+\ell^-$}},
  \href{https://arxiv.org/abs/2011.09813}{{\ttfamily 2011.09813}}.

\bibitem{Alonso:2014csa}
R.~Alonso, B.~Grinstein and J.~Martin~Camalich, \emph{{$SU(2)\times U(1)$ gauge
  invariance and the shape of new physics in rare $B$ decays}},
  \href{https://doi.org/10.1103/PhysRevLett.113.241802}{\emph{Phys. Rev. Lett.}
  {\bfseries 113} (2014) 241802}
  [\href{https://arxiv.org/abs/1407.7044}{{\ttfamily 1407.7044}}].

\bibitem{Cata:2015lta}
O.~Cat\`a and M.~Jung, \emph{{Signatures of a nonstandard Higgs boson from
  flavor physics}},
  \href{https://doi.org/10.1103/PhysRevD.92.055018}{\emph{Phys. Rev. D}
  {\bfseries 92} (2015) 055018}
  [\href{https://arxiv.org/abs/1505.05804}{{\ttfamily 1505.05804}}].

\bibitem{Descotes-Genon:2018foz}
S.~Descotes-Genon, A.~Falkowski, M.~Fedele, M.~Gonz\'alez-Alonso and J.~Virto,
  \emph{{The CKM parameters in the SMEFT}},
  \href{https://doi.org/10.1007/JHEP05(2019)172}{\emph{JHEP} {\bfseries 05}
  (2019) 172} [\href{https://arxiv.org/abs/1812.08163}{{\ttfamily
  1812.08163}}].

\bibitem{Aebischer:2018bkb}
J.~Aebischer, J.~Kumar and D.~M. Straub, \emph{{Wilson: a Python package for
  the running and matching of Wilson coefficients above and below the
  electroweak scale}},
  \href{https://doi.org/10.1140/epjc/s10052-018-6492-7}{\emph{Eur. Phys. J. C}
  {\bfseries 78} (2018) 1026}
  [\href{https://arxiv.org/abs/1804.05033}{{\ttfamily 1804.05033}}].

\bibitem{Bobeth:1999mk}
C.~Bobeth, M.~Misiak and J.~Urban, \emph{{Photonic penguins at two loops and
  $m_t$ dependence of $BR[B \to X_s l^+ l^-]$}},
  \href{https://doi.org/10.1016/S0550-3213(00)00007-9}{\emph{Nucl. Phys. B}
  {\bfseries 574} (2000) 291}
  [\href{https://arxiv.org/abs/hep-ph/9910220}{{\ttfamily hep-ph/9910220}}].

\bibitem{Bobeth:2003at}
C.~Bobeth, P.~Gambino, M.~Gorbahn and U.~Haisch, \emph{{Complete NNLO QCD
  analysis of $\bar{B} \to X_s\ell^+\ell^-$ and higher order electroweak
  effects}}, \href{https://doi.org/10.1088/1126-6708/2004/04/071}{\emph{JHEP}
  {\bfseries 04} (2004) 071}
  [\href{https://arxiv.org/abs/hep-ph/0312090}{{\ttfamily hep-ph/0312090}}].

\bibitem{Huber:2005ig}
T.~Huber, E.~Lunghi, M.~Misiak and D.~Wyler, \emph{{Electromagnetic logarithms
  in $\bar B \to X_s l^+ l^-$}},
  \href{https://doi.org/10.1016/j.nuclphysb.2006.01.037}{\emph{Nucl. Phys. B}
  {\bfseries 740} (2006) 105}
  [\href{https://arxiv.org/abs/hep-ph/0512066}{{\ttfamily hep-ph/0512066}}].

\bibitem{Bobeth:2013uxa}
C.~Bobeth, M.~Gorbahn, T.~Hermann, M.~Misiak, E.~Stamou and M.~Steinhauser,
  \emph{{$B_{s,d} \to l^+ l^-$ in the Standard Model with Reduced Theoretical
  Uncertainty}},
  \href{https://doi.org/10.1103/PhysRevLett.112.101801}{\emph{Phys. Rev. Lett.}
  {\bfseries 112} (2014) 101801}
  [\href{https://arxiv.org/abs/1311.0903}{{\ttfamily 1311.0903}}].

\bibitem{Durieux:2014xla}
G.~Durieux, F.~Maltoni and C.~Zhang, \emph{{Global approach to top-quark
  flavor-changing interactions}},
  \href{https://doi.org/10.1103/PhysRevD.91.074017}{\emph{Phys. Rev. D}
  {\bfseries 91} (2015) 074017}
  [\href{https://arxiv.org/abs/1412.7166}{{\ttfamily 1412.7166}}].

\bibitem{Degrande:2018fog}
C.~Degrande, F.~Maltoni, K.~Mimasu, E.~Vryonidou and C.~Zhang,
  \emph{{Single-top associated production with a $Z$ or $H$ boson at the LHC:
  the SMEFT interpretation}},
  \href{https://doi.org/10.1007/JHEP10(2018)005}{\emph{JHEP} {\bfseries 10}
  (2018) 005} [\href{https://arxiv.org/abs/1804.07773}{{\ttfamily
  1804.07773}}].

\bibitem{Rosello:2015sck}
M.~P. Rosello and M.~Vos, \emph{{Constraints on four-fermion interactions from
  the $t\bar{t}$ charge asymmetry at hadron colliders}},
  \href{https://doi.org/10.1140/epjc/s10052-016-4040-x}{\emph{Eur. Phys. J. C}
  {\bfseries 76} (2016) 200}
  [\href{https://arxiv.org/abs/1512.07542}{{\ttfamily 1512.07542}}].

\bibitem{Berge:2012rc}
S.~Berge and S.~Westhoff, \emph{{Top-Quark Charge Asymmetry with a Jet
  Handle}}, \href{https://doi.org/10.1103/PhysRevD.86.094036}{\emph{Phys. Rev.
  D} {\bfseries 86} (2012) 094036}
  [\href{https://arxiv.org/abs/1208.4104}{{\ttfamily 1208.4104}}].

\bibitem{Basan:2020btr}
A.~Basan, P.~Berta, L.~Masetti, E.~Vryonidou and S.~Westhoff, \emph{{Measuring
  the top energy asymmetry at the LHC: QCD and SMEFT interpretations}},
  \href{https://doi.org/10.1007/JHEP03(2020)184}{\emph{JHEP} {\bfseries 03}
  (2020) 184} [\href{https://arxiv.org/abs/2001.07225}{{\ttfamily
  2001.07225}}].

\bibitem{Lafaye:2004cn}
R.~Lafaye, T.~Plehn and D.~Zerwas, \emph{{SFITTER: SUSY parameter analysis at
  LHC and LC}},  \href{https://arxiv.org/abs/hep-ph/0404282}{{\ttfamily
  hep-ph/0404282}}.

\bibitem{Lafaye:2007vs}
R.~Lafaye, T.~Plehn, M.~Rauch and D.~Zerwas, \emph{{Measuring Supersymmetry}},
  \href{https://doi.org/10.1140/epjc/s10052-008-0548-z}{\emph{Eur. Phys. J. C}
  {\bfseries 54} (2008) 617} [\href{https://arxiv.org/abs/0709.3985}{{\ttfamily
  0709.3985}}].

\bibitem{Lafaye:2009vr}
R.~Lafaye, T.~Plehn, M.~Rauch, D.~Zerwas and M.~Duhrssen, \emph{{Measuring the
  Higgs Sector}},
  \href{https://doi.org/10.1088/1126-6708/2009/08/009}{\emph{JHEP} {\bfseries
  08} (2009) 009} [\href{https://arxiv.org/abs/0904.3866}{{\ttfamily
  0904.3866}}].

\bibitem{Klute:2012pu}
M.~Klute, R.~Lafaye, T.~Plehn, M.~Rauch and D.~Zerwas, \emph{{Measuring Higgs
  Couplings from LHC Data}},
  \href{https://doi.org/10.1103/PhysRevLett.109.101801}{\emph{Phys. Rev. Lett.}
  {\bfseries 109} (2012) 101801}
  [\href{https://arxiv.org/abs/1205.2699}{{\ttfamily 1205.2699}}].

\bibitem{Corbett:2015ksa}
T.~Corbett, O.~J.~P. Eboli, D.~Goncalves, J.~Gonzalez-Fraile, T.~Plehn and
  M.~Rauch, \emph{{The Higgs Legacy of the LHC Run I}},
  \href{https://doi.org/10.1007/JHEP08(2015)156}{\emph{JHEP} {\bfseries 08}
  (2015) 156} [\href{https://arxiv.org/abs/1505.05516}{{\ttfamily
  1505.05516}}].

\bibitem{Butter:2016cvz}
A.~Butter, O.~J.~P. \'Eboli, J.~Gonzalez-Fraile, M.~Gonzalez-Garcia, T.~Plehn
  and M.~Rauch, \emph{{The Gauge-Higgs Legacy of the LHC Run I}},
  \href{https://doi.org/10.1007/JHEP07(2016)152}{\emph{JHEP} {\bfseries 07}
  (2016) 152} [\href{https://arxiv.org/abs/1604.03105}{{\ttfamily
  1604.03105}}].

\bibitem{Hocker:2001xe}
A.~Hocker, H.~Lacker, S.~Laplace and F.~Le~Diberder, \emph{{A New approach to a
  global fit of the CKM matrix}},
  \href{https://doi.org/10.1007/s100520100729}{\emph{Eur. Phys. J. C}
  {\bfseries 21} (2001) 225}
  [\href{https://arxiv.org/abs/hep-ph/0104062}{{\ttfamily hep-ph/0104062}}].

\bibitem{Aaboud:2018mst}
{\scshape ATLAS} collaboration, \emph{{Study of the rare decays of $B^0_s$ and
  $B^0$ mesons into muon pairs using data collected during 2015 and 2016 with
  the ATLAS detector}},
  \href{https://doi.org/10.1007/JHEP04(2019)098}{\emph{JHEP} {\bfseries 04}
  (2019) 098} [\href{https://arxiv.org/abs/1812.03017}{{\ttfamily
  1812.03017}}].

\bibitem{Sirunyan:2019xdu}
{\scshape CMS} collaboration, \emph{{Measurement of properties of
  B$^0_\mathrm{s}\to\mu^+\mu^-$ decays and search for B$^0\to\mu^+\mu^-$ with
  the CMS experiment}},
  \href{https://doi.org/10.1007/JHEP04(2020)188}{\emph{JHEP} {\bfseries 04}
  (2020) 188} [\href{https://arxiv.org/abs/1910.12127}{{\ttfamily
  1910.12127}}].

\bibitem{Aaij:2017vad}
{\scshape LHCb} collaboration, \emph{{Measurement of the $B^0_s\to\mu^+\mu^-$
  branching fraction and effective lifetime and search for $B^0\to\mu^+\mu^-$
  decays}}, \href{https://doi.org/10.1103/PhysRevLett.118.191801}{\emph{Phys.
  Rev. Lett.} {\bfseries 118} (2017) 191801}
  [\href{https://arxiv.org/abs/1703.05747}{{\ttfamily 1703.05747}}].

\bibitem{LHCb-CONF-2020-002}
{\scshape LHCb Collaboration} collaboration, \emph{{Combination of the ATLAS,
  CMS and LHCb results on the $B^0_{(s)} \to \mu^+ \mu^-$ decays}},  Tech. Rep.
  LHCb-CONF-2020-002. CERN-LHCb-CONF-2020-002, CERN, Geneva, Aug, 2020.

\bibitem{Aubert:2007my}
{\scshape BaBar} collaboration, \emph{{Measurement of the $B \to X_s \gamma$
  branching fraction and photon energy spectrum using the recoil method}},
  \href{https://doi.org/10.1103/PhysRevD.77.051103}{\emph{Phys. Rev. D}
  {\bfseries 77} (2008) 051103}
  [\href{https://arxiv.org/abs/0711.4889}{{\ttfamily 0711.4889}}].

\bibitem{Lees:2012ym}
{\scshape BaBar} collaboration, \emph{{Precision Measurement of the $B \to X_s
  \gamma$ Photon Energy Spectrum, Branching Fraction, and Direct CP Asymmetry
  $A_{CP}(B \to X_{s+d}\gamma)$}},
  \href{https://doi.org/10.1103/PhysRevLett.109.191801}{\emph{Phys. Rev. Lett.}
  {\bfseries 109} (2012) 191801}
  [\href{https://arxiv.org/abs/1207.2690}{{\ttfamily 1207.2690}}].

\bibitem{Lees:2012wg}
{\scshape BaBar} collaboration, \emph{{Exclusive Measurements of $b \to
  s\gamma$ Transition Rate and Photon Energy Spectrum}},
  \href{https://doi.org/10.1103/PhysRevD.86.052012}{\emph{Phys. Rev. D}
  {\bfseries 86} (2012) 052012}
  [\href{https://arxiv.org/abs/1207.2520}{{\ttfamily 1207.2520}}].

\bibitem{Limosani:2009qg}
{\scshape Belle} collaboration, \emph{{Measurement of Inclusive Radiative
  B-meson Decays with a Photon Energy Threshold of 1.7-GeV}},
  \href{https://doi.org/10.1103/PhysRevLett.103.241801}{\emph{Phys. Rev. Lett.}
  {\bfseries 103} (2009) 241801}
  [\href{https://arxiv.org/abs/0907.1384}{{\ttfamily 0907.1384}}].

\bibitem{Saito:2014das}
{\scshape Belle} collaboration, \emph{{Measurement of the $\bar{B} \rightarrow
  X_s \gamma$ Branching Fraction with a Sum of Exclusive Decays}},
  \href{https://doi.org/10.1103/PhysRevD.91.052004}{\emph{Phys. Rev. D}
  {\bfseries 91} (2015) 052004}
  [\href{https://arxiv.org/abs/1411.7198}{{\ttfamily 1411.7198}}].

\bibitem{Chen:2001fja}
{\scshape CLEO} collaboration, \emph{{Branching fraction and photon energy
  spectrum for $b \to s \gamma$}},
  \href{https://doi.org/10.1103/PhysRevLett.87.251807}{\emph{Phys. Rev. Lett.}
  {\bfseries 87} (2001) 251807}
  [\href{https://arxiv.org/abs/hep-ex/0108032}{{\ttfamily hep-ex/0108032}}].

\bibitem{Amhis:2019ckw}
{\scshape HFLAV} collaboration, \emph{{Averages of $b$-hadron, $c$-hadron, and
  $\tau$-lepton properties as of 2018}},
  \href{https://arxiv.org/abs/1909.12524}{{\ttfamily 1909.12524}}.

\bibitem{Misiak:2020vlo}
M.~Misiak, A.~Rehman and M.~Steinhauser, \emph{{Towards $ \overline{B}\to
  {X}_s\gamma $ at the NNLO in QCD without interpolation in m$_{c}$}},
  \href{https://doi.org/10.1007/JHEP06(2020)175}{\emph{JHEP} {\bfseries 06}
  (2020) 175} [\href{https://arxiv.org/abs/2002.01548}{{\ttfamily
  2002.01548}}].

\bibitem{Zhang:2017mls}
C.~Zhang, \emph{{Constraining $qqtt$ operators from four-top production: a case
  for enhanced EFT sensitivity}},
  \href{https://doi.org/10.1088/1674-1137/42/2/023104}{\emph{Chin. Phys. C}
  {\bfseries 42} (2018) 023104}
  [\href{https://arxiv.org/abs/1708.05928}{{\ttfamily 1708.05928}}].

\bibitem{Banelli:2020iau}
G.~Banelli, E.~Salvioni, J.~Serra, T.~Theil and A.~Weiler, \emph{{The Present
  and Future of Four Tops}},
  \href{https://arxiv.org/abs/2010.05915}{{\ttfamily 2010.05915}}.

\bibitem{Alte:2017pme}
S.~Alte, M.~K\"onig and W.~Shepherd, \emph{{Consistent Searches for SMEFT
  Effects in Non-Resonant Dijet Events}},
  \href{https://doi.org/10.1007/JHEP01(2018)094}{\emph{JHEP} {\bfseries 01}
  (2018) 094} [\href{https://arxiv.org/abs/1711.07484}{{\ttfamily
  1711.07484}}].

\bibitem{Ellis:2018gqa}
J.~Ellis, C.~W. Murphy, V.~Sanz and T.~You, \emph{{Updated Global SMEFT Fit to
  Higgs, Diboson and Electroweak Data}},
  \href{https://doi.org/10.1007/JHEP06(2018)146}{\emph{JHEP} {\bfseries 06}
  (2018) 146} [\href{https://arxiv.org/abs/1803.03252}{{\ttfamily
  1803.03252}}].

\bibitem{Hermann:2013kca}
T.~Hermann, M.~Misiak and M.~Steinhauser, \emph{{Three-loop QCD corrections to
  $B_s \to \mu^+ \mu^-$}},
  \href{https://doi.org/10.1007/JHEP12(2013)097}{\emph{JHEP} {\bfseries 12}
  (2013) 097} [\href{https://arxiv.org/abs/1311.1347}{{\ttfamily 1311.1347}}].

\bibitem{Bobeth:2013tba}
C.~Bobeth, M.~Gorbahn and E.~Stamou, \emph{{Electroweak Corrections to $B_{s,d}
  \to \ell^+ \ell^-$}},
  \href{https://doi.org/10.1103/PhysRevD.89.034023}{\emph{Phys. Rev. D}
  {\bfseries 89} (2014) 034023}
  [\href{https://arxiv.org/abs/1311.1348}{{\ttfamily 1311.1348}}].

\bibitem{Bazavov:2017lyh}
A.~Bazavov et~al., \emph{{$B$- and $D$-meson leptonic decay constants from
  four-flavor lattice QCD}},
  \href{https://doi.org/10.1103/PhysRevD.98.074512}{\emph{Phys. Rev. D}
  {\bfseries 98} (2018) 074512}
  [\href{https://arxiv.org/abs/1712.09262}{{\ttfamily 1712.09262}}].

\bibitem{Bona:2006ah}
{\scshape UTfit} collaboration, \emph{{The Unitarity Triangle Fit in the
  Standard Model and Hadronic Parameters from Lattice QCD: A Reappraisal after
  the Measurements of $\Delta m_s$ and $\mathcal{B}(B \to \tau
  \bar{\nu}_\tau$)}},
  \href{https://doi.org/10.1088/1126-6708/2006/10/081}{\emph{JHEP} {\bfseries
  10} (2006) 081} [\href{https://arxiv.org/abs/hep-ph/0606167}{{\ttfamily
  hep-ph/0606167}}].

\bibitem{Ali:1991fy}
A.~Ali and T.~Mannel, \emph{{Exclusive rare B decays in the heavy quark
  limit}}, \href{https://doi.org/10.1016/0370-2693(91)90376-2}{\emph{Phys.
  Lett. B} {\bfseries 264} (1991) 447}.

\bibitem{Gunawardana:2019gep}
A.~Gunawardana and G.~Paz, \emph{{Reevaluating uncertainties in $\overline{B}
  \to X_{s}\gamma$ decay}},
  \href{https://doi.org/10.1007/JHEP11(2019)141}{\emph{JHEP} {\bfseries 11}
  (2019) 141} [\href{https://arxiv.org/abs/1908.02812}{{\ttfamily
  1908.02812}}].

\bibitem{Dimou:2012un}
M.~Dimou, J.~Lyon and R.~Zwicky, \emph{{Exclusive Chromomagnetism in
  heavy-to-light FCNCs}},
  \href{https://doi.org/10.1103/PhysRevD.87.074008}{\emph{Phys. Rev. D}
  {\bfseries 87} (2013) 074008}
  [\href{https://arxiv.org/abs/1212.2242}{{\ttfamily 1212.2242}}].

\bibitem{Lyon:2013gba}
J.~Lyon and R.~Zwicky, \emph{{Isospin asymmetries in
  $B\to(K^*,\rho)\gamma/l^+l^-$ and $B\to Kl^+l^-$ in and beyond the standard
  model}}, \href{https://doi.org/10.1103/PhysRevD.88.094004}{\emph{Phys. Rev.
  D} {\bfseries 88} (2013) 094004}
  [\href{https://arxiv.org/abs/1305.4797}{{\ttfamily 1305.4797}}].

\bibitem{Beneke:2004dp}
M.~Beneke, T.~Feldmann and D.~Seidel, \emph{{Exclusive radiative and
  electroweak $b \to d$ and $b \to s$ penguin decays at NLO}},
  \href{https://doi.org/10.1140/epjc/s2005-02181-5}{\emph{Eur. Phys. J. C}
  {\bfseries 41} (2005) 173}
  [\href{https://arxiv.org/abs/hep-ph/0412400}{{\ttfamily hep-ph/0412400}}].

\bibitem{Khodjamirian:2010vf}
A.~Khodjamirian, T.~Mannel, A.~Pivovarov and Y.-M. Wang, \emph{{Charm-loop
  effect in $B \to K^{(*)} \ell^{+} \ell^{-}$ and $B\to K^*\gamma$}},
  \href{https://doi.org/10.1007/JHEP09(2010)089}{\emph{JHEP} {\bfseries 09}
  (2010) 089} [\href{https://arxiv.org/abs/1006.4945}{{\ttfamily 1006.4945}}].

\bibitem{Khodjamirian:2012rm}
A.~Khodjamirian, T.~Mannel and Y.~Wang, \emph{{$B \to K \ell^{+}\ell^{-}$ decay
  at large hadronic recoil}},
  \href{https://doi.org/10.1007/JHEP02(2013)010}{\emph{JHEP} {\bfseries 02}
  (2013) 010} [\href{https://arxiv.org/abs/1211.0234}{{\ttfamily 1211.0234}}].

\end{thebibliography}\endgroup


\end{document}